\documentclass[english,prb,preprint]{revtex4-1}
\usepackage[T1]{fontenc}
\usepackage[latin9]{inputenc}
\usepackage{color}
\usepackage{units}
\usepackage{textcomp}
\usepackage{amsmath}
\usepackage{graphicx}

\makeatletter

\providecommand{\tabularnewline}{\\}

\@ifundefined{showcaptionsetup}{}{%
 \PassOptionsToPackage{caption=false}{subfig}}
\usepackage{subfig}
\makeatother

\usepackage{babel}
\begin{document}

\title{Electron correlation effects and two-photon absorption in diamond-shaped
graphene quantum dots}

\author{Tista Basak}

\affiliation{Mukesh Patel School of Technology Management and Engineering, NMIMS
University, Mumbai 400056, India}
\email{Tista.Basak@nmims.edu}

\author{Tushima Basak }

\affiliation{Department of Physics, Mithibai College, Mumbai 400056, India}
\email{Tushima.Basak@mithibai.ac.in}

\author{Alok Shukla}

\affiliation{Department of Physics, School of Engineering and Applied Sciences,
Bennett University, Plot No. 8-{}-11, TechZone II, Greater Noida 201310,
Uttar Pradesh, India }

\affiliation{Permanent Address: Department of Physics, Indian Institute of Technology
Bombay, Powai, Mumbai 400076, India}
\email{shukla@phy.iitb.ac.in}

\begin{abstract}
In quasi-1D $\pi$-conjugated polymers such as \emph{trans}-polyacetylene
and polyenes, electron correlation effects determine the ``reversed''
excited state ordering in which the lowest two-photon $2A_{g}$ state
lies below the lowest one-photon $1B_{u}$ state. In this work, we
present conclusive theoretical evidence of reversed excited state
ordering in fairly 2D $\pi$-conjugated systems, namely, diamond-shaped
graphene quantum dots (DQDs). Our electron correlated calculations
show that DQDs begin to exhibit reversed excited ordering with increasing
size, in disagreement with independent-particle picture. This signals
the onset of strong correlation effects which renders them nonluminescent.
Further, we calculate and analyze the two-photon absorption (TPA)
spectra as well as photoinduced absorption (PA) spectra of these systems
and find excellent agreement with the available experimental results.
Our investigations demonstrate that unlike a strictly 1D system like\emph{
trans}-polyacetylene, the non-linear and excited state absorptions
in DQDs are highly intricate, with several even parity states responsible
for strong absorptions. Our results could play an important role in
the design of graphene-based non-linear optical devices.
\end{abstract}

\maketitle

\section{Introduction}

The intricate role played by electron correlations in describing the
photophysics of carbon-based systems exhibiting $sp^{2}$ hybridization
has been an active area of research \cite{sumit-springer-chapter,barford-book}.
The extensive focus of earlier experimental \cite{Wang838_Experimental_Correlation,Gabor1367}
and theoretical studies \cite{SONY2005316,Sony_PhysRevB.75.155208,Tista_PRB92,Brus_doi:10.1021/ar500175h}
was to determine the influence of electron correlations on one-photon
allowed optical states. However, as per electric dipole selection
rules in centrosymmetric systems, the optical states accessible by
one-photon excitations are distinct from those obtained by two-photon
excitations. Existing literature have provided exclusive evidence
of the fact that one-photon allowed optical states are chiefly due
to excitations comprising of one electron and one hole, while two-electron
two-hole excitations can have a major contribution in the description
of two-photon states \cite{McWilliams_PhysRevB.43.9777,Guo_PhysRevB.49.10102,Dixit-polyenes-PhysRevB.43.6781}.
This signifies that the influence of electron correlation effects
is more complicated in two-photon allowed states, as compared to the
one-photon allowed optical states. Earlier studies on quasi-1D $\pi$-electron
material \emph{trans}-polyacetylene, and its oligomers, polyenes,
have shown that the inclusion of electron correlation effects is responsible
for the occurrence of the lowest energy two-photon \textcolor{black}{$2$$A_{g}$
state}\textcolor{red}{{} }\textcolor{black}{below the lowest energy
one-photon 1$B_{u}$ state, in contrast to that expected within the
one-electron Hückel and mean-field Hartree-Fock theories \cite{Christensen_polyenes_jp8060202,Dixit-polyenes-PhysRevB.43.6781}.
}However, such ``reversed'' excited state ordering as well as significant
contribution of two-electron two-hole excitations in the description
of two-photon states is not expected in fairly 2D and 3D systems,
as the strength of electron correlations decreases with increase in
the dimension of the system considered.

The study of non-linear optical properties of graphene and its fragments,
in its own right, forms an extremely important field of research,
with a fairly large number of experimental,\cite{Hendry_Exp_NLO_Graphene_PhysRevLett.105.097401,Hasan_graphene_SA_ADMA:ADMA200901122,Sun_SA_graphene_doi:10.1021/nn901703e,Demetriou:16_Exp_graphene,Lim_exp_graphene_nlo,Lim_exp_graphene_nlo_ADMA:ADMA200803616,Cheng_exp_graphene_nlo_:13,Dragonman_Frequency_Multiplication_graphene_exp_doi:10.1063/1.3483872,Chen_TPA_graphene_exp_doi:10.1021/acs.jpcc.5b03819,Dean_SHG_PhysRevB.82.125411,CHU_CPL_Kerr_Effect,Zhenyu_Sun_oxy_func_grp}
and theoretical \cite{Yoneda2009278,Brinkley_theory_graphene_0953-8984-28-36-365001,Mikhail_plasmon_Theory_graphene_PhysRevB.84.045432,Yang_TPA_graphene_theory_doi:10.1021/nl200587h,Cox_quant_chem_cal,Yamijala,Mikhail_TPA_Theory_graphene_PhysRevB.93.085403,Feng_oe_analy_exp}
studies dedicated to the subject. An earlier study by our group on
polyaromatic hydrocarbons revealed that the strength of electron correlations
becomes dominant as the symmetry of molecules decreases from $D_{6h}$
to $D_{2h}$ point group.\cite{Aryanpour_electron} This has motivated
us to consider graphene quantum dots possessing $D_{2h}$ point group.
Our previous investigation on fairly 2D systems, namely diamond-shaped
graphene quantum dots (DQDs), demonstrated that electron correlations
are important in determining their linear optical properties.\cite{Tista_PRB92}
In addition, an earlier study from Kaxiras and co-workers,\cite{dqd-kaxiras-prb-2010}
and from our own group,\cite{Tista_PRB93} predicted that suitably
aligned external electric field can induce energy-level shifts of
spin-ordered edge states, giving rise to phase transition between
the well-defined magnetic states of DQDs. Given the profound influence
of an external electric-field on the electronic, magnetic, and linear
optical properties of DQDs, it is but natural to explore the nonlinear
response of DQDs to incident optical radiation. This has motivated
us not just to compute the nonlinear optical spectra of these structures,
but also to explore the influence of electron correlations on them,
which, to the best of our knowledge has not been done till date.  

\textcolor{black}{In this work, we report a computational study of
two-photon absorption spectra of DQDs, which provides exclusive evidence
of strong electron correlations resulting in the aforesaid reversed
excited state ordering.} Like polyenes, DQDs are also centro-symmetric
systems, but with a higher point group symmetry $D_{2h}$. Thus, its
even parity states belonging to irreducible representations (irreps)
$A_{g}$ and $B_{1g}$ are invisible in linear spectroscopy, which
can only detect its odd-parity excited states belonging to irreps
$B_{2u}$ and $B_{3u}$ \cite{Tista_PRB92}. In order to obtain a
better understanding of the even parity states, and non-linear optical
response of DQDs, we compute their two-photon absorption (TPA), and
photoinduced absorption (PA) spectra, at the independent-particle
(tight-binding), and configuration interaction (CI) levels. The CI
results demonstrate that the DQDs exhibit reversed excited state ordering
like polyenes, with increasing size, in complete disagreement with
the independent-particle predictions. This highlights the presence
of strong electron correlations in the DQDs. For the case of DQD-16
(DQD consisting of 16 carbon atoms), we compare its calculated TPA
and PA spectra with the experiments for its hydrogen passivated analog
pyrene \cite{SALVI1983206}, and obtain excellent agreement. We also
calculate and analyze the TPA and PA spectra of DQD-30 (DQD consisting
of 30 carbon atoms), and PA spectrum of DQD-48 (DQD consisting of
48 carbon atoms), for which, at present, no experimental or theoretical
results exist. Our studies reveal that, due to structural anisotropy,
the nonlinear optical response of DQDs is much more complicated as
compared to that of 1D systems. A further extension of this work will
be to analyze the implication of electron correlation effects in extended
2D carbon-based system in the thermodynamic limit, i.e., graphene.
According to the Hückel tight-binding $\pi$-electron theory, 2D graphene
is a semimetal. However, several studies \cite{Elias_NaturePhysics}
have proposed that the inclusion of electron-electron interactions
is reponsible for considerable modification of graphene's linear spectrum
leading to an opening of energy gap, in contrast to the results obtained
from the single-particle theory. 

The organization of the rest of this paper is as follows. A brief
description of the computational methodology is presented in Sect.
II, followed by results and discussion in Sect. III. Finally, the
conclusions are given in Sect. IV.

\section{$ $Computational Methodology}

\begin{figure}
\subfloat[DQD-16]{\includegraphics[width=1.8cm]{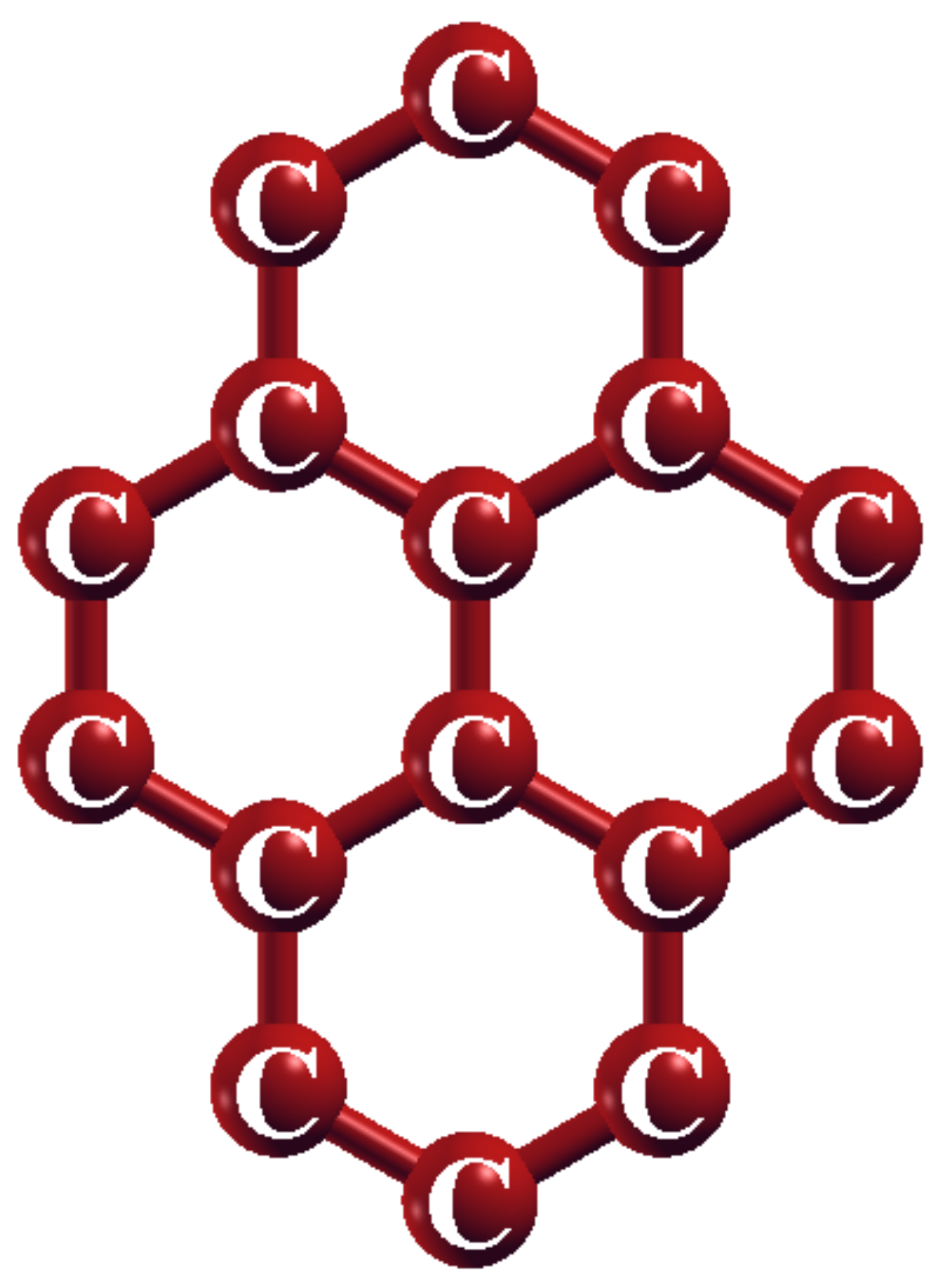}

}\subfloat[DQD-30]{

\includegraphics[width=2cm]{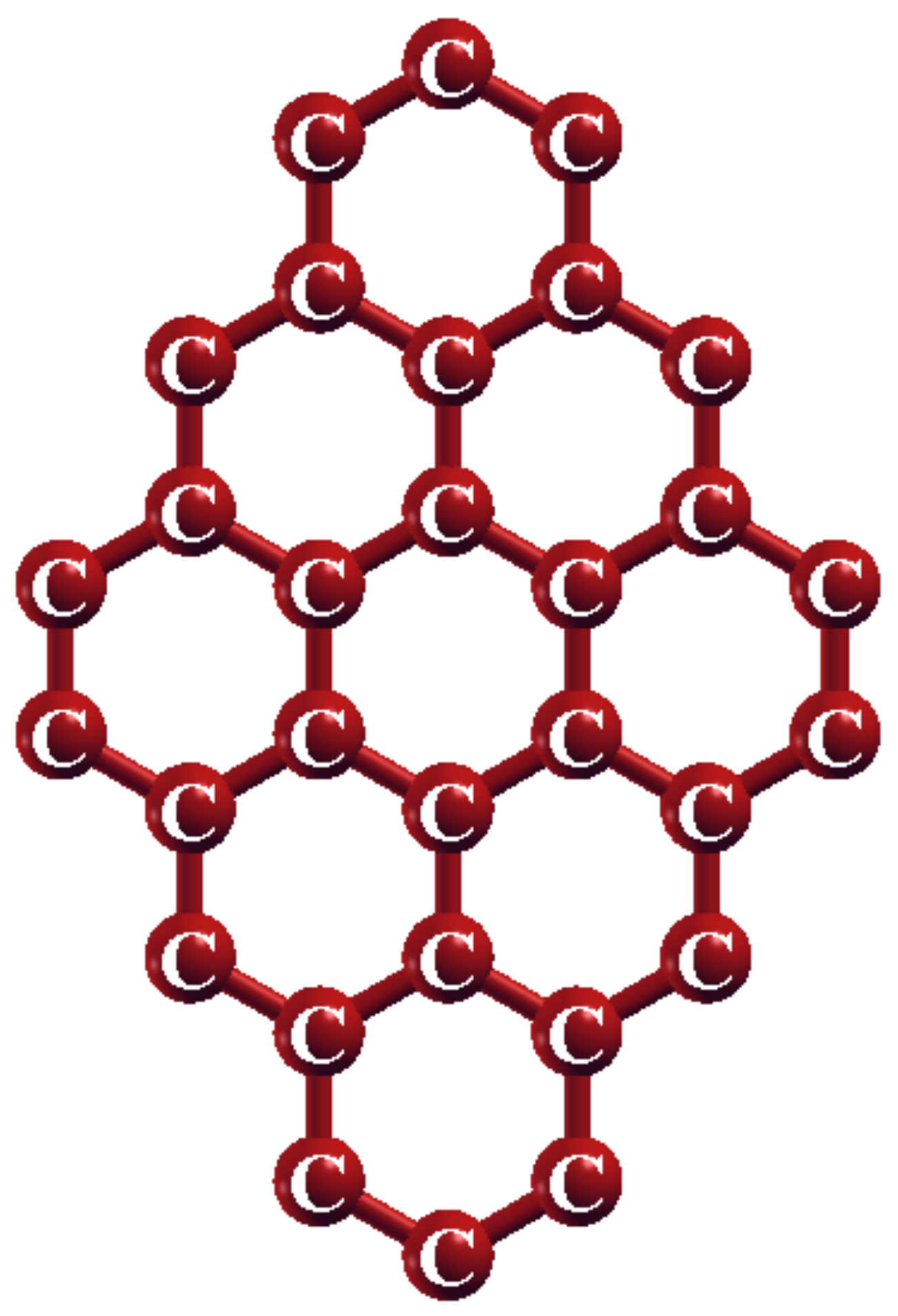}

}\subfloat[DQD-48]{

\includegraphics[width=2.5cm]{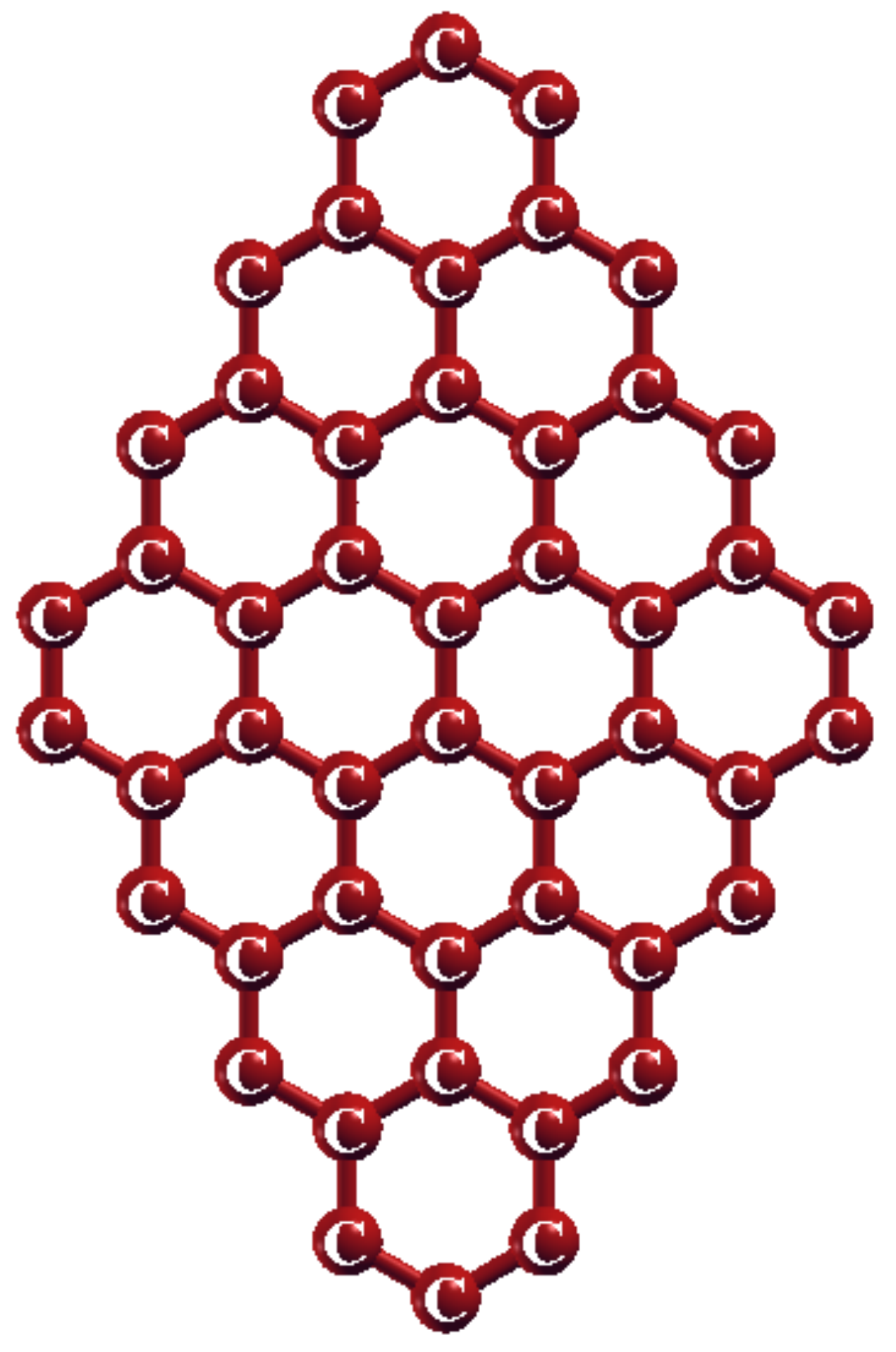}

}\caption{Geometric \label{fig:Geometric-configuration-of}configuration of
DQDs considered in this work: (a) DQD-16 (b) DQD-30 (c) DQD-48.}
\end{figure}

\textcolor{black}{The geometric configurations of different DQDs conside}red
in this work are given in fig.\ref{fig:Geometric-configuration-of}.
The DQDs are assumed to lie in the $xy$ plane, with the $y$-axis
along the longer diagonal, and the $x$-axis along the shorter one.
We have chosen uniform carbon-carbon bond lengths of 1.40 \AA, and
bond angles of 120\textdegree, so that each DQD has $D_{2h}$ symmetry.
The symmetries of the relevant one-photon and two-photon excited states
are $B_{2u}$, $B_{3u}$ and $A_{g}$, $B_{1g}$, respectively, as
predicted by the electric-dipole selection rules. 

For performing computations, we employed Pariser-Parr-Pople (PPP)
model Hamiltonian \cite{ppp-pople,ppp-pariser-parr} 

\begin{align}
H & \mbox{\ensuremath{=}\ensuremath{-}\ensuremath{\sum_{<i,j>,\sigma}t_{0}\left(c_{i\sigma}^{\dagger}c_{j\sigma}+c_{j\sigma}^{\dagger}c_{i\sigma}\right)}\ensuremath{+}\ensuremath{U\sum_{i}n_{i\uparrow}n_{i\downarrow}}}\nonumber \\
\mbox{\mbox{\mbox{}}} & \mbox{+\ensuremath{\sum_{i<j}V_{ij}(n_{i}-1)(n_{j}-1),}}\label{eq:ppp}
\end{align}
where $c_{i\sigma}^{\dagger}($c$_{i\sigma})$ creates (annihilates)
an electron of spin $\sigma$, in the $p_{z}$ orbital located on
the $i$-th carbon atom, while $n$$_{i}=\sum_{\sigma}c_{i\sigma}^{\dagger}c_{i\sigma}$
denotes the total number of electrons on the atom. The first term
in Eq. \ref{eq:ppp} denotes the hopping between $i$-th and $j$-th
atoms which are nearest neighbors, with $t_{0}=2.4$ eV, for the nearest
neighbor distance 1.40 \AA. The second and the third terms in Eq.
\ref{eq:ppp} represent the electron repulsion, with parameters $U$,
and $V_{ij}$, denoting the on-site, and the long-range Coulomb interactions,
respectively. The distance-dependence of $V_{ij}$ is assumed as per
Ohno relationship \cite{Theor.chim.act.2Ohno}, modified to include
the screening effects \cite{PhysRevB.55.1497Chandross}

\begin{equation}
V_{ij}=U/\kappa_{i,j}(1+0.6117R_{i,j}^{2})^{\nicefrac{1}{2}},\label{eq:ohno}
\end{equation}

where $\kappa_{i,j}$ is the dielectric constant of the system, included
to take into account the screening effects, and $R_{i,j}$ is the
distance (in \AA) between the $i$th and $j$th carbon atoms. In
the present set of calculations we have used the ``screened parameters''
of Chandross and Mazumdar \cite{PhysRevB.55.1497Chandross} with $U=8.0$
eV, $\kappa_{i,j}=2.0(i\neq j)$, and $\kappa_{i,i}=1.0$. We have
used these Coulomb parameters, coupled with the hopping value $t_{0}=2.4$
eV, extensively in the past for conjugated polymers \cite{PhysRevB.65.125204Shukla65,PhysRevB.69.165218Shukla69,PhysRevB.71.165204Priya_t0,:/content/aip/journal/jcp/131/1/10.1063/1.3159670Priyaanthracene,doi:10.1021/jp408535u,himanshu-triplet,Sony_PhysRevB.75.155208},
polyaromatic hydrocarbons \cite{Aryanpour_Subgap,Aryanpour_electron},
and graphene quantum dots \cite{Tista_PRB92,Tista_PRB93}. In case
of a charge-neutral DQD consisting of $N$ carbon atoms (DQD-$N$,
with $N$ always even), each carbon atom contributes one $\pi$ electron,
and the $p_{z}$ orbital of each of the carbon atom is part of the
basis set. Thus, for each DQD-$N$ treated within the PPP model, we
have $N$ basis functions, and $N$ electrons, making it a half-filled
system. Our calculations are initiated at the restricted Hartree-Fock
(RHF) level using a code developed in our group \cite{Sony2010821}.
The mean-field approximation at the RHF level results in $N/2$ doubly-occupied
orbitals filled in accordance with the aufbau principle, and $N/2$
unoccupied molecular orbitals (MOs). Pictorial representations of
few of the frontier molecular orbitals contributing significantly
to the non-linear properties for DQD-16, DQD-30 and DQD-48 and the
Hartree-Fock wave-function are given in Figs. S1--S4 of the Supporting
Information\cite{Supp_Info}. Thereafter, the PPP Hamiltonian is transformed
from the site representation to the MO representation, and used to
perform configuration interaction (CI) calculations to include electron-correlation
effects. The electron-correlations at the CI level are incorporated
by considering virtual excitations from the occupied to unoccupied
orbitals (see Fig. S5 of Supporting Information\cite{Supp_Info},
for a pictorial representation), leading to configuration mixing.
The CI calculations are performed at the quadruple configuration interaction
(QCI), or the multi-reference singles-doubles configuration interaction
(MRSDCI), level depending on the size of the system. The QCI method
was employed for the $A_{g}$ and the $B_{2u}$ states of DQD-16,
while the MRSDCI approach was adopted for the $B_{3u}$ and the $B_{1g}$
symmetries of DQD-16. Further, the MRSDCI methodology was considered
for optimizing all the symmetries of the larger systems DQD-30 and
-48. The increase in the number of molecular orbitals, with increase
in the size of DQD, leads to an enormous increase in the dimension
of the CI matrix. Hence, the frozen orbital approximation was implemented
for DQD-48, in order to make the computations feasible. According
to this approximation, the lowest two occupied orbitals were frozen
and highest two virtual orbitals were deleted so as to retain the
particle-hole symmetry \cite{Tista_PRB92}. The many-particle wave
functions thus obtained are used to compute the transition electric
dipole moments between various states, which in turn are employed
to calculate their TPA or PA spectra. In particular, the TPA spectrum
is computed as the imaginary part of the third-order non-linear susceptibility
$\chi_{ijkl}^{(3)}\mathrm{(\omega,\omega,\omega,-\omega)}$ ($i,j,k,$
and $l$ denote Cartesian directions) using the sum-over-states (SOS)
formula of Orr and Ward \cite{Orr_Ward}. Henceforth, we adopt the
shorthand notation $TPA_{ijkl}=\mbox{Im}(\chi_{ijkl}^{(3)}(\omega,\omega,\omega,-\omega)$),
and in this work we present results on $TPA_{xxxx}$, $TPA_{yyyy}$,
and $TPA_{xyxy}$, for DQD-16, and DQD-30. This PPP-CI computational
methodology has been extensively utilized in our earlier works on
optical properties of $\pi$-conjugated systems \cite{PhysRevB.65.125204Shukla65,PhysRevB.69.165218Shukla69,Sony_PhysRevB.75.155208,PhysRevB.71.165204Priya_t0,:/content/aip/journal/jcp/131/1/10.1063/1.3159670Priyaanthracene,doi:10.1021/jp408535u,himanshu-triplet,Tista_PRB92},
including their TPA spectra \cite{shukla-pdpa-tpa,Aryanpour_electron}. 

The states with $A_{g}$ symmetry give rise to two-photon resonances
in the $TPA_{xxxx}$ and $TPA_{yyyy}$ spectra for these DQDs. Similarly,
the two-photon resonances in the $TPA_{xyxy}$ spectrum corresponds
to states having $B_{1g}$ symmetry. In addition, the PA spectrum
with respect to the first optically excited $1B_{2u}$ state has been
computed for DQD-48, for which CI level TPA calculations were too
time consuming. The PA spectrum for DQD-16 and DQD-30 has also been
computed. It is observed that the salient features of the non-linear
optical properties obtained from the PA spectrum for DQD-16 and DQD-30
are qualitatively in excellent agreement with those derived from their
TPA spectra, validating our approach for the prediction of the non-linear
characteristics of DQD-48 from its PA spectrum. 

\section{Results and Discussion}

In this section, we discuss the excited state ordering in the DQDs,
and compare it with that observed in the polyenes. Furthermore, striking
characteristics of PA, as well as TPA, spectra for the DQDs of different
sizes are analyzed based upon the Hückel (tight-binding), and the
CI level calculations.

\subsection{\emph{Excited State Ordering}\textmd{\normalsize{} }}

For comparison with polyenes, we will refer to the one/two-photon
states of DQDs generically as $B_{u}$ / $A_{g}$ - type states. For
accurate determination of excited state ordering in DQDs, large-scale
CI calculations were performed for the first two lowest energy two-photon
states ($1A_{g}$ ground state and $2A_{g}$), and first one-photon
($1B_{u}$) state. The dimensions of CI matrices considered for the
computation of excited state ordering are given in the table S1 of
the Supporting Information \cite{Supp_Info}.\textcolor{blue}{{} }

\begin{figure}

\includegraphics[scale=0.6]{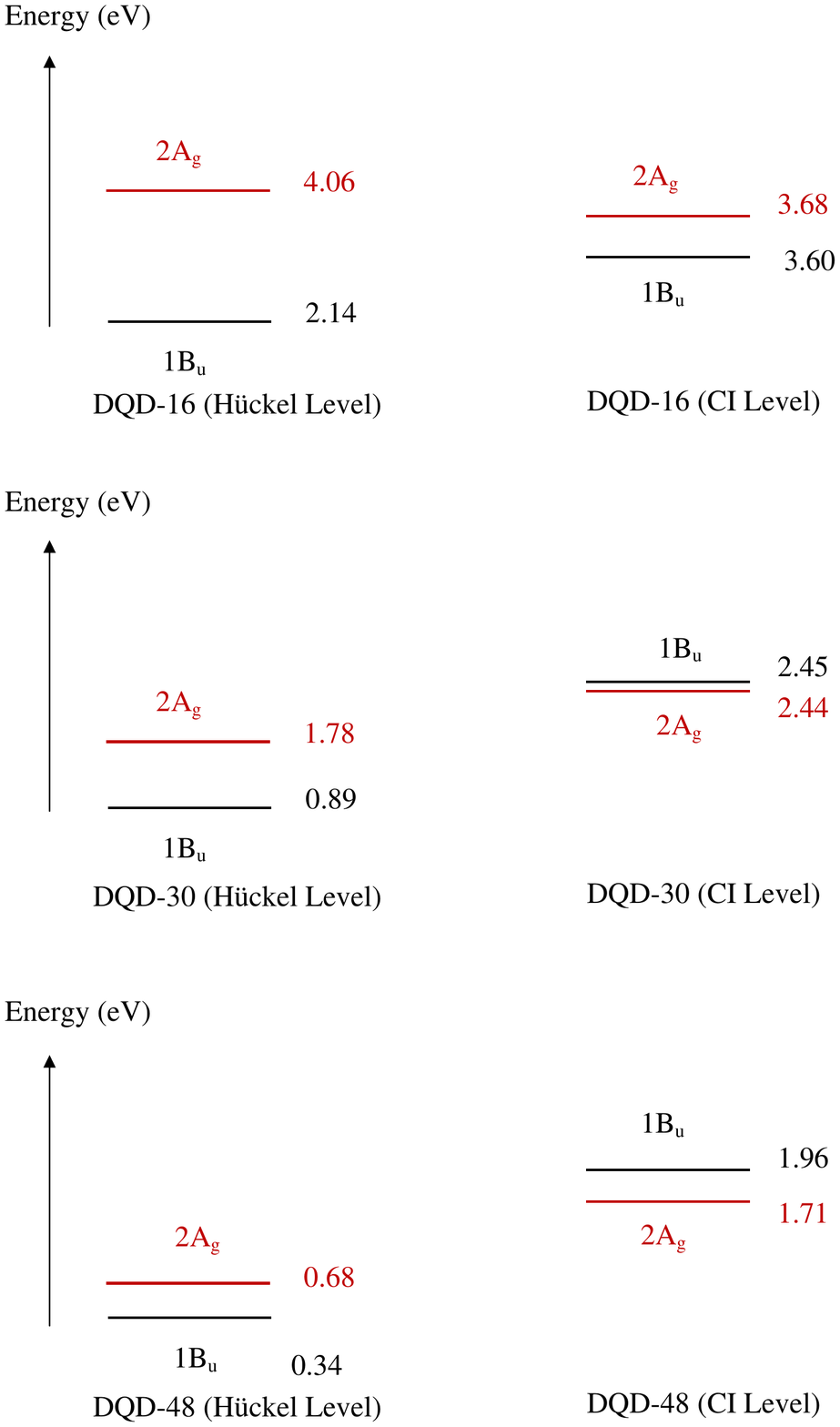}\caption{Schematic diagram representing the energies \label{fig:Schematic-diagram-representing energies of 1Bu}of
$1B_{u}$ and $2A_{g}$ states of DQD-16, DQD-30 and DQD-48, respectively,
at the Hückel and the CI levels.}

\end{figure}

Table \ref{tab:Computed-energies-of} presents the computed excitation
energies of the $2A_{g}$ and $1B_{u}$ states of DQD-16, DQD-30 and
DQD-48, respectively, at the \textcolor{black}{Hückel} (tight-binding),
and CI levels. The computed energies of the $2A_{g}$ and $1B_{u}$
states at the full configuration interaction (FCI) level of 1-D $n$-polyene
($n$ = 4, 6 and 8 carbon atoms equal to the number of carbon atoms
along the longer diagonal of DQD-16, DQD-30 and DQD-48, respectively)
are also given in table \ref{tab:Computed-energies-of}, for a qualitative
comparison of the nature of excited state ordering between 1-D and
2-D systems. In addition, theoretical values of the energies of $1B_{u}$
state obtained by the hybrid Hartree-Fock/density functional theory
B3LYP method by Yumura \emph{et al}.\cite{Yumura_HOMO-LUMO_Band-gap}
are also reported for comparison with our results. Figure \ref{fig:Schematic-diagram-representing energies of 1Bu}
represents the energies of $1B_{u}$ and $2A_{g}$ states of DQD-16,
DQD-30 and DQD-48, respectively, at the Hückel and CI level.

\begin{table}
\begin{tabular}{ccc}
\hline 
DQD & $1B_{u}$ state (eV) & $2A_{g}$ state (eV)\tabularnewline
\cline{2-3} 
 & Hückel/B3LYP Theory\cite{Yumura_HOMO-LUMO_Band-gap} (Ref. \cite{Yumura_HOMO-LUMO_Band-gap}) & Hückel\tabularnewline
\hline 
DQD-16 & 2.14/$\sim3.8$ & 4.06\tabularnewline
DQD-30 & 0.89/ $\sim2.1$ & 1.78\tabularnewline
DQD-48 & 0.34/ $\sim1.1$ & 0.68\tabularnewline
\hline 
 & $1B_{u}$ state (eV) & $2A_{g}$ state (eV)\tabularnewline
\cline{2-3} 
 & PPP-CI & PPP-CI\tabularnewline
\hline 
DQD-16 & 3.60 & 3.68\tabularnewline
DQD-30 & 2.45 & 2.44\tabularnewline
DQD-48 & 1.96 & 1.71\tabularnewline
4-polyene & 6.08 & 4.83\tabularnewline
6-polyene & 5.02 & 3.90\tabularnewline
8-polyene & 4.43 & 3.32\tabularnewline
\hline 
\end{tabular}

\caption{Computed energies \label{tab:Computed-energies-of}of $1B_{u}$ and
$2A_{g}$ states of DQD-16, DQD-30 and DQD-48, respectively, at the
\textcolor{black}{Hückel} and the CI levels. The computed energies
of these states for 4-polyene, 6-polyene and 8-polyene are also given
at the CI level. Theoretical values of the energies of $1B_{u}$ state
obtained by Yumura \emph{et al.,} \cite{Yumura_HOMO-LUMO_Band-gap}using
the density functional theory (B3LYP approach), are also reported
for comparison. }
\end{table}

The computed CI energies of $1B_{u}$ state of DQD-16 and DQD-30 at
3.60 eV and 2.45 eV, respectively, are in excellent agreement with
the earlier experimental values of 3.69 eV \cite{BASURAY_pyrene_2006248}
and 2.55 eV \cite{CLAR_DQD_30_19783219} for pyrene and dibenzo{[}bc,kl{]}coronene
(hydrogen passivated counterpart of DQD-30), respectively, marking
the high precision of our calculated results. The accuracy of the
results obtained by the B3LYP method\cite{Yumura_HOMO-LUMO_Band-gap}
is more than that obtained at the Hückel level, but less than those
obtained at the CI level. Because of the alternant nature of the systems
considered (DQDs and polyenes), H\"uckel model always predicts $1B_{u}$
state to be below $2A_{g}$. However, it is observed that in case
of 1-D polyenes, PPP-CI calculations correctly capture the expected
reversed excited state ordering with the $2A_{g}$ state below $1B_{u}$.
By contrast, the PPP-CI calculations for DQD-16 predict that the $2A_{g}$
state lies very slightly above the $1B_{u}$ state, while for DQD-30
it is almost degenerate with the $1B_{u}$ state. However, for DQD-48,
the $2A_{g}$ state is lowered significantly below the $1B_{u}$ state.
Thus, we note that, at the CI level, reversed excited state ordering
is realized, with increasing size of DQDs, in complete contradiction
with the Hückel model results. Furthermore, the wave function of the
$2$$A_{g}$ states of DQD-16, DQD-30 and DQD-48, are dominated by
the double excitation $|H\rightarrow L;H\rightarrow L\rangle$, where
$H$ and $L$ represent the highest occupied molecular orbital (HOMO),
and the lowest unoccupied molecular orbital (LUMO), respectively.\textcolor{red}{{}
}Both these facts clearly indicate the presence of strong electron
correlations in DQDs.

A plausible qualitative explanation of the lowering of $2A_{g}$ state
below $1B_{u}$, with increasing size of DQDs is as follows. DQD-16
has a 2D character with almost equal size in $x$ and $y$ directions
resulting in relatively weaker correlation effects, rendering $1B_{u}$
below $2A_{g}$. But with increasing size, DQDs approach a more 1D
character with their size along the $y$ direction becoming larger
compared to $x$ direction, leading to stronger correlation effects,
and hence, a polyene like excited state ordering with the $2A_{g}$
state below $1B_{u}$. 

\subsection{\emph{Nonlinear Optical Properties }}

In this section, we discuss the computed PA and TPA spectra of DQDs.
As per SOS formulas for PA and TPA spectra,\cite{Orr_Ward} the widths
of the intense peaks are dependent on the line-widths of various states,
which in turn depend upon carrier-lifetime, temperature, impurities
etc.\textcolor{red}{{} }The heights of different peaks, on the other
hand, are determined by the strengths of transition dipole operator
between various states of the system, as well as the magnitude of
the energy difference term in the denominator close to resonance.
For computing the photoinduced absorption and two-photon absorption
spectra, the maximum energy limit of the excited states was restricted
to 10 eV for CI calculations. The dimensions of CI matrices of $A_{g}$,
$B_{1g}$, $B_{2u}$ and $B_{3u}$ symmetries considered for the computation
of TPA and PA spectra of DQD-16, DQD-30 and DQD-48 are given in the
table S2 of the Supporting Information \cite{Supp_Info}. 

For the purpose of validating our CI (MRSDCI and QCI) approach, we
compare the calculated energies of even parity states giving rise
to peaks in the computed TPA and PA spectra of DQD-16, with the earlier
experimental results of its hydrogen-saturated counterpart pyrene
\cite{SALVI1983206} in Table \ref{tab:Computed-energies-of-1}. The
dominant wave-functions of the excited states contributing to these
peaks are also given in the table. It is observed that our computed
energies of the two-photon peaks of DQD-16 are in excellent quantitative
agreement with the experimental data for pyrene for a number of two-photon
states. The $A_{g}$ peaks contributing to the $TPA_{xxxx}$ component
are not observed in the experiment, as their intensities are much
less as compared to the $TPA_{yyyy}$ component. The computed $A_{g}$
state with energy 6.98 eV, giving rise to a two-photon peak at 3.49
eV, is the most intense peak in the $TPA_{yyyy}$ spectrum (fig. \ref{TPA-CI-DQDs}).
Wave function of this state is dominated by double excitations $|H-1\rightarrow L;H-1\rightarrow L\rangle+c.c$.
where the abbreviation ``c.c.'' represents coefficient of charge
conjugate of a given singly excited configuration while the sign (+/\textminus )
preceding \textquotedblleft c.c.\textquotedblright{} implies that
the two coefficients have (same/opposite) signs. This state is also
responsible for the second most intense peak ($IX_{y})$ in the computed
PA spectrum (fig. \ref{fig:Computed-photoinduced-absorption-DQD-16-DQD-30-DQD-48}).
The computed $A_{g}$ state with energy 9.04 eV gives rise to an intense
peak in the PA spectrum. However, it does not contribute to the TPA
spectrum. Hence, the energetic ordering as well as symmetry assignment
of our computed peaks are in perfect agreement with the experimental
data, demonstrating the extremely high precision of our correlated
electron approach. Thus, it lends credibility to our TPA/PA calculations
on DQD-30 and DQD-48, for which we could not find any experimental
data.\textcolor{blue}{{} }In addition, it is observed that both the
TPA and PA spectroscopies are efficient methods for detecting the
two-photon states of this system. \index{}

\begin{table}
\begin{tabular}{ccc}
\hline 
Peak & CI Results/ & Dominant Configurations\tabularnewline
 & Experiment\cite{SALVI1983206} & \tabularnewline
 & (eV) & \tabularnewline
\hline 
$A_{g}$ & 6.98/6.93 & $|H-1\rightarrow L;H-1\rightarrow L\rangle$$+c.c.(0.3367)$\tabularnewline
 &  & $|H\rightarrow L;H-1\rightarrow L+1\rangle$$(0.2960)$ \tabularnewline
$B_{1g}$  & 8.30/8.23 & $|H-6\rightarrow L+3\rangle$$+c.c.(0.2486)$\tabularnewline
 &  & $|H-7\rightarrow L\rangle$$+c.c.(0.1678)$\tabularnewline
$B_{1g}$  & 8.54/8.57 & $|H\rightarrow L+2;H-3\rightarrow L\rangle$ $c.c.$$(0.2619)$ \tabularnewline
 &  & $|H-4\rightarrow L;H-1\rightarrow L+2\rangle$ $c.c.$$(0.2559)$ \tabularnewline
$A_{g}$ & 9.04/9.08 & $|H-3\rightarrow L+4;H\rightarrow L+1\rangle$$-c.c.(0.2192)$\tabularnewline
 &  & $|H-5\rightarrow L;H\rightarrow L\rangle$$+c.c.(0.1885)$\tabularnewline
$A_{g}$ & 9.26/9.22 & $|H\rightarrow L;H-4\rightarrow L+2\rangle$ $-c.c.$$(0.2015)$\tabularnewline
 &  & $|H-1\rightarrow L+1;H-1\rightarrow L+1\rangle$$(0.1639)$\tabularnewline
\hline 
\end{tabular}

\caption{Comparison of the calculated excitation energies (in eV) of the even-parity
excited states of DQD-16, contributing to peaks in its TPA and PA
spectra,\textcolor{red}{{} }\label{tab:Computed-energies-of-1}with
the experimental data for pyrene \cite{SALVI1983206}. Note that the
two-photon resonances in the absorption spectra are located at half
the excitation energies. The calculations were performed at the CI
level (see text for details), and the dominant configurations contributing
to the many-particle wave functions of these states are also presented.}
\end{table}

\subsubsection{Photoinduced Absorption}

\begin{figure}

\includegraphics[scale=0.3]{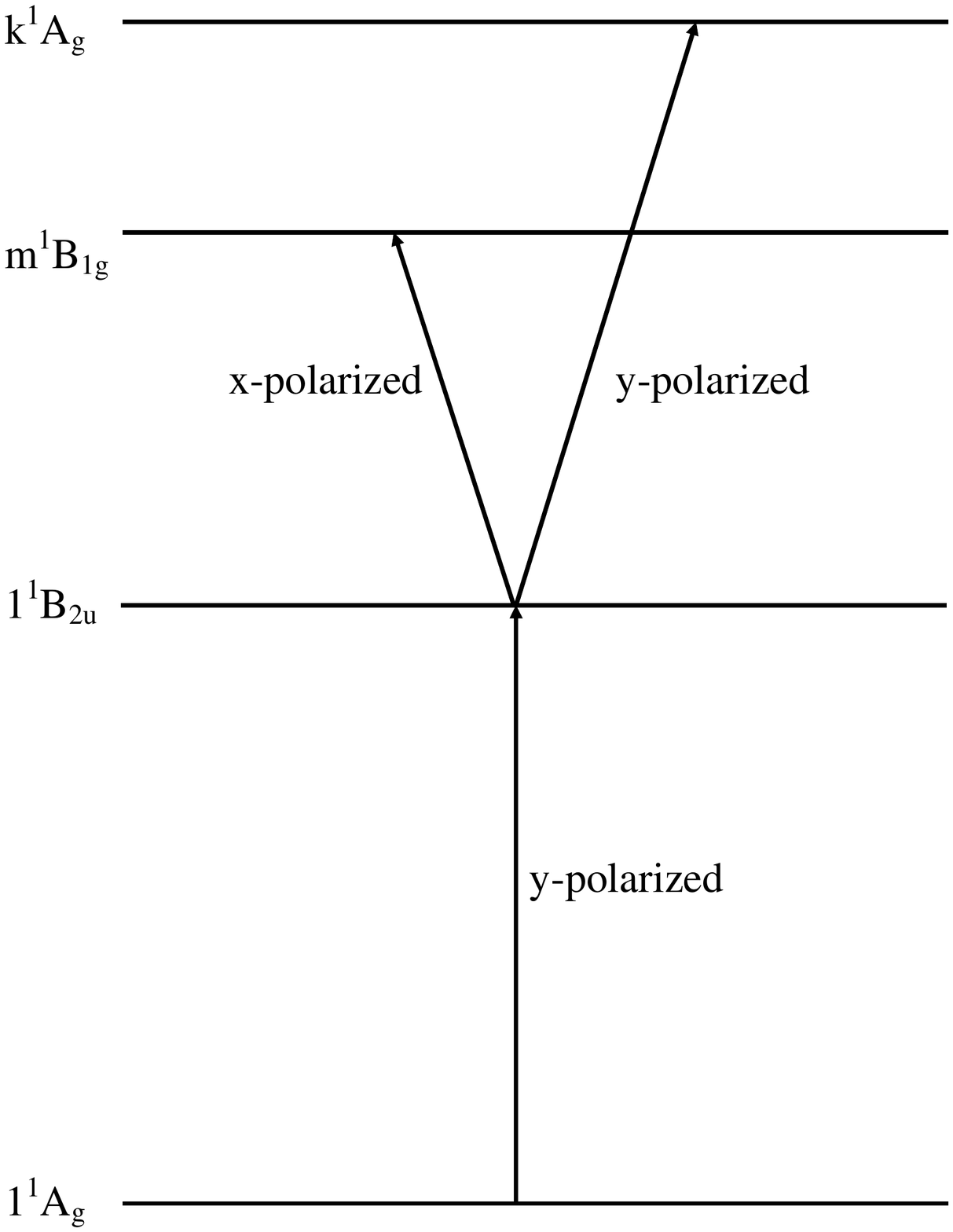}\caption{Schematic diagram describing the process of photoinduced absorption,
along with the states \label{fig:Schematic_PA}involved in it. An
arrow between two states denotes optical transition, with the polarization
direction of the absorbed photon indicated next to it.}

\end{figure}

The schematic diagram describing the concept of photoinduced absorption
(PA) is shown in Fig. \ref{fig:Schematic_PA} . The system initially
absorbs $y$-polarized photons from a pump laser, and gets excited
from the $1^{1}A\textrm{\ensuremath{_{g}}}$ ground state to the first
excited $1^{1}B\textrm{\ensuremath{_{2u}}}$ state. Thereafter, from
a probe laser, it absorbs another photon polarized either in $x$
or $y$ directions to make a transition from the $1^{1}B\textrm{\ensuremath{_{2u}}}$
state to higher energy $m^{1}B\textrm{\ensuremath{_{1g}}}$ or $k^{1}A\textrm{\ensuremath{_{g}}}$
excited states, respectively. The spectrum corresponding to the probe
absorption is measured, and is called the PA spectrum. Thus, the peak
locations in PA spectrum correspond to the excitation energies of
the even-parity states, measured with respect to the $1^{1}B\textrm{\ensuremath{_{2u}}}$
state. Hence, using oriented samples of DQDs, one can determine the
symmetries of these even parity states, based upon the polarization
of the absorbed probe photons.

\begin{figure}
\includegraphics[scale=0.4]{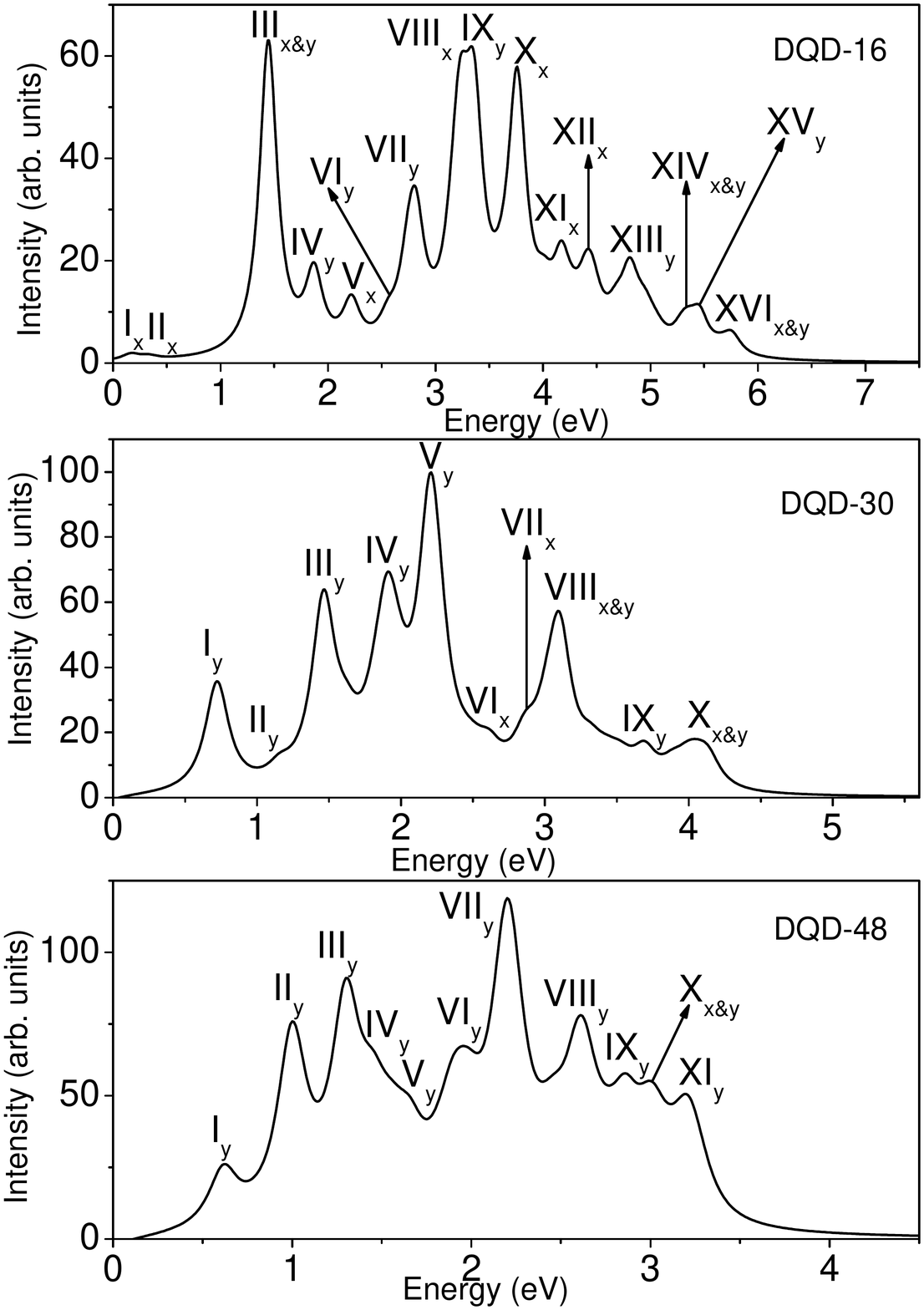}\caption{Computed photoinduced absorption spectrum of DQD-16, DQD-30 and DQD-48
with respect to the $1B_{2u}$ state. The subscripts on the peak labels
indicate the polarization direction of the absorbed photon. \label{fig:Computed-photoinduced-absorption-DQD-16-DQD-30-DQD-48}}
\end{figure}

In Fig. \ref{fig:Computed-photoinduced-absorption-DQD-16-DQD-30-DQD-48},
we present the computed PA spectrum with respect to the first optically
excited $1B_{2u}$ state for DQD-16, DQD-30 and DQD-48, respectively.
Detailed quantitative information about various excited states contributing
to their PA spectra is presented in the tables S9-S11 of the Supporting
Information \cite{Supp_Info}. Next, we summarize the important features
of the PA spectra of the DQDs, computed at the CI level: (i) with
the increasing size of the DQDs, the PA spectrum gets red-shifted,
(ii) the $2A_{g}$ state is not observed in the PA spectra of DQD-16,
DQD-30 and DQD-48 due to its weak dipole coupling to the $1B_{2u}$
state, and also because of its proximity to it, (iii) there are several
high intensity peaks contributing to the PA spectra. For DQD-16, peaks
$III_{x\&y}$, $VII_{y}$, $VIII_{x}$, $IX_{y}$ and $X_{x}$ make
intense contribution to its PA spectrum. Of these, the most intense
peak is $III_{x\&y}$ which corresponds to almost degenerate higher
energy even parity states, $4A_{g}$ and $4B_{1g},$ having excitation
energies 5.06 and 5.07 eV, respectively. The double excitation $|H\rightarrow L;H\rightarrow L\rangle$,
and the single excitations $|H-1\rightarrow L+3\rangle-c.c$ make
dominant contributions to the many-body wave functions of these states.
In case of DQD-30, the peaks $III_{y}$, $IV_{y}$, $V_{y}$ and $VIII_{x\&y}$
are quite intense, with the $V_{y}$ peak carrying the maximum intensity.
This peak corresponds to the $10A_{g}$ state, with the excitation
energy 4.81 eV, and the double excitation $|H-1\rightarrow L+1;H\rightarrow L\rangle$
contributing mainly to its wave function. Similarly, for DQD-48, there
are several peaks, namely, $II_{y}$, $III_{y}$, $VII_{y}$, and
$VIII_{y}$ with strong intensity. Of these, the most intense peak
is $VII_{y}$, corresponding to the $16A_{g}$ state with excitation
energy 4.19 eV, and single excitations $|H\rightarrow L+8\rangle-c.c.$
contributing dominantly to its wave function. Hence, it is observed
that several states of $A_{g}$ and $B_{1g}$ symmetries, having energies
much greater than the $2A_{g}$ state, give rise to strong resonances
in the PA spectra of these DQDs. It is to be noted that most of the
peaks become $y$-polarized with increase in size of DQD. This is
because greater system size in the $y$ direction for larger DQDs
leads to increased dominance of $y$ components of the transition
dipole matrix elements, over $x$ components. With the increasing
size of the DQDs, the PA spectrum overall becomes more intense, due
to the larger number of $\pi$-electrons in the system. Also noteworthy
is that doubly-excited configurations make large contributions to
the wave functions of the excited states giving rise to the peaks
in the PA spectra, signifying the importance of electron-correlation
effects in the description of even-parity states. 

\subsubsection{Two-photon absorption}

\begin{figure}
\includegraphics[scale=0.4]{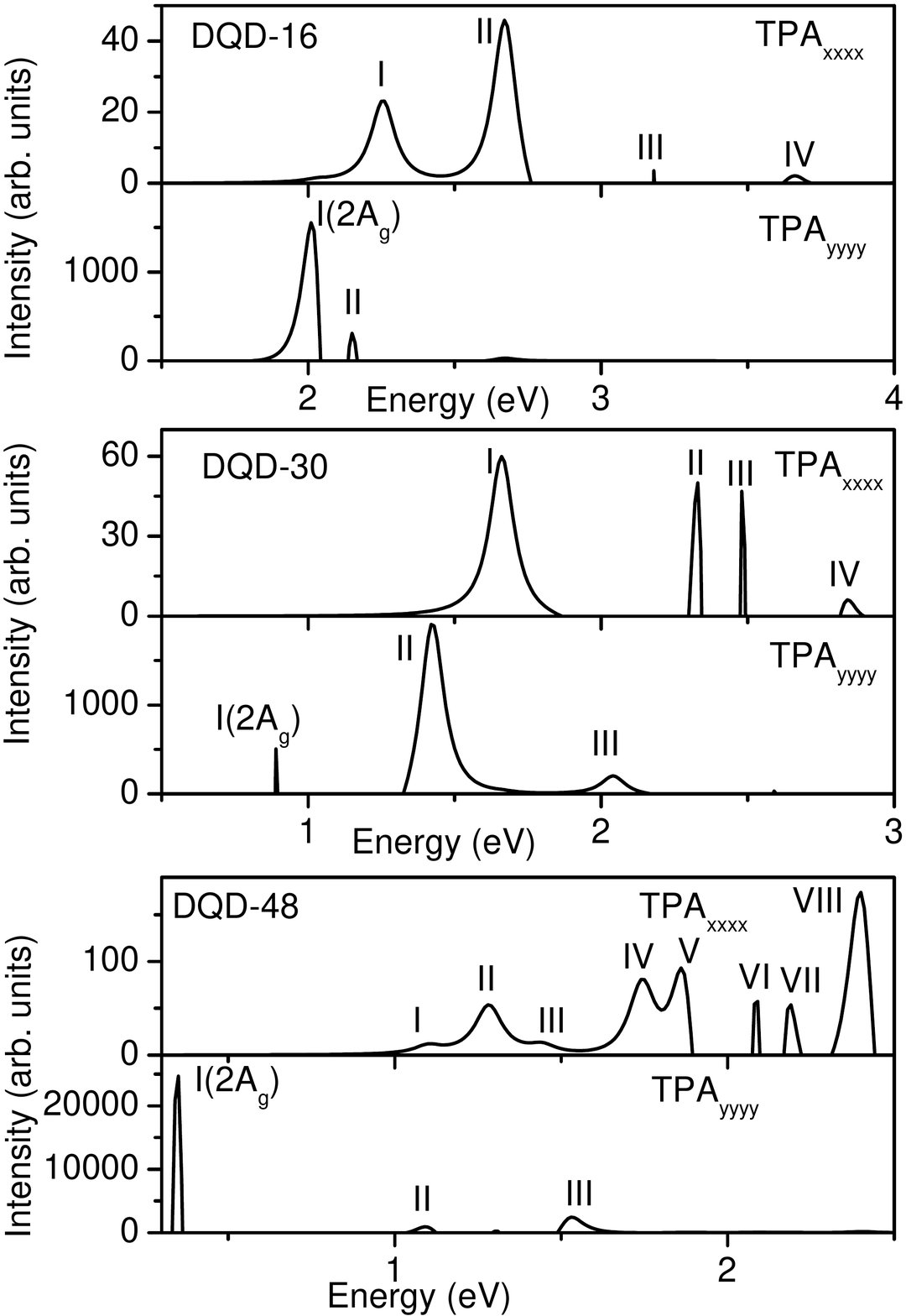}\caption{Computed $TPA_{xxxx}$ and $TPA_{yyyy}$ \label{fig:TPA_PPP}spectra
of DQD-16, DQD-30 and DQD-48, at the \textcolor{black}{Hückel} level.}
\end{figure}

Fig. \ref{fig:TPA_PPP} represents the two-photon absorption spectra
$TPA_{xxxx}$ and $TPA_{yyyy}$ of DQD-16, DQD-30, and DQD-48 at the
\textcolor{black}{one-electron (Hückel model)} level. The following
salient features are observed in  the TPA spectra: (i) The TPA spectrum
is red-shifted with increasing size of DQDs, analogous to the behavior
exhibited by the PA spectrum, (ii) the $TPA_{xxxx}$ component of
the spectrum is significantly different as compared to the $TPA_{yyyy}$
component, both qualitatively, and quantitatively, (iii) the resonant
intensities of the $TPA_{yyyy}$ component are much larger compared
to those of the $TPA_{xxxx}$ component. The reason behind this is
that $y$ components of the transition dipole matrix elements are
much larger than $x$ components, because of the larger system size
in the $y$ direction, (iv) resonant absorption in $TPA_{yyyy}$ occurs
at lower energies as compared to the $TPA_{xxxx}$ spectrum. The lowest
energy two-photon peak corresponding to the \textcolor{black}{$2A_{g}$}
state occurs in the $TPA_{yyyy}$ spectrum. For DQD-16, the \textcolor{black}{$2A_{g}$}
peak is due to degenerate single excitations $|H\rightarrow L+3\rangle$
and $|H-3\rightarrow L\rangle$, while in case of DQD-30 and DQD-48,
it corresponds to double excitation $|H\rightarrow L;H\rightarrow L\rangle$,\textcolor{red}{{}
}(v)\textcolor{black}{{} the $2A_{g}$ peak has the highest intensity}
for \textcolor{black}{DQD-16, and DQD-48,} while \textcolor{black}{the
most intense peak for DQD-30, in the }$TPA_{yyyy}$ \textcolor{black}{component,
corresponds to an $mA_{g}$ state (a unique state having a strong
dipole coupling with $1B_{u}$ state) located higher than the $2A_{g}$
state. The $mA_{g}$ peak of DQD-30 is due to degenerate single excitations
}$|H\rightarrow L+3\rangle$ and $|H-3\rightarrow L\rangle$, and
(vi) the higher energy peaks in the $TPA_{xxxx}$ component become
more intense, with increasing size of DQD. 

Now, we present the energies obtained by incorporating electron correlation
effects at single configuration interaction (SCI) level with the experimental
data for pyrene in Table \ref{tab:Computed-energies-SCI}. In SCI
approach, only one-electron one-hole, i.e., singly-excited configurations
with respect to the closed-shell RHF wave function are included. The
dominant configurations contributing to the SCI wave functions of
important states are given in Table \ref{tab:Computed-energies-SCI},
and it is observed that the energies obtained from single excitations
are in reasonable agreement with the experimental data. However, we
note that we obtained much better agreement with the experiments when
two-electron two-hole doubly-excited configurations were included
at the MRSDCI, and the QCI level (Table. \ref{tab:Computed-energies-of-1})
for DQD-16. Futhermore, as the electron correlation effects are getting
stronger with the increasing sizes of the DQDs, as is evident from
the relative 2$A_{g}-1B_{u}$ ordering, the SCI method will become
progressively worse for these systems as far as the description of
two-photon states is concerned. This implies that the two-electron
two-hole excitations are essential for quantitatively accurate description
of two-photon states. 

\begin{table}
\begin{tabular}{ccc}
\hline 
Peak & SCI Results/ & Dominant Configurations\tabularnewline
 & Experiment\cite{SALVI1983206} & \tabularnewline
 & (eV) & \tabularnewline
\hline 
$A_{g}$ & 6.46/6.93 & $|H-1\rightarrow L+4\rangle$$+c.c.(0.6995)$\tabularnewline
$B_{1g}$  & 8.02/8.23 & $|H-7\rightarrow L\rangle$$+c.c.(0.5698)$\tabularnewline
$B_{1g}$  & 8.76/8.57 & $|H-5\rightarrow L+4\rangle$$-$ $c.c.$$(0.5738)$ \tabularnewline
 &  & $|H-3\rightarrow L+6\rangle$ $+c.c.$$(0.4034)$ \tabularnewline
$A_{g}$ & 9.01/9.08 & $|H-1\rightarrow L+7\rangle$$+c.c.(0.6438)$\tabularnewline
$A_{g}$ & 9.60/9.22 & $|H-6\rightarrow L+4\rangle$ $-c.c.$$(0.6347)$\tabularnewline
\hline 
\end{tabular}\caption{Comparison of the calculated excitation energies (in eV) of the even-parity
excited states of DQD-16, contributing to peaks in its TPA spectra,
\label{tab:Computed-energies-SCI}with the experimental data for pyrene
\cite{SALVI1983206}. Note that the two-photon resonances in the absorption
spectra are located at half the excitation energies. The calculations
were performed at the SCI level, and the dominant configurations contributing
to the one electron-hole excitations are also presented.}
\end{table}

\begin{figure}
\includegraphics[scale=0.4]{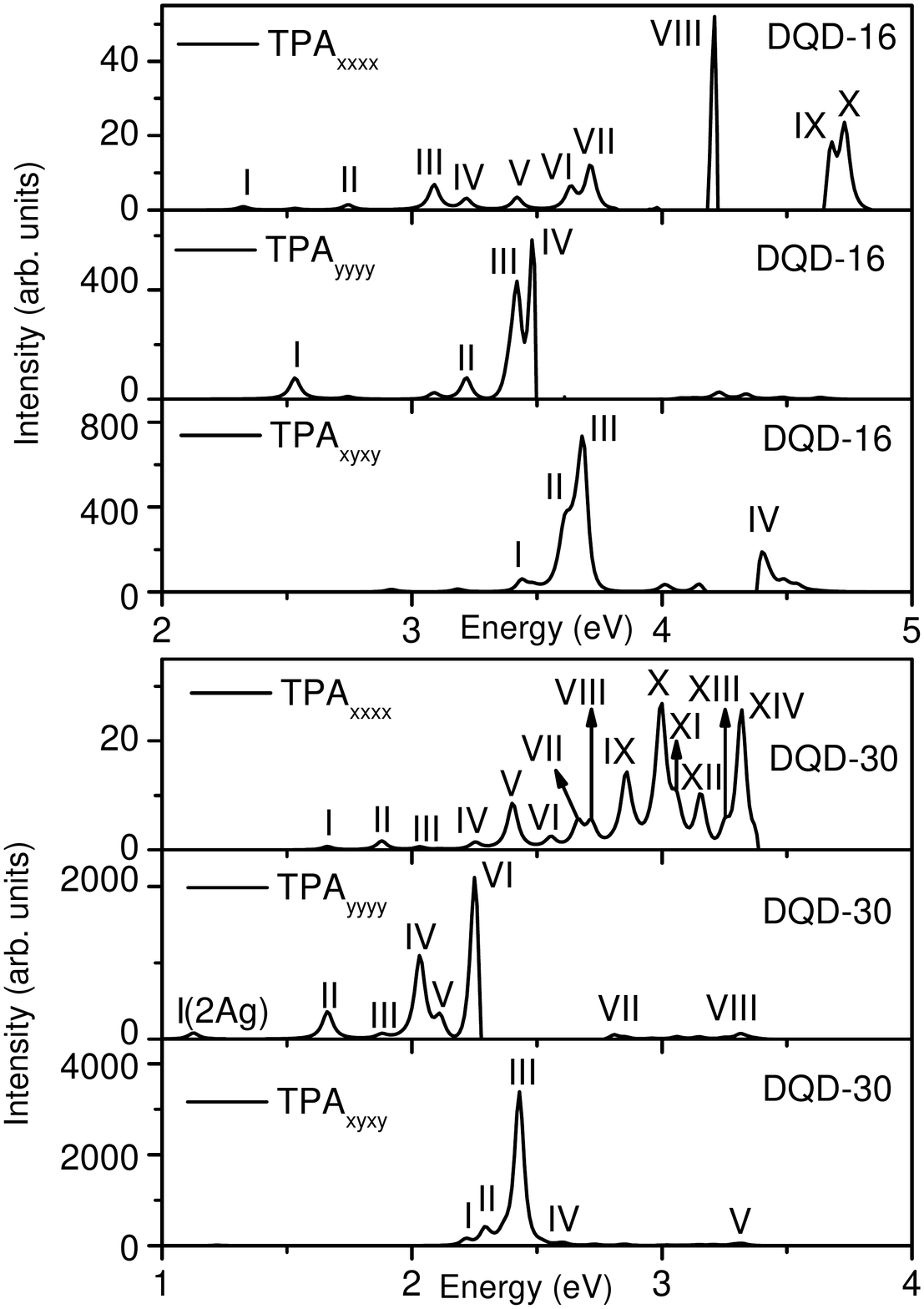}\caption{Calculated MRSDCI level $TPA_{xxxx}$, $TPA_{yyyy}$ and $TPA_{xyxy}$
\label{TPA-CI-DQDs}spectra of DQD-16 and DQD-30. }

\end{figure}

Next, in fig. \ref{TPA-CI-DQDs} we present the two-photon absorptions
$TPA_{xxxx}$, $TPA_{yyyy}$ and $TPA_{xyxy}$ of DQD-16 and DQD-30,
respectively, which include the electron-correlation effects at the
MRSDCI level. The detailed quantitative information of the various
excited states contributing to the TPA spectra of DQD-16 and DQD-30
are listed in the tables S3-S8 of the Supporting Information \cite{Supp_Info}.
Next, we summarize the important features of the TPA/PA spectra of
the DQDs, computed at the CI level. 

(i) With increasing size of DQDs, the TPA spectrum gets red-shifted,
in agreement to the predictions of independent electron level, indicative
of band formation. Similarly, the intensity and energy profiles of
the peaks are significantly different for $TPA_{xxxx}$, $TPA_{yyyy}$
and $TPA_{xyxy}$ components, due to the fact that different intermediate
states contribute to these spectra, (ii) in disagreement with H\"uckel
model results, for DQD-16, the contribution of $2$$A_{g}$ state
to non-linear susceptibilities is insignificant due to its weak dipole
coupling to the $B_{u}$ state. This is in complete accordance with
the PA results discussed earlier. For DQD-30, the $2$$A_{g}$ state
contributes a very weak peak in the $TPA_{yyyy}$ component of the
spectrum, but higher energy $A_{g}$ states have more intensity. Furthermore,
the wave function of the $2$$A_{g}$ states are mainly dominated
by the double excitation $|H\rightarrow L;H\rightarrow L\rangle$,
(iii) the most intense two-photon peaks for DQD-16 and DQD-30 appear
in the $TPA_{xyxy}$ component, and correspond to higher energy even
parity states with excitation energies 7.38 eV and 4.86 eV, respectively.
These states have $B_{1g}$ symmetry, and the double excitations $|H\rightarrow L+3;H\rightarrow L+2\rangle-c.c.$
and single excitation $|H-3\rightarrow L+2\rangle+c.c.$ contribute
mainly to their many-body wave functions.\textcolor{black}{{} }

Several years back Dixit \emph{et al.}, \cite{Dixit-polyenes-PhysRevB.43.6781}
in the context of\emph{ trans}-polyacetylene, had argued that the
maximum two-photon intensity is not because of the $2A_{g}$ state,
but due to a higher energy even parity state, which they named $mA_{g}$,
and which has a strong dipole coupling with the $1B_{u}$ state. Therefore,
such a state will give rise to intense peaks both in TPA as well as
in $1B_{u}$ PA spectra of the system. For other polymers, with more
complex structures due to side groups, several states were shown to
satisfy $mA_{g}$-like properties, \cite{shukla-pdpa-tpa,PhysRevB.71.165204Priya_t0,shukla-nlo-ppv}.\textcolor{red}{{}
}In case of DQDs, different TPA components (depending on the selection
rules) have either $A_{g}$ or $B_{1g}$ type two-photon peaks having
high intensity, a consequence of their more 2D structure. In order
to examine the presence of such generically called $mA_{g}$ states
in DQDs, we do a detailed comparison of their TPA and PA spectra.
In case of DQD-16, the states $4A_{g}$, $5A_{g}$, $8A_{g}$, $9A_{g}$,
$10B_{1g}$, $13A_{g}$, $13B_{1g}$, $24A_{g}$ and $38A_{g}$ give
rise to intense peaks in PA as well TPA spectra. In particular, the
most intense peak in the TPA spectrum is due to $13B_{1g}$ state
which also contributes to the strong peak $X_{x}$ peak in the PA
spectrum. In addition, the $13A_{g}$ state is not only responsible
for the most intense peak in the $TPA_{yyyy}$ spectrum, but also
gives rise to the second most intense $IX_{y}$ peak in the PA spectrum.
Similarly, the $24A_{g}$ state contributes to the highest intensity
peak along $TPA_{xxxx}$ component, and the strong $XIII_{y}$ peak
in the PA spectra. For DQD-30, the states $3A_{g}$, $4A_{g}$, $5A_{g}$,
$8A_{g}$, $10A_{g}$, $12B_{1g}$, $19A_{g}$, $28A_{g}$, $34A_{g}$,
$36A_{g}$ and $34B_{1g}$ lead to strong peaks both in the PA, and
the TPA spectra. The $10A_{g}$ state is responsible for the most
intense peak $V_{y}$ in the PA spectrum, and the fifth strong peak
in the $TPA_{xxxx}$ component. Hence, our studies demonstrate that
there are many states responsible for strong peaks in the TPA and
PA spectra, confirming the existence of several $mA_{g}$ states in
DQDs. In addition, our investigations reveal that unlike a strictly
1D system like\emph{ trans}-polyacetylene, the two-photon and photoinduced
absorptions in DQDs lead to enhanced nonlinear optical response, with
several even parity states contributing to strong absorptions. Based
upon the preceding observations for DQD-16 and DQD-30, one can say
with confidence that in case of DQD-48 also, the peaks contributing
to the PA spectra will also contribute to various components of its
TPA spectrum. This shows that TPA and PA spectroscopies are both efficient
approaches for probing the two-photon states of these systems.

\section{Conclusions}

In conclusion, our large scale correlated computations employing the
PPP Hamiltonian have demonstrated the dominant influence of electron
correlation and finite-size effects on the non-linear optical properties
of DQDs. Our investigations indicate that the presence of strong electron
correlation effects, incorporated within configuration interaction
(CI) method, gives rise to reversed excited state ordering in fairly
2D diamond-shaped graphene quantum dots, in contradiction to the results
obtained by non-interacting single particle theories. In addition,
the role played by electron correlation becomes more conspicuous with
increasing size of DQD. However, unlike a strictly 1D system like\emph{
trans}-polyacetylene, the non-linear and excited state absorptions
in DQDs lead to richer optical properties, with several even parity
states contributing to strong absorptions. Therefore, our investigation
reveals that both TPA and PA spectroscopies can be utilized to probe
the two-photon states of these systems, which are invisible in the
linear absorption. We hope that these results will lead to experimental
investigations of these systems, paving the way to rational design
of graphene-based nonlinear optical devices. 

\section*{Acknowledgements}

Work of Tista Basak was supported by DST-SERB,  India (Grant No. ECR/2016/000793),
while that of A.S. was supported by the DST-SERB, India grant SB/S2/CMP-066/2013.

\bibliographystyle{apsrev4-1}
\addcontentsline{toc}{section}{\refname}\bibliography{NLO}

\begin{thebibliography}{56}%
\makeatletter
\providecommand \@ifxundefined [1]{%
 \@ifx{#1\undefined}
}%
\providecommand \@ifnum [1]{%
 \ifnum #1\expandafter \@firstoftwo
 \else \expandafter \@secondoftwo
 \fi
}%
\providecommand \@ifx [1]{%
 \ifx #1\expandafter \@firstoftwo
 \else \expandafter \@secondoftwo
 \fi
}%
\providecommand \natexlab [1]{#1}%
\providecommand \enquote  [1]{``#1''}%
\providecommand \bibnamefont  [1]{#1}%
\providecommand \bibfnamefont [1]{#1}%
\providecommand \citenamefont [1]{#1}%
\providecommand \href@noop [0]{\@secondoftwo}%
\providecommand \href [0]{\begingroup \@sanitize@url \@href}%
\providecommand \@href[1]{\@@startlink{#1}\@@href}%
\providecommand \@@href[1]{\endgroup#1\@@endlink}%
\providecommand \@sanitize@url [0]{\catcode `\\12\catcode `\$12\catcode
  `\&12\catcode `\#12\catcode `\^12\catcode `\_12\catcode `\%12\relax}%
\providecommand \@@startlink[1]{}%
\providecommand \@@endlink[0]{}%
\providecommand \url  [0]{\begingroup\@sanitize@url \@url }%
\providecommand \@url [1]{\endgroup\@href {#1}{\urlprefix }}%
\providecommand \urlprefix  [0]{URL }%
\providecommand \Eprint [0]{\href }%
\providecommand \doibase [0]{http://dx.doi.org/}%
\providecommand \selectlanguage [0]{\@gobble}%
\providecommand \bibinfo  [0]{\@secondoftwo}%
\providecommand \bibfield  [0]{\@secondoftwo}%
\providecommand \translation [1]{[#1]}%
\providecommand \BibitemOpen [0]{}%
\providecommand \bibitemStop [0]{}%
\providecommand \bibitemNoStop [0]{.\EOS\space}%
\providecommand \EOS [0]{\spacefactor3000\relax}%
\providecommand \BibitemShut  [1]{\csname bibitem#1\endcsname}%
\let\auto@bib@innerbib\@empty
\bibitem [{\citenamefont {Baeriswyl}\ \emph {et~al.}(1992)\citenamefont
  {Baeriswyl}, \citenamefont {Campbell},\ and\ \citenamefont
  {Mazumdar}}]{sumit-springer-chapter}%
  \BibitemOpen
  \bibfield  {author} {\bibinfo {author} {\bibfnamefont {D.}~\bibnamefont
  {Baeriswyl}}, \bibinfo {author} {\bibfnamefont {D.~K.}\ \bibnamefont
  {Campbell}}, \ and\ \bibinfo {author} {\bibfnamefont {S.}~\bibnamefont
  {Mazumdar}},\ }\enquote {\bibinfo {title} {An overview of the theory of
  $\pi$-conjugated polymers},}\ in\ \href {\doibase
  10.1007/978-3-642-46729-5\_2} {\emph {\bibinfo {booktitle} {Conjugated
  Conducting Polymers. Springer Series in Solid-State Sciences}}},\ \bibinfo
  {editor} {edited by\ \bibinfo {editor} {\bibfnamefont {H.}~\bibnamefont
  {Kiess}}}\ (\bibinfo  {publisher} {Springer Berlin, Heidelberg},\ \bibinfo
  {year} {1992})\ pp.\ \bibinfo {pages} {7--133}\BibitemShut {NoStop}%
\bibitem [{\citenamefont {Barford}(2013)}]{barford-book}%
  \BibitemOpen
  \bibfield  {author} {\bibinfo {author} {\bibfnamefont {W.}~\bibnamefont
  {Barford}},\ }\href {\doibase 10.1093/acprof:oso/9780199677467.001.0001}
  {\emph {\bibinfo {title} {Electronic and Optical Properties of Conjugated
  Polymers}}},\ \bibinfo {edition} {2nd}\ ed.\ (\bibinfo  {publisher} {Oxford
  Univeristy Press},\ \bibinfo {year} {2013})\BibitemShut {NoStop}%
\bibitem [{\citenamefont {Wang}\ \emph {et~al.}(2005)\citenamefont {Wang},
  \citenamefont {Dukovic}, \citenamefont {Brus},\ and\ \citenamefont
  {Heinz}}]{Wang838_Experimental_Correlation}%
  \BibitemOpen
  \bibfield  {author} {\bibinfo {author} {\bibfnamefont {F.}~\bibnamefont
  {Wang}}, \bibinfo {author} {\bibfnamefont {G.}~\bibnamefont {Dukovic}},
  \bibinfo {author} {\bibfnamefont {L.~E.}\ \bibnamefont {Brus}}, \ and\
  \bibinfo {author} {\bibfnamefont {T.~F.}\ \bibnamefont {Heinz}},\ }\href
  {\doibase 10.1126/science.1110265} {\bibfield  {journal} {\bibinfo  {journal}
  {Science}\ }\textbf {\bibinfo {volume} {308}},\ \bibinfo {pages} {838}
  (\bibinfo {year} {2005})}\BibitemShut {NoStop}%
\bibitem [{\citenamefont {Gabor}\ \emph {et~al.}(2009)\citenamefont {Gabor},
  \citenamefont {Zhong}, \citenamefont {Bosnick}, \citenamefont {Park},\ and\
  \citenamefont {McEuen}}]{Gabor1367}%
  \BibitemOpen
  \bibfield  {author} {\bibinfo {author} {\bibfnamefont {N.~M.}\ \bibnamefont
  {Gabor}}, \bibinfo {author} {\bibfnamefont {Z.}~\bibnamefont {Zhong}},
  \bibinfo {author} {\bibfnamefont {K.}~\bibnamefont {Bosnick}}, \bibinfo
  {author} {\bibfnamefont {J.}~\bibnamefont {Park}}, \ and\ \bibinfo {author}
  {\bibfnamefont {P.~L.}\ \bibnamefont {McEuen}},\ }\href {\doibase
  10.1126/science.1176112} {\bibfield  {journal} {\bibinfo  {journal}
  {Science}\ }\textbf {\bibinfo {volume} {325}},\ \bibinfo {pages} {1367}
  (\bibinfo {year} {2009})}\BibitemShut {NoStop}%
\bibitem [{\citenamefont {Sony}\ and\ \citenamefont
  {Shukla}(2005{\natexlab{a}})}]{SONY2005316}%
  \BibitemOpen
  \bibfield  {author} {\bibinfo {author} {\bibfnamefont {P.}~\bibnamefont
  {Sony}}\ and\ \bibinfo {author} {\bibfnamefont {A.}~\bibnamefont {Shukla}},\
  }\href {\doibase http://dx.doi.org/10.1016/j.synthmet.2005.01.036} {\bibfield
   {journal} {\bibinfo  {journal} {Synthetic Metals}\ }\textbf {\bibinfo
  {volume} {155}},\ \bibinfo {pages} {316 } (\bibinfo {year}
  {2005}{\natexlab{a}})},\ \bibinfo {note} {proceedings of the Sixth
  International Topical Conference on Optical Probes of Conjugated Polymers and
  Biosystems, Bangalore-INDIA, January 4-8th, 2005}\BibitemShut {NoStop}%
\bibitem [{\citenamefont {Sony}\ and\ \citenamefont
  {Shukla}(2007)}]{Sony_PhysRevB.75.155208}%
  \BibitemOpen
  \bibfield  {author} {\bibinfo {author} {\bibfnamefont {P.}~\bibnamefont
  {Sony}}\ and\ \bibinfo {author} {\bibfnamefont {A.}~\bibnamefont {Shukla}},\
  }\href {\doibase 10.1103/PhysRevB.75.155208} {\bibfield  {journal} {\bibinfo
  {journal} {Phys. Rev. B}\ }\textbf {\bibinfo {volume} {75}},\ \bibinfo
  {pages} {155208} (\bibinfo {year} {2007})}\BibitemShut {NoStop}%
\bibitem [{\citenamefont {Basak}\ \emph {et~al.}(2015)\citenamefont {Basak},
  \citenamefont {Chakraborty},\ and\ \citenamefont {Shukla}}]{Tista_PRB92}%
  \BibitemOpen
  \bibfield  {author} {\bibinfo {author} {\bibfnamefont {T.}~\bibnamefont
  {Basak}}, \bibinfo {author} {\bibfnamefont {H.}~\bibnamefont {Chakraborty}},
  \ and\ \bibinfo {author} {\bibfnamefont {A.}~\bibnamefont {Shukla}},\ }\href
  {\doibase 10.1103/PhysRevB.92.205404} {\bibfield  {journal} {\bibinfo
  {journal} {Phys. Rev. B}\ }\textbf {\bibinfo {volume} {92}},\ \bibinfo
  {pages} {205404} (\bibinfo {year} {2015})}\BibitemShut {NoStop}%
\bibitem [{\citenamefont {Brus}(2014)}]{Brus_doi:10.1021/ar500175h}%
  \BibitemOpen
  \bibfield  {author} {\bibinfo {author} {\bibfnamefont {L.}~\bibnamefont
  {Brus}},\ }\href {\doibase 10.1021/ar500175h} {\bibfield  {journal} {\bibinfo
   {journal} {Accounts of Chemical Research}\ }\textbf {\bibinfo {volume}
  {47}},\ \bibinfo {pages} {2951} (\bibinfo {year} {2014})},\ \bibinfo {note}
  {pMID: 25120173}\BibitemShut {NoStop}%
\bibitem [{\citenamefont {McWilliams}\ \emph {et~al.}(1991)\citenamefont
  {McWilliams}, \citenamefont {Hayden},\ and\ \citenamefont
  {Soos}}]{McWilliams_PhysRevB.43.9777}%
  \BibitemOpen
  \bibfield  {author} {\bibinfo {author} {\bibfnamefont {P.~C.~M.}\
  \bibnamefont {McWilliams}}, \bibinfo {author} {\bibfnamefont {G.~W.}\
  \bibnamefont {Hayden}}, \ and\ \bibinfo {author} {\bibfnamefont {Z.~G.}\
  \bibnamefont {Soos}},\ }\href {\doibase 10.1103/PhysRevB.43.9777} {\bibfield
  {journal} {\bibinfo  {journal} {Phys. Rev. B}\ }\textbf {\bibinfo {volume}
  {43}},\ \bibinfo {pages} {9777} (\bibinfo {year} {1991})}\BibitemShut
  {NoStop}%
\bibitem [{\citenamefont {Guo}\ \emph {et~al.}(1994)\citenamefont {Guo},
  \citenamefont {Guo},\ and\ \citenamefont {Mazumdar}}]{Guo_PhysRevB.49.10102}%
  \BibitemOpen
  \bibfield  {author} {\bibinfo {author} {\bibfnamefont {F.}~\bibnamefont
  {Guo}}, \bibinfo {author} {\bibfnamefont {D.}~\bibnamefont {Guo}}, \ and\
  \bibinfo {author} {\bibfnamefont {S.}~\bibnamefont {Mazumdar}},\ }\href
  {\doibase 10.1103/PhysRevB.49.10102} {\bibfield  {journal} {\bibinfo
  {journal} {Phys. Rev. B}\ }\textbf {\bibinfo {volume} {49}},\ \bibinfo
  {pages} {10102} (\bibinfo {year} {1994})}\BibitemShut {NoStop}%
\bibitem [{\citenamefont {Dixit}\ \emph {et~al.}(1991)\citenamefont {Dixit},
  \citenamefont {Guo},\ and\ \citenamefont
  {Mazumdar}}]{Dixit-polyenes-PhysRevB.43.6781}%
  \BibitemOpen
  \bibfield  {author} {\bibinfo {author} {\bibfnamefont {S.~N.}\ \bibnamefont
  {Dixit}}, \bibinfo {author} {\bibfnamefont {D.}~\bibnamefont {Guo}}, \ and\
  \bibinfo {author} {\bibfnamefont {S.}~\bibnamefont {Mazumdar}},\ }\href
  {\doibase 10.1103/PhysRevB.43.6781} {\bibfield  {journal} {\bibinfo
  {journal} {Phys. Rev. B}\ }\textbf {\bibinfo {volume} {43}},\ \bibinfo
  {pages} {6781} (\bibinfo {year} {1991})}\BibitemShut {NoStop}%
\bibitem [{\citenamefont {Christensen}\ \emph {et~al.}(2008)\citenamefont
  {Christensen}, \citenamefont {Galinato}, \citenamefont {Chu}, \citenamefont
  {Howard}, \citenamefont {Broene},\ and\ \citenamefont
  {Frank}}]{Christensen_polyenes_jp8060202}%
  \BibitemOpen
  \bibfield  {author} {\bibinfo {author} {\bibfnamefont {R.~L.}\ \bibnamefont
  {Christensen}}, \bibinfo {author} {\bibfnamefont {M.~G.~I.}\ \bibnamefont
  {Galinato}}, \bibinfo {author} {\bibfnamefont {E.~F.}\ \bibnamefont {Chu}},
  \bibinfo {author} {\bibfnamefont {J.~N.}\ \bibnamefont {Howard}}, \bibinfo
  {author} {\bibfnamefont {R.~D.}\ \bibnamefont {Broene}}, \ and\ \bibinfo
  {author} {\bibfnamefont {H.~A.}\ \bibnamefont {Frank}},\ }\href {\doibase
  10.1021/jp8060202} {\bibfield  {journal} {\bibinfo  {journal} {The Journal of
  Physical Chemistry A}\ }\textbf {\bibinfo {volume} {112}},\ \bibinfo {pages}
  {12629} (\bibinfo {year} {2008})},\ \bibinfo {note} {pMID:
  19007144}\BibitemShut {NoStop}%
\bibitem [{\citenamefont {Hendry}\ \emph {et~al.}(2010)\citenamefont {Hendry},
  \citenamefont {Hale}, \citenamefont {Moger}, \citenamefont {Savchenko},\ and\
  \citenamefont {Mikhailov}}]{Hendry_Exp_NLO_Graphene_PhysRevLett.105.097401}%
  \BibitemOpen
  \bibfield  {author} {\bibinfo {author} {\bibfnamefont {E.}~\bibnamefont
  {Hendry}}, \bibinfo {author} {\bibfnamefont {P.~J.}\ \bibnamefont {Hale}},
  \bibinfo {author} {\bibfnamefont {J.}~\bibnamefont {Moger}}, \bibinfo
  {author} {\bibfnamefont {A.~K.}\ \bibnamefont {Savchenko}}, \ and\ \bibinfo
  {author} {\bibfnamefont {S.~A.}\ \bibnamefont {Mikhailov}},\ }\href {\doibase
  10.1103/PhysRevLett.105.097401} {\bibfield  {journal} {\bibinfo  {journal}
  {Phys. Rev. Lett.}\ }\textbf {\bibinfo {volume} {105}},\ \bibinfo {pages}
  {097401} (\bibinfo {year} {2010})}\BibitemShut {NoStop}%
\bibitem [{\citenamefont {Hasan}\ \emph {et~al.}(2009)\citenamefont {Hasan},
  \citenamefont {Sun}, \citenamefont {Wang}, \citenamefont {Bonaccorso},
  \citenamefont {Tan}, \citenamefont {Rozhin},\ and\ \citenamefont
  {Ferrari}}]{Hasan_graphene_SA_ADMA:ADMA200901122}%
  \BibitemOpen
  \bibfield  {author} {\bibinfo {author} {\bibfnamefont {T.}~\bibnamefont
  {Hasan}}, \bibinfo {author} {\bibfnamefont {Z.}~\bibnamefont {Sun}}, \bibinfo
  {author} {\bibfnamefont {F.}~\bibnamefont {Wang}}, \bibinfo {author}
  {\bibfnamefont {F.}~\bibnamefont {Bonaccorso}}, \bibinfo {author}
  {\bibfnamefont {P.~H.}\ \bibnamefont {Tan}}, \bibinfo {author} {\bibfnamefont
  {A.~G.}\ \bibnamefont {Rozhin}}, \ and\ \bibinfo {author} {\bibfnamefont
  {A.~C.}\ \bibnamefont {Ferrari}},\ }\href {\doibase 10.1002/adma.200901122}
  {\bibfield  {journal} {\bibinfo  {journal} {Advanced Materials}\ }\textbf
  {\bibinfo {volume} {21}},\ \bibinfo {pages} {3874} (\bibinfo {year}
  {2009})}\BibitemShut {NoStop}%
\bibitem [{\citenamefont {Sun}\ \emph {et~al.}(2010)\citenamefont {Sun},
  \citenamefont {Hasan}, \citenamefont {Torrisi}, \citenamefont {Popa},
  \citenamefont {Privitera}, \citenamefont {Wang}, \citenamefont {Bonaccorso},
  \citenamefont {Basko},\ and\ \citenamefont
  {Ferrari}}]{Sun_SA_graphene_doi:10.1021/nn901703e}%
  \BibitemOpen
  \bibfield  {author} {\bibinfo {author} {\bibfnamefont {Z.}~\bibnamefont
  {Sun}}, \bibinfo {author} {\bibfnamefont {T.}~\bibnamefont {Hasan}}, \bibinfo
  {author} {\bibfnamefont {F.}~\bibnamefont {Torrisi}}, \bibinfo {author}
  {\bibfnamefont {D.}~\bibnamefont {Popa}}, \bibinfo {author} {\bibfnamefont
  {G.}~\bibnamefont {Privitera}}, \bibinfo {author} {\bibfnamefont
  {F.}~\bibnamefont {Wang}}, \bibinfo {author} {\bibfnamefont {F.}~\bibnamefont
  {Bonaccorso}}, \bibinfo {author} {\bibfnamefont {D.~M.}\ \bibnamefont
  {Basko}}, \ and\ \bibinfo {author} {\bibfnamefont {A.~C.}\ \bibnamefont
  {Ferrari}},\ }\href {\doibase 10.1021/nn901703e} {\bibfield  {journal}
  {\bibinfo  {journal} {ACS Nano}\ }\textbf {\bibinfo {volume} {4}},\ \bibinfo
  {pages} {803} (\bibinfo {year} {2010})},\ \bibinfo {note} {pMID:
  20099874}\BibitemShut {NoStop}%
\bibitem [{\citenamefont {Demetriou}\ \emph {et~al.}(2016)\citenamefont
  {Demetriou}, \citenamefont {Bookey}, \citenamefont {Biancalana},
  \citenamefont {Abraham}, \citenamefont {Wang}, \citenamefont {Ji},\ and\
  \citenamefont {Kar}}]{Demetriou:16_Exp_graphene}%
  \BibitemOpen
  \bibfield  {author} {\bibinfo {author} {\bibfnamefont {G.}~\bibnamefont
  {Demetriou}}, \bibinfo {author} {\bibfnamefont {H.~T.}\ \bibnamefont
  {Bookey}}, \bibinfo {author} {\bibfnamefont {F.}~\bibnamefont {Biancalana}},
  \bibinfo {author} {\bibfnamefont {E.}~\bibnamefont {Abraham}}, \bibinfo
  {author} {\bibfnamefont {Y.}~\bibnamefont {Wang}}, \bibinfo {author}
  {\bibfnamefont {W.}~\bibnamefont {Ji}}, \ and\ \bibinfo {author}
  {\bibfnamefont {A.~K.}\ \bibnamefont {Kar}},\ }\href {\doibase
  10.1364/OE.24.013033} {\bibfield  {journal} {\bibinfo  {journal} {Opt.
  Express}\ }\textbf {\bibinfo {volume} {24}},\ \bibinfo {pages} {13033}
  (\bibinfo {year} {2016})}\BibitemShut {NoStop}%
\bibitem [{\citenamefont {Geok-Kieng}\ \emph {et~al.}(2011)\citenamefont
  {Geok-Kieng}, \citenamefont {Zhi-Li}, \citenamefont {Jenny}, \citenamefont
  {S.}, \citenamefont {Wee-Hao}, \citenamefont {Hong-Wee}, \citenamefont {H.},
  \citenamefont {H.},\ and\ \citenamefont {Lay-Lay}}]{Lim_exp_graphene_nlo}%
  \BibitemOpen
  \bibfield  {author} {\bibinfo {author} {\bibfnamefont {L.}~\bibnamefont
  {Geok-Kieng}}, \bibinfo {author} {\bibfnamefont {C.}~\bibnamefont {Zhi-Li}},
  \bibinfo {author} {\bibfnamefont {C.}~\bibnamefont {Jenny}}, \bibinfo
  {author} {\bibfnamefont {G.~R.~G.}\ \bibnamefont {S.}}, \bibinfo {author}
  {\bibfnamefont {N.}~\bibnamefont {Wee-Hao}}, \bibinfo {author} {\bibfnamefont
  {T.}~\bibnamefont {Hong-Wee}}, \bibinfo {author} {\bibfnamefont {F.~R.}\
  \bibnamefont {H.}}, \bibinfo {author} {\bibfnamefont {H.~P.~K.}\ \bibnamefont
  {H.}}, \ and\ \bibinfo {author} {\bibfnamefont {C.}~\bibnamefont {Lay-Lay}},\
  }\href {\doibase 10.1038/nphoton.2011.177} {\bibfield  {journal} {\bibinfo
  {journal} {Nat Photon}\ }\textbf {\bibinfo {volume} {5}},\ \bibinfo {pages}
  {554} (\bibinfo {year} {2011})}\BibitemShut {NoStop}%
\bibitem [{\citenamefont {Wang}\ \emph {et~al.}(2009)\citenamefont {Wang},
  \citenamefont {Hernandez}, \citenamefont {Lotya}, \citenamefont {Coleman},\
  and\ \citenamefont {Blau}}]{Lim_exp_graphene_nlo_ADMA:ADMA200803616}%
  \BibitemOpen
  \bibfield  {author} {\bibinfo {author} {\bibfnamefont {J.}~\bibnamefont
  {Wang}}, \bibinfo {author} {\bibfnamefont {Y.}~\bibnamefont {Hernandez}},
  \bibinfo {author} {\bibfnamefont {M.}~\bibnamefont {Lotya}}, \bibinfo
  {author} {\bibfnamefont {J.~N.}\ \bibnamefont {Coleman}}, \ and\ \bibinfo
  {author} {\bibfnamefont {W.~J.}\ \bibnamefont {Blau}},\ }\href {\doibase
  10.1002/adma.200803616} {\bibfield  {journal} {\bibinfo  {journal} {Advanced
  Materials}\ }\textbf {\bibinfo {volume} {21}},\ \bibinfo {pages} {2430}
  (\bibinfo {year} {2009})}\BibitemShut {NoStop}%
\bibitem [{\citenamefont {Cheng}\ \emph {et~al.}(2013)\citenamefont {Cheng},
  \citenamefont {Dong}, \citenamefont {Li}, \citenamefont {Zhang},
  \citenamefont {Zhang}, \citenamefont {Jiao}, \citenamefont {Blau},
  \citenamefont {Zhang},\ and\ \citenamefont
  {Wang}}]{Cheng_exp_graphene_nlo_:13}%
  \BibitemOpen
  \bibfield  {author} {\bibinfo {author} {\bibfnamefont {X.}~\bibnamefont
  {Cheng}}, \bibinfo {author} {\bibfnamefont {N.}~\bibnamefont {Dong}},
  \bibinfo {author} {\bibfnamefont {B.}~\bibnamefont {Li}}, \bibinfo {author}
  {\bibfnamefont {X.}~\bibnamefont {Zhang}}, \bibinfo {author} {\bibfnamefont
  {S.}~\bibnamefont {Zhang}}, \bibinfo {author} {\bibfnamefont
  {J.}~\bibnamefont {Jiao}}, \bibinfo {author} {\bibfnamefont {W.~J.}\
  \bibnamefont {Blau}}, \bibinfo {author} {\bibfnamefont {L.}~\bibnamefont
  {Zhang}}, \ and\ \bibinfo {author} {\bibfnamefont {J.}~\bibnamefont {Wang}},\
  }\href {\doibase 10.1364/OE.21.016486} {\bibfield  {journal} {\bibinfo
  {journal} {Opt. Express}\ }\textbf {\bibinfo {volume} {21}},\ \bibinfo
  {pages} {16486} (\bibinfo {year} {2013})}\BibitemShut {NoStop}%
\bibitem [{\citenamefont {Dragoman}\ \emph {et~al.}(2010)\citenamefont
  {Dragoman}, \citenamefont {Neculoiu}, \citenamefont {Deligeorgis},
  \citenamefont {Konstantinidis}, \citenamefont {Dragoman}, \citenamefont
  {Cismaru}, \citenamefont {Muller},\ and\ \citenamefont
  {Plana}}]{Dragonman_Frequency_Multiplication_graphene_exp_doi:10.1063/1.3483872}%
  \BibitemOpen
  \bibfield  {author} {\bibinfo {author} {\bibfnamefont {M.}~\bibnamefont
  {Dragoman}}, \bibinfo {author} {\bibfnamefont {D.}~\bibnamefont {Neculoiu}},
  \bibinfo {author} {\bibfnamefont {G.}~\bibnamefont {Deligeorgis}}, \bibinfo
  {author} {\bibfnamefont {G.}~\bibnamefont {Konstantinidis}}, \bibinfo
  {author} {\bibfnamefont {D.}~\bibnamefont {Dragoman}}, \bibinfo {author}
  {\bibfnamefont {A.}~\bibnamefont {Cismaru}}, \bibinfo {author} {\bibfnamefont
  {A.~A.}\ \bibnamefont {Muller}}, \ and\ \bibinfo {author} {\bibfnamefont
  {R.}~\bibnamefont {Plana}},\ }\href {\doibase 10.1063/1.3483872} {\bibfield
  {journal} {\bibinfo  {journal} {Applied Physics Letters}\ }\textbf {\bibinfo
  {volume} {97}},\ \bibinfo {pages} {093101} (\bibinfo {year}
  {2010})}\BibitemShut {NoStop}%
\bibitem [{\citenamefont {Chen}\ \emph {et~al.}(2015)\citenamefont {Chen},
  \citenamefont {Wang},\ and\ \citenamefont
  {Ji}}]{Chen_TPA_graphene_exp_doi:10.1021/acs.jpcc.5b03819}%
  \BibitemOpen
  \bibfield  {author} {\bibinfo {author} {\bibfnamefont {W.}~\bibnamefont
  {Chen}}, \bibinfo {author} {\bibfnamefont {Y.}~\bibnamefont {Wang}}, \ and\
  \bibinfo {author} {\bibfnamefont {W.}~\bibnamefont {Ji}},\ }\href {\doibase
  10.1021/acs.jpcc.5b03819} {\bibfield  {journal} {\bibinfo  {journal} {The
  Journal of Physical Chemistry C}\ }\textbf {\bibinfo {volume} {119}},\
  \bibinfo {pages} {16954} (\bibinfo {year} {2015})}\BibitemShut {NoStop}%
\bibitem [{\citenamefont {Dean}\ and\ \citenamefont {van
  Driel}(2010)}]{Dean_SHG_PhysRevB.82.125411}%
  \BibitemOpen
  \bibfield  {author} {\bibinfo {author} {\bibfnamefont {J.~J.}\ \bibnamefont
  {Dean}}\ and\ \bibinfo {author} {\bibfnamefont {H.~M.}\ \bibnamefont {van
  Driel}},\ }\href {\doibase 10.1103/PhysRevB.82.125411} {\bibfield  {journal}
  {\bibinfo  {journal} {Phys. Rev. B}\ }\textbf {\bibinfo {volume} {82}},\
  \bibinfo {pages} {125411} (\bibinfo {year} {2010})}\BibitemShut {NoStop}%
\bibitem [{\citenamefont {Chu}\ \emph {et~al.}(2012)\citenamefont {Chu},
  \citenamefont {Wang},\ and\ \citenamefont {Gong}}]{CHU_CPL_Kerr_Effect}%
  \BibitemOpen
  \bibfield  {author} {\bibinfo {author} {\bibfnamefont {S.}~\bibnamefont
  {Chu}}, \bibinfo {author} {\bibfnamefont {S.}~\bibnamefont {Wang}}, \ and\
  \bibinfo {author} {\bibfnamefont {Q.}~\bibnamefont {Gong}},\ }\href {\doibase
  http://dx.doi.org/10.1016/j.cplett.2011.12.024} {\bibfield  {journal}
  {\bibinfo  {journal} {Chemical Physics Letters}\ }\textbf {\bibinfo {volume}
  {523}},\ \bibinfo {pages} {104 } (\bibinfo {year} {2012})}\BibitemShut
  {NoStop}%
\bibitem [{\citenamefont {Sun}\ \emph {et~al.}(2013)\citenamefont {Sun},
  \citenamefont {Dong}, \citenamefont {Xie}, \citenamefont {Xia}, \citenamefont
  {K{\"o}nig}, \citenamefont {Nagaiah}, \citenamefont {S{\'a}nchez},
  \citenamefont {Ebbinghaus}, \citenamefont {Erbe}, \citenamefont {Zhang},
  \citenamefont {Ludwig}, \citenamefont {Schuhmann}, \citenamefont {Wang},\
  and\ \citenamefont {Muhler}}]{Zhenyu_Sun_oxy_func_grp}%
  \BibitemOpen
  \bibfield  {author} {\bibinfo {author} {\bibfnamefont {Z.}~\bibnamefont
  {Sun}}, \bibinfo {author} {\bibfnamefont {N.}~\bibnamefont {Dong}}, \bibinfo
  {author} {\bibfnamefont {K.}~\bibnamefont {Xie}}, \bibinfo {author}
  {\bibfnamefont {W.}~\bibnamefont {Xia}}, \bibinfo {author} {\bibfnamefont
  {D.}~\bibnamefont {K{\"o}nig}}, \bibinfo {author} {\bibfnamefont {T.~C.}\
  \bibnamefont {Nagaiah}}, \bibinfo {author} {\bibfnamefont {M.~D.}\
  \bibnamefont {S{\'a}nchez}}, \bibinfo {author} {\bibfnamefont
  {P.}~\bibnamefont {Ebbinghaus}}, \bibinfo {author} {\bibfnamefont
  {A.}~\bibnamefont {Erbe}}, \bibinfo {author} {\bibfnamefont {X.}~\bibnamefont
  {Zhang}}, \bibinfo {author} {\bibfnamefont {A.}~\bibnamefont {Ludwig}},
  \bibinfo {author} {\bibfnamefont {W.}~\bibnamefont {Schuhmann}}, \bibinfo
  {author} {\bibfnamefont {J.}~\bibnamefont {Wang}}, \ and\ \bibinfo {author}
  {\bibfnamefont {M.}~\bibnamefont {Muhler}},\ }\href {\doibase
  10.1021/jp401736n} {\bibfield  {journal} {\bibinfo  {journal} {The Journal of
  Physical Chemistry C}\ }\textbf {\bibinfo {volume} {117}},\ \bibinfo {pages}
  {11811} (\bibinfo {year} {2013})}\BibitemShut {NoStop}%
\bibitem [{\citenamefont {Yoneda}\ \emph {et~al.}(2009)\citenamefont {Yoneda},
  \citenamefont {Nakano}, \citenamefont {Kishi}, \citenamefont {Takahashi},
  \citenamefont {Shimizu}, \citenamefont {Kubo}, \citenamefont {Kamada},
  \citenamefont {Ohta}, \citenamefont {Champagne},\ and\ \citenamefont
  {Botek}}]{Yoneda2009278}%
  \BibitemOpen
  \bibfield  {author} {\bibinfo {author} {\bibfnamefont {K.}~\bibnamefont
  {Yoneda}}, \bibinfo {author} {\bibfnamefont {M.}~\bibnamefont {Nakano}},
  \bibinfo {author} {\bibfnamefont {R.}~\bibnamefont {Kishi}}, \bibinfo
  {author} {\bibfnamefont {H.}~\bibnamefont {Takahashi}}, \bibinfo {author}
  {\bibfnamefont {A.}~\bibnamefont {Shimizu}}, \bibinfo {author} {\bibfnamefont
  {T.}~\bibnamefont {Kubo}}, \bibinfo {author} {\bibfnamefont {K.}~\bibnamefont
  {Kamada}}, \bibinfo {author} {\bibfnamefont {K.}~\bibnamefont {Ohta}},
  \bibinfo {author} {\bibfnamefont {B.}~\bibnamefont {Champagne}}, \ and\
  \bibinfo {author} {\bibfnamefont {E.}~\bibnamefont {Botek}},\ }\href
  {\doibase http://dx.doi.org/10.1016/j.cplett.2009.09.047} {\bibfield
  {journal} {\bibinfo  {journal} {Chemical Physics Letters}\ }\textbf {\bibinfo
  {volume} {480}},\ \bibinfo {pages} {278 } (\bibinfo {year}
  {2009})}\BibitemShut {NoStop}%
\bibitem [{\citenamefont {Brinkley}\ \emph {et~al.}(2016)\citenamefont
  {Brinkley}, \citenamefont {Abergel},\ and\ \citenamefont
  {Clader}}]{Brinkley_theory_graphene_0953-8984-28-36-365001}%
  \BibitemOpen
  \bibfield  {author} {\bibinfo {author} {\bibfnamefont {M.~K.}\ \bibnamefont
  {Brinkley}}, \bibinfo {author} {\bibfnamefont {D.~S.~L.}\ \bibnamefont
  {Abergel}}, \ and\ \bibinfo {author} {\bibfnamefont {B.~D.}\ \bibnamefont
  {Clader}},\ }\href@noop {} {\bibfield  {journal} {\bibinfo  {journal}
  {Journal of Physics: Condensed Matter}\ }\textbf {\bibinfo {volume} {28}},\
  \bibinfo {pages} {365001} (\bibinfo {year} {2016})}\BibitemShut {NoStop}%
\bibitem [{\citenamefont
  {Mikhailov}(2011)}]{Mikhail_plasmon_Theory_graphene_PhysRevB.84.045432}%
  \BibitemOpen
  \bibfield  {author} {\bibinfo {author} {\bibfnamefont {S.~A.}\ \bibnamefont
  {Mikhailov}},\ }\href {\doibase 10.1103/PhysRevB.84.045432} {\bibfield
  {journal} {\bibinfo  {journal} {Phys. Rev. B}\ }\textbf {\bibinfo {volume}
  {84}},\ \bibinfo {pages} {045432} (\bibinfo {year} {2011})}\BibitemShut
  {NoStop}%
\bibitem [{\citenamefont {Yang}\ \emph {et~al.}(2011)\citenamefont {Yang},
  \citenamefont {Feng}, \citenamefont {Wang}, \citenamefont {Huang},
  \citenamefont {Chen}, \citenamefont {Wee},\ and\ \citenamefont
  {Ji}}]{Yang_TPA_graphene_theory_doi:10.1021/nl200587h}%
  \BibitemOpen
  \bibfield  {author} {\bibinfo {author} {\bibfnamefont {H.}~\bibnamefont
  {Yang}}, \bibinfo {author} {\bibfnamefont {X.}~\bibnamefont {Feng}}, \bibinfo
  {author} {\bibfnamefont {Q.}~\bibnamefont {Wang}}, \bibinfo {author}
  {\bibfnamefont {H.}~\bibnamefont {Huang}}, \bibinfo {author} {\bibfnamefont
  {W.}~\bibnamefont {Chen}}, \bibinfo {author} {\bibfnamefont {A.~T.~S.}\
  \bibnamefont {Wee}}, \ and\ \bibinfo {author} {\bibfnamefont
  {W.}~\bibnamefont {Ji}},\ }\href {\doibase 10.1021/nl200587h} {\bibfield
  {journal} {\bibinfo  {journal} {Nano Letters}\ }\textbf {\bibinfo {volume}
  {11}},\ \bibinfo {pages} {2622} (\bibinfo {year} {2011})}\BibitemShut
  {NoStop}%
\bibitem [{\citenamefont {Cox}\ \emph {et~al.}(2013)\citenamefont {Cox},
  \citenamefont {Singh}, \citenamefont {Ant\'{o}n},\ and\ \citenamefont
  {Carre{\~n}o}}]{Cox_quant_chem_cal}%
  \BibitemOpen
  \bibfield  {author} {\bibinfo {author} {\bibfnamefont {J.~D.}\ \bibnamefont
  {Cox}}, \bibinfo {author} {\bibfnamefont {M.~R.}\ \bibnamefont {Singh}},
  \bibinfo {author} {\bibfnamefont {M.~A.}\ \bibnamefont {Ant\'{o}n}}, \ and\
  \bibinfo {author} {\bibfnamefont {F.}~\bibnamefont {Carre{\~n}o}},\
  }\href@noop {} {\bibfield  {journal} {\bibinfo  {journal} {Journal of
  Physics: Condensed Matter}\ }\textbf {\bibinfo {volume} {25}},\ \bibinfo
  {pages} {385302} (\bibinfo {year} {2013})}\BibitemShut {NoStop}%
\bibitem [{\citenamefont {Yamijala}\ \emph {et~al.}(2015)\citenamefont
  {Yamijala}, \citenamefont {Mukhopadhyay},\ and\ \citenamefont
  {Pati}}]{Yamijala}%
  \BibitemOpen
  \bibfield  {author} {\bibinfo {author} {\bibfnamefont {S.~S. R. K.~C.}\
  \bibnamefont {Yamijala}}, \bibinfo {author} {\bibfnamefont {M.}~\bibnamefont
  {Mukhopadhyay}}, \ and\ \bibinfo {author} {\bibfnamefont {S.~K.}\
  \bibnamefont {Pati}},\ }\href {\doibase 10.1021/acs.jpcc.5b03531} {\bibfield
  {journal} {\bibinfo  {journal} {The Journal of Physical Chemistry C}\
  }\textbf {\bibinfo {volume} {119}},\ \bibinfo {pages} {12079} (\bibinfo
  {year} {2015})}\BibitemShut {NoStop}%
\bibitem [{\citenamefont
  {Mikhailov}(2016)}]{Mikhail_TPA_Theory_graphene_PhysRevB.93.085403}%
  \BibitemOpen
  \bibfield  {author} {\bibinfo {author} {\bibfnamefont {S.~A.}\ \bibnamefont
  {Mikhailov}},\ }\href {\doibase 10.1103/PhysRevB.93.085403} {\bibfield
  {journal} {\bibinfo  {journal} {Phys. Rev. B}\ }\textbf {\bibinfo {volume}
  {93}},\ \bibinfo {pages} {085403} (\bibinfo {year} {2016})}\BibitemShut
  {NoStop}%
\bibitem [{\citenamefont {Feng}\ \emph {et~al.}(2016)\citenamefont {Feng},
  \citenamefont {Li}, \citenamefont {Li},\ and\ \citenamefont
  {Liu}}]{Feng_oe_analy_exp}%
  \BibitemOpen
  \bibfield  {author} {\bibinfo {author} {\bibfnamefont {X.}~\bibnamefont
  {Feng}}, \bibinfo {author} {\bibfnamefont {X.}~\bibnamefont {Li}}, \bibinfo
  {author} {\bibfnamefont {Z.}~\bibnamefont {Li}}, \ and\ \bibinfo {author}
  {\bibfnamefont {Y.}~\bibnamefont {Liu}},\ }\href {\doibase
  10.1364/OE.24.002877} {\bibfield  {journal} {\bibinfo  {journal} {Opt.
  Express}\ }\textbf {\bibinfo {volume} {24}},\ \bibinfo {pages} {2877}
  (\bibinfo {year} {2016})}\BibitemShut {NoStop}%
\bibitem [{\citenamefont {Aryanpour}\ \emph
  {et~al.}(2014{\natexlab{a}})\citenamefont {Aryanpour}, \citenamefont
  {Shukla},\ and\ \citenamefont {Mazumdar}}]{Aryanpour_electron}%
  \BibitemOpen
  \bibfield  {author} {\bibinfo {author} {\bibfnamefont {K.}~\bibnamefont
  {Aryanpour}}, \bibinfo {author} {\bibfnamefont {A.}~\bibnamefont {Shukla}}, \
  and\ \bibinfo {author} {\bibfnamefont {S.}~\bibnamefont {Mazumdar}},\ }\href
  {\doibase http://dx.doi.org/10.1063/1.4867363} {\bibfield  {journal}
  {\bibinfo  {journal} {The Journal of Chemical Physics}\ }\textbf {\bibinfo
  {volume} {140}},\ \bibinfo {eid} {104301} (\bibinfo {year}
  {2014}{\natexlab{a}})}\BibitemShut {NoStop}%
\bibitem [{\citenamefont {Agapito}\ \emph {et~al.}(2010)\citenamefont
  {Agapito}, \citenamefont {Kioussis},\ and\ \citenamefont
  {Kaxiras}}]{dqd-kaxiras-prb-2010}%
  \BibitemOpen
  \bibfield  {author} {\bibinfo {author} {\bibfnamefont {L.~A.}\ \bibnamefont
  {Agapito}}, \bibinfo {author} {\bibfnamefont {N.}~\bibnamefont {Kioussis}}, \
  and\ \bibinfo {author} {\bibfnamefont {E.}~\bibnamefont {Kaxiras}},\ }\href
  {\doibase 10.1103/PhysRevB.82.201411} {\bibfield  {journal} {\bibinfo
  {journal} {Phys. Rev. B}\ }\textbf {\bibinfo {volume} {82}},\ \bibinfo
  {pages} {201411} (\bibinfo {year} {2010})}\BibitemShut {NoStop}%
\bibitem [{\citenamefont {Basak}\ and\ \citenamefont
  {Shukla}(2016)}]{Tista_PRB93}%
  \BibitemOpen
  \bibfield  {author} {\bibinfo {author} {\bibfnamefont {T.}~\bibnamefont
  {Basak}}\ and\ \bibinfo {author} {\bibfnamefont {A.}~\bibnamefont {Shukla}},\
  }\href {\doibase 10.1103/PhysRevB.93.235432} {\bibfield  {journal} {\bibinfo
  {journal} {Phys. Rev. B}\ }\textbf {\bibinfo {volume} {93}},\ \bibinfo
  {pages} {235432} (\bibinfo {year} {2016})}\BibitemShut {NoStop}%
\bibitem [{\citenamefont {Salvi}\ \emph {et~al.}(1983)\citenamefont {Salvi},
  \citenamefont {Foggi},\ and\ \citenamefont {Castellucci}}]{SALVI1983206}%
  \BibitemOpen
  \bibfield  {author} {\bibinfo {author} {\bibfnamefont {P.}~\bibnamefont
  {Salvi}}, \bibinfo {author} {\bibfnamefont {P.}~\bibnamefont {Foggi}}, \ and\
  \bibinfo {author} {\bibfnamefont {E.}~\bibnamefont {Castellucci}},\ }\href
  {\doibase http://dx.doi.org/10.1016/0009-2614(83)87151-6} {\bibfield
  {journal} {\bibinfo  {journal} {Chemical Physics Letters}\ }\textbf {\bibinfo
  {volume} {98}},\ \bibinfo {pages} {206 } (\bibinfo {year}
  {1983})}\BibitemShut {NoStop}%
\bibitem [{\citenamefont {Elias}\ \emph {et~al.}(2011)\citenamefont {Elias},
  \citenamefont {Gorbachev}, \citenamefont {Mayorov}, \citenamefont {Morozov},
  \citenamefont {Zhukov}, \citenamefont {Blake}, \citenamefont {Ponomarenko},
  \citenamefont {Grigorieva}, \citenamefont {Novoselov}, \citenamefont
  {Guinea},\ and\ \citenamefont {Geim}}]{Elias_NaturePhysics}%
  \BibitemOpen
  \bibfield  {author} {\bibinfo {author} {\bibfnamefont {D.~C.}\ \bibnamefont
  {Elias}}, \bibinfo {author} {\bibfnamefont {R.~V.}\ \bibnamefont
  {Gorbachev}}, \bibinfo {author} {\bibfnamefont {A.~S.}\ \bibnamefont
  {Mayorov}}, \bibinfo {author} {\bibfnamefont {S.~V.}\ \bibnamefont
  {Morozov}}, \bibinfo {author} {\bibfnamefont {A.~A.}\ \bibnamefont {Zhukov}},
  \bibinfo {author} {\bibfnamefont {P.}~\bibnamefont {Blake}}, \bibinfo
  {author} {\bibfnamefont {L.~A.}\ \bibnamefont {Ponomarenko}}, \bibinfo
  {author} {\bibfnamefont {I.~V.}\ \bibnamefont {Grigorieva}}, \bibinfo
  {author} {\bibfnamefont {K.~S.}\ \bibnamefont {Novoselov}}, \bibinfo {author}
  {\bibfnamefont {F.}~\bibnamefont {Guinea}}, \ and\ \bibinfo {author}
  {\bibfnamefont {A.~K.}\ \bibnamefont {Geim}},\ }\href {\doibase
  10.1038/nphys2049} {\bibfield  {journal} {\bibinfo  {journal} {Nature
  Physics}\ }\textbf {\bibinfo {volume} {7}},\ \bibinfo {pages} {701} (\bibinfo
  {year} {2011})}\BibitemShut {NoStop}%
\bibitem [{\citenamefont {Pople}(1953)}]{ppp-pople}%
  \BibitemOpen
  \bibfield  {author} {\bibinfo {author} {\bibfnamefont {J.~A.}\ \bibnamefont
  {Pople}},\ }\href {\doibase 10.1039/TF9534901375} {\bibfield  {journal}
  {\bibinfo  {journal} {Trans. Faraday Soc.}\ }\textbf {\bibinfo {volume}
  {49}},\ \bibinfo {pages} {1375} (\bibinfo {year} {1953})}\BibitemShut
  {NoStop}%
\bibitem [{\citenamefont {Pariser}\ and\ \citenamefont
  {Parr}(1953)}]{ppp-pariser-parr}%
  \BibitemOpen
  \bibfield  {author} {\bibinfo {author} {\bibfnamefont {R.}~\bibnamefont
  {Pariser}}\ and\ \bibinfo {author} {\bibfnamefont {R.~G.}\ \bibnamefont
  {Parr}},\ }\href {\doibase http://dx.doi.org/10.1063/1.1699030} {\bibfield
  {journal} {\bibinfo  {journal} {J. Chem. Phys.}\ }\textbf {\bibinfo {volume}
  {21}},\ \bibinfo {pages} {767} (\bibinfo {year} {1953})}\BibitemShut
  {NoStop}%
\bibitem [{\citenamefont {Ohno}(1964)}]{Theor.chim.act.2Ohno}%
  \BibitemOpen
  \bibfield  {author} {\bibinfo {author} {\bibfnamefont {K.}~\bibnamefont
  {Ohno}},\ }\href {\doibase 10.1007/BF00528281} {\bibfield  {journal}
  {\bibinfo  {journal} {Theoretica chimica acta}\ }\textbf {\bibinfo {volume}
  {2}},\ \bibinfo {pages} {219} (\bibinfo {year} {1964})}\BibitemShut {NoStop}%
\bibitem [{\citenamefont {Chandross}\ and\ \citenamefont
  {Mazumdar}(1997)}]{PhysRevB.55.1497Chandross}%
  \BibitemOpen
  \bibfield  {author} {\bibinfo {author} {\bibfnamefont {M.}~\bibnamefont
  {Chandross}}\ and\ \bibinfo {author} {\bibfnamefont {S.}~\bibnamefont
  {Mazumdar}},\ }\href {\doibase 10.1103/PhysRevB.55.1497} {\bibfield
  {journal} {\bibinfo  {journal} {Phys. Rev. B}\ }\textbf {\bibinfo {volume}
  {55}},\ \bibinfo {pages} {1497} (\bibinfo {year} {1997})}\BibitemShut
  {NoStop}%
\bibitem [{\citenamefont {Shukla}(2002)}]{PhysRevB.65.125204Shukla65}%
  \BibitemOpen
  \bibfield  {author} {\bibinfo {author} {\bibfnamefont {A.}~\bibnamefont
  {Shukla}},\ }\href {\doibase 10.1103/PhysRevB.65.125204} {\bibfield
  {journal} {\bibinfo  {journal} {Phys. Rev. B}\ }\textbf {\bibinfo {volume}
  {65}},\ \bibinfo {pages} {125204} (\bibinfo {year} {2002})}\BibitemShut
  {NoStop}%
\bibitem [{\citenamefont
  {Shukla}(2004{\natexlab{a}})}]{PhysRevB.69.165218Shukla69}%
  \BibitemOpen
  \bibfield  {author} {\bibinfo {author} {\bibfnamefont {A.}~\bibnamefont
  {Shukla}},\ }\href {\doibase 10.1103/PhysRevB.69.165218} {\bibfield
  {journal} {\bibinfo  {journal} {Phys. Rev. B}\ }\textbf {\bibinfo {volume}
  {69}},\ \bibinfo {pages} {165218} (\bibinfo {year}
  {2004}{\natexlab{a}})}\BibitemShut {NoStop}%
\bibitem [{\citenamefont {Sony}\ and\ \citenamefont
  {Shukla}(2005{\natexlab{b}})}]{PhysRevB.71.165204Priya_t0}%
  \BibitemOpen
  \bibfield  {author} {\bibinfo {author} {\bibfnamefont {P.}~\bibnamefont
  {Sony}}\ and\ \bibinfo {author} {\bibfnamefont {A.}~\bibnamefont {Shukla}},\
  }\href {\doibase 10.1103/PhysRevB.71.165204} {\bibfield  {journal} {\bibinfo
  {journal} {Phys. Rev. B}\ }\textbf {\bibinfo {volume} {71}},\ \bibinfo
  {pages} {165204} (\bibinfo {year} {2005}{\natexlab{b}})}\BibitemShut
  {NoStop}%
\bibitem [{\citenamefont {Sony}\ and\ \citenamefont
  {Shukla}(2009)}]{:/content/aip/journal/jcp/131/1/10.1063/1.3159670Priyaanthracene}%
  \BibitemOpen
  \bibfield  {author} {\bibinfo {author} {\bibfnamefont {P.}~\bibnamefont
  {Sony}}\ and\ \bibinfo {author} {\bibfnamefont {A.}~\bibnamefont {Shukla}},\
  }\href {\doibase http://dx.doi.org/10.1063/1.3159670} {\bibfield  {journal}
  {\bibinfo  {journal} {The Journal of Chemical Physics}\ }\textbf {\bibinfo
  {volume} {131}},\ \bibinfo {eid} {014302} (\bibinfo {year}
  {2009})}\BibitemShut {NoStop}%
\bibitem [{\citenamefont {Chakraborty}\ and\ \citenamefont
  {Shukla}(2013)}]{doi:10.1021/jp408535u}%
  \BibitemOpen
  \bibfield  {author} {\bibinfo {author} {\bibfnamefont {H.}~\bibnamefont
  {Chakraborty}}\ and\ \bibinfo {author} {\bibfnamefont {A.}~\bibnamefont
  {Shukla}},\ }\href {\doibase 10.1021/jp408535u} {\bibfield  {journal}
  {\bibinfo  {journal} {The Journal of Physical Chemistry A}\ }\textbf
  {\bibinfo {volume} {117}},\ \bibinfo {pages} {14220} (\bibinfo {year}
  {2013})}\BibitemShut {NoStop}%
\bibitem [{\citenamefont {Chakraborty}\ and\ \citenamefont
  {Shukla}(2014)}]{himanshu-triplet}%
  \BibitemOpen
  \bibfield  {author} {\bibinfo {author} {\bibfnamefont {H.}~\bibnamefont
  {Chakraborty}}\ and\ \bibinfo {author} {\bibfnamefont {A.}~\bibnamefont
  {Shukla}},\ }\href {\doibase http://dx.doi.org/10.1063/1.4897955} {\bibfield
  {journal} {\bibinfo  {journal} {The Journal of Chemical Physics}\ }\textbf
  {\bibinfo {volume} {141}},\ \bibinfo {eid} {164301} (\bibinfo {year}
  {2014})}\BibitemShut {NoStop}%
\bibitem [{\citenamefont {Aryanpour}\ \emph
  {et~al.}(2014{\natexlab{b}})\citenamefont {Aryanpour}, \citenamefont
  {Roberts}, \citenamefont {Sandhu}, \citenamefont {Rathore}, \citenamefont
  {Shukla},\ and\ \citenamefont {Mazumdar}}]{Aryanpour_Subgap}%
  \BibitemOpen
  \bibfield  {author} {\bibinfo {author} {\bibfnamefont {K.}~\bibnamefont
  {Aryanpour}}, \bibinfo {author} {\bibfnamefont {A.}~\bibnamefont {Roberts}},
  \bibinfo {author} {\bibfnamefont {A.}~\bibnamefont {Sandhu}}, \bibinfo
  {author} {\bibfnamefont {R.}~\bibnamefont {Rathore}}, \bibinfo {author}
  {\bibfnamefont {A.}~\bibnamefont {Shukla}}, \ and\ \bibinfo {author}
  {\bibfnamefont {S.}~\bibnamefont {Mazumdar}},\ }\href {\doibase
  10.1021/jp410793r} {\bibfield  {journal} {\bibinfo  {journal} {The Journal of
  Physical Chemistry C}\ }\textbf {\bibinfo {volume} {118}},\ \bibinfo {pages}
  {3331} (\bibinfo {year} {2014}{\natexlab{b}})}\BibitemShut {NoStop}%
\bibitem [{\citenamefont {Sony}\ and\ \citenamefont
  {Shukla}(2010)}]{Sony2010821}%
  \BibitemOpen
  \bibfield  {author} {\bibinfo {author} {\bibfnamefont {P.}~\bibnamefont
  {Sony}}\ and\ \bibinfo {author} {\bibfnamefont {A.}~\bibnamefont {Shukla}},\
  }\href {\doibase http://dx.doi.org/10.1016/j.cpc.2009.12.015} {\bibfield
  {journal} {\bibinfo  {journal} {Computer Physics Communications}\ }\textbf
  {\bibinfo {volume} {181}},\ \bibinfo {pages} {821 } (\bibinfo {year}
  {2010})}\BibitemShut {NoStop}%
\bibitem [{Sup()}]{Supp_Info}%
  \BibitemOpen
  \href@noop {} {\bibinfo  {journal} {Supporting Information}\ }\BibitemShut
  {NoStop}%
\bibitem [{\citenamefont {Orr}\ and\ \citenamefont {Ward}(1971)}]{Orr_Ward}%
  \BibitemOpen
\bibfield  {journal} {  }\bibfield  {author} {\bibinfo {author} {\bibfnamefont
  {B.}~\bibnamefont {Orr}}\ and\ \bibinfo {author} {\bibfnamefont
  {J.}~\bibnamefont {Ward}},\ }\href {\doibase 10.1080/00268977100100481}
  {\bibfield  {journal} {\bibinfo  {journal} {Molecular Physics}\ }\textbf
  {\bibinfo {volume} {20}},\ \bibinfo {pages} {513} (\bibinfo {year}
  {1971})}\BibitemShut {NoStop}%
\bibitem [{\citenamefont {Shukla}(2004{\natexlab{b}})}]{shukla-pdpa-tpa}%
  \BibitemOpen
  \bibfield  {author} {\bibinfo {author} {\bibfnamefont {A.}~\bibnamefont
  {Shukla}},\ }\href {\doibase
  http://dx.doi.org/10.1016/j.chemphys.2004.02.004} {\bibfield  {journal}
  {\bibinfo  {journal} {Chemical Physics}\ }\textbf {\bibinfo {volume} {300}},\
  \bibinfo {pages} {177 } (\bibinfo {year} {2004}{\natexlab{b}})}\BibitemShut
  {NoStop}%
\bibitem [{\citenamefont {Yumura}\ \emph {et~al.}(2009)\citenamefont {Yumura},
  \citenamefont {Kimura}, \citenamefont {Kobayashi}, \citenamefont {Tanaka},
  \citenamefont {Okumura},\ and\ \citenamefont
  {Yamabe}}]{Yumura_HOMO-LUMO_Band-gap}%
  \BibitemOpen
  \bibfield  {author} {\bibinfo {author} {\bibfnamefont {T.}~\bibnamefont
  {Yumura}}, \bibinfo {author} {\bibfnamefont {K.}~\bibnamefont {Kimura}},
  \bibinfo {author} {\bibfnamefont {H.}~\bibnamefont {Kobayashi}}, \bibinfo
  {author} {\bibfnamefont {R.}~\bibnamefont {Tanaka}}, \bibinfo {author}
  {\bibfnamefont {N.}~\bibnamefont {Okumura}}, \ and\ \bibinfo {author}
  {\bibfnamefont {T.}~\bibnamefont {Yamabe}},\ }\href {\doibase
  10.1039/B905866D} {\bibfield  {journal} {\bibinfo  {journal} {Phys. Chem.
  Chem. Phys.}\ }\textbf {\bibinfo {volume} {11}},\ \bibinfo {pages} {8275}
  (\bibinfo {year} {2009})}\BibitemShut {NoStop}%
\bibitem [{\citenamefont {Ray}\ \emph {et~al.}(2006)\citenamefont {Ray},
  \citenamefont {Chakraborty},\ and\ \citenamefont
  {Moulik}}]{BASURAY_pyrene_2006248}%
  \BibitemOpen
  \bibfield  {author} {\bibinfo {author} {\bibfnamefont {G.~B.}\ \bibnamefont
  {Ray}}, \bibinfo {author} {\bibfnamefont {I.}~\bibnamefont {Chakraborty}}, \
  and\ \bibinfo {author} {\bibfnamefont {S.~P.}\ \bibnamefont {Moulik}},\
  }\href {\doibase http://dx.doi.org/10.1016/j.jcis.2005.07.006} {\bibfield
  {journal} {\bibinfo  {journal} {Journal of Colloid and Interface Science}\
  }\textbf {\bibinfo {volume} {294}},\ \bibinfo {pages} {248 } (\bibinfo {year}
  {2006})}\BibitemShut {NoStop}%
\bibitem [{\citenamefont {Clar}\ and\ \citenamefont
  {Schmidt}(1978)}]{CLAR_DQD_30_19783219}%
  \BibitemOpen
  \bibfield  {author} {\bibinfo {author} {\bibfnamefont {E.}~\bibnamefont
  {Clar}}\ and\ \bibinfo {author} {\bibfnamefont {W.}~\bibnamefont {Schmidt}},\
  }\href {\doibase http://dx.doi.org/10.1016/0040-4020(78)87020-3} {\bibfield
  {journal} {\bibinfo  {journal} {Tetrahedron}\ }\textbf {\bibinfo {volume}
  {34}},\ \bibinfo {pages} {3219 } (\bibinfo {year} {1978})}\BibitemShut
  {NoStop}%
\bibitem [{\citenamefont {Shukla}\ \emph {et~al.}(2003)\citenamefont {Shukla},
  \citenamefont {Ghosh},\ and\ \citenamefont {Mazumdar}}]{shukla-nlo-ppv}%
  \BibitemOpen
  \bibfield  {author} {\bibinfo {author} {\bibfnamefont {A.}~\bibnamefont
  {Shukla}}, \bibinfo {author} {\bibfnamefont {H.}~\bibnamefont {Ghosh}}, \
  and\ \bibinfo {author} {\bibfnamefont {S.}~\bibnamefont {Mazumdar}},\ }\href
  {\doibase 10.1103/PhysRevB.67.245203} {\bibfield  {journal} {\bibinfo
  {journal} {Phys. Rev. B}\ }\textbf {\bibinfo {volume} {67}},\ \bibinfo
  {pages} {245203} (\bibinfo {year} {2003})}\BibitemShut {NoStop}%
\end{thebibliography}%

\end{document}


\begin{frontmatter}{}

\title{Supporting Information: Electron correlation effects and two-photon
absorption in diamond-shaped graphene quantum dots}

\author{Tista Basak$^{a}$}

\address[1]{Mukesh Patel School of Technology Management and Engineering, NMIMS
University, Mumbai 400056, India}

\ead{Tista.Basak@nmims.edu}

\cortext[1]{Tista Basak}

\author{Tushima Basak$^{b}$}

\address[2]{Department of Physics, Mithibai College, Mumbai 400056, India}

\ead{Tushima.Basak@mithibai.ac.in}

\cortext[2]{Tushima Basak}

\author{Alok Shukla$^{c,d}$}

\address[3]{Department of Physics, School of Engineering and Applied Sciences,
Bennett University, Plot No. 8-{}-11, TechZone II, Greater Noida 201310,
Uttar Pradesh, India }

\address[4]{Permanent Address: Department of Physics, Indian Institute of Technology
Bombay, Powai, Mumbai 400076, India}

\ead{shukla@phy.iitb.ac.in}

\cortext[3,4]{Alok Shukla }

\end{frontmatter}{}

\section{Pìctorial representation of frontier orbitals}

In this section, the pictorial plots of the frontier molecular orbitals
H, H-1, H-2, H-3 and H-4 contributing significantly to the non-linear
optical properties of DQD-16, DQD-30 and DQD-48 are presented in Figs.
S1, S2, and S3, respectively. The expansion coefficients of virtual
molecular orbitals L, L+1, L+2, L+3 and L+4 have the same magnitudes
(although different relative phases) as those of H, H-1, H-2, H-3
and H-4, respectively, on account of electron-hole symmetry, and hence
have not been plotted. The symbols H and L represent the highest occupied
molecular orbital (HOMO), and the lowest unoccupied molecular orbital
(LUMO), respectively. 

\begin{figure}[H]
\subfloat[H]{\includegraphics[width=2cm]{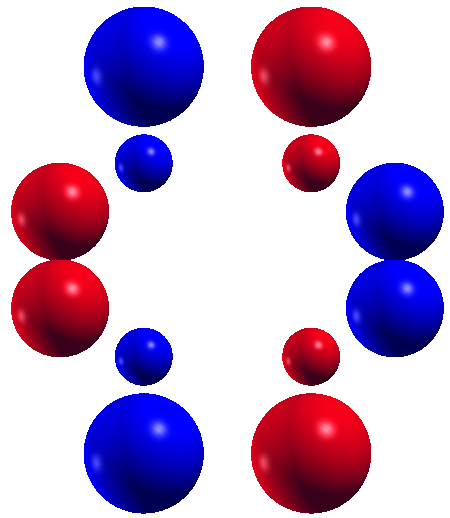}

} \subfloat[H-1]{\includegraphics[width=2cm]{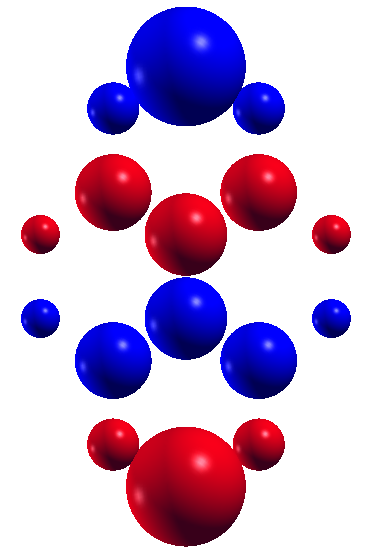}

} \subfloat[H-2]{\includegraphics[width=2cm]{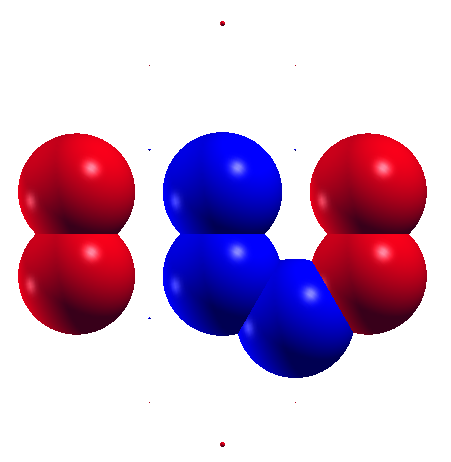}} \subfloat[H-3]{\includegraphics[width=2cm]{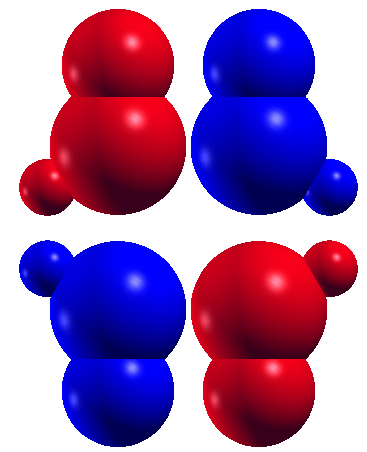}}
\subfloat[H-4]{\includegraphics[width=2cm]{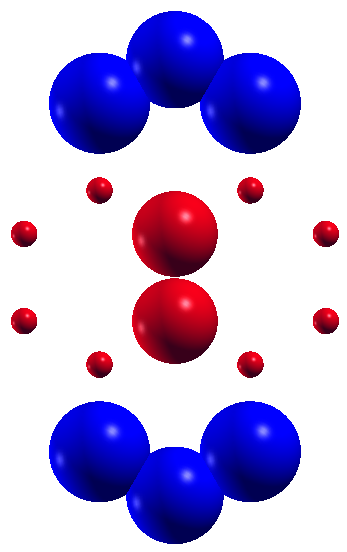}}

\caption{Pictorial representation of the frontier molecular orbitals contributing
significantly to the non-linear optical properties of DQD-16. The
size of a bubble is proportional to the magnitude of the expansion
coefficient at that site (carbon atom), while red/blue colors imply
opposite signs of the coefficients. The symbol H represents the highest
occupied molecular orbital (HOMO).}

\end{figure}

\begin{figure}[H]
\subfloat[H]{\includegraphics[width=2cm]{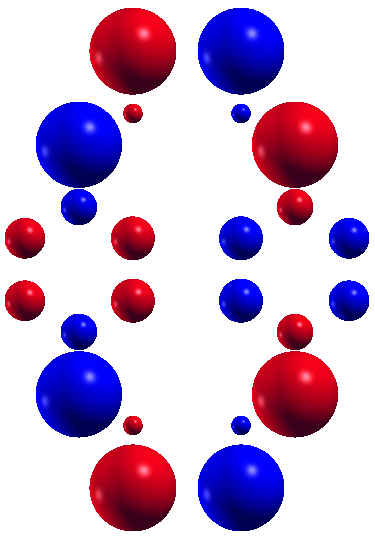}

} \subfloat[H-1]{\includegraphics[width=2cm]{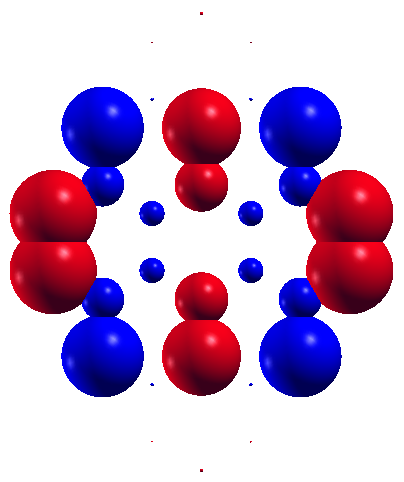}

} \subfloat[H-2]{\includegraphics[width=2cm]{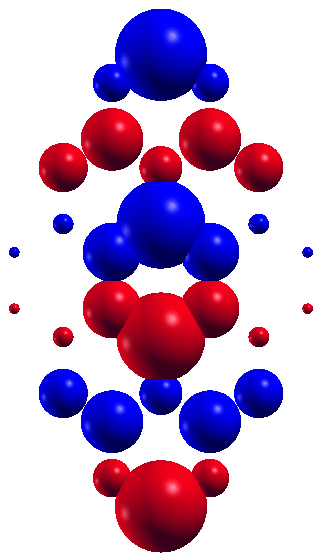}

} \subfloat[H-3]{\includegraphics[width=2cm]{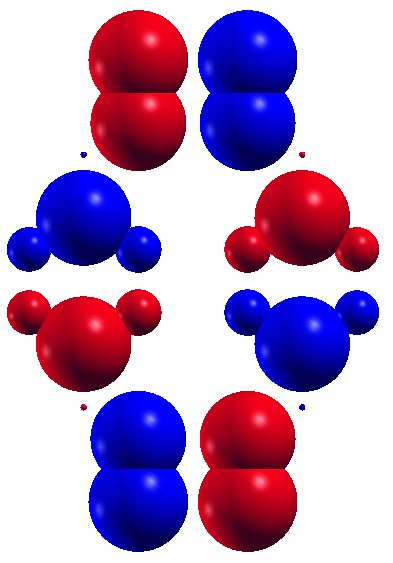}

} \subfloat[H-4]{\includegraphics[width=2cm]{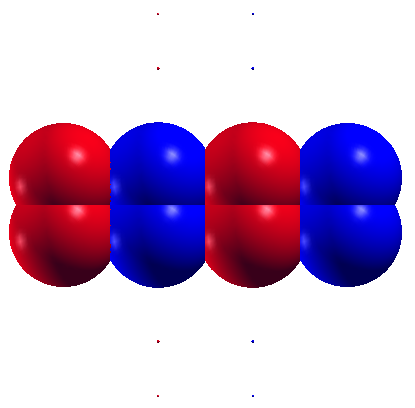}

}\caption{Pictorial representation of the frontier molecular orbitals contributing
significantly to the non-linear optical properties of DQD-30. The
size of a bubble is proportional to the magnitude of the expansion
coefficient at that site (carbon atom), while red/blue colors imply
opposite signs of the coefficients. The symbol H represents the highest
occupied molecular orbital (HOMO).}
\end{figure}

\begin{figure}[H]

\subfloat[H]{\includegraphics[width=2cm]{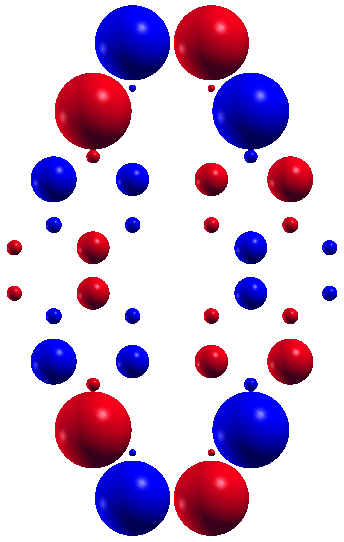}

} \subfloat[H-1]{

\includegraphics[width=2cm]{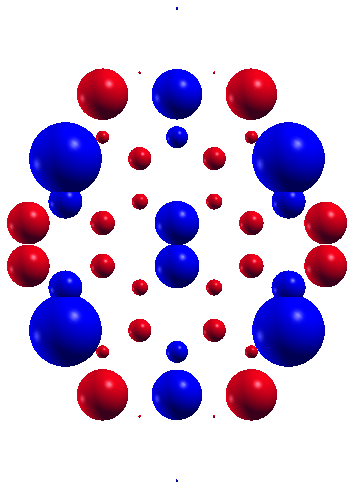}

} \subfloat[H-2]{

\includegraphics[width=2cm]{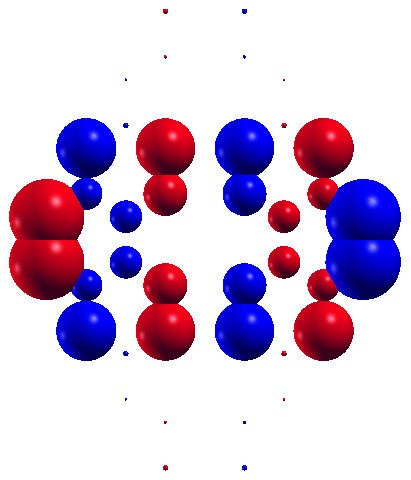}

} \subfloat[H-3]{

\includegraphics[width=2cm]{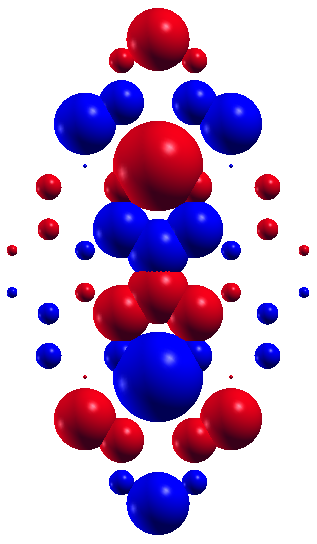}} \subfloat[H-4]{

\includegraphics[width=2cm]{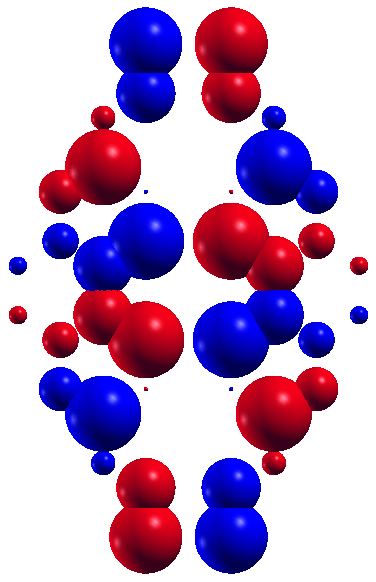}

}\caption{Pictorial representation of the frontier molecular orbitals contributing
significantly to the non-linear optical properties of DQD-48. The
size of a bubble is proportional to the magnitude of the expansion
coefficient at that site (carbon atom), while red/blue colors imply
opposite signs of the coefficients. The symbol H represents the highest
occupied molecular orbital (HOMO).}

\end{figure}

In addition, pictorial representation of a closed-shell restricted
Hartree-Fock wave-function for a general system is presented in Fig.
S4. The mean-field approximation at the the restricted Hartree-Fock
(RHF) level for a DQD consisting of N carbon atoms (DQD-N), results
in N/2 doubly occupied, and N/2 unoccupied (virtual) molecular orbitals
(MOs). A similar graphical representation of configurations contributing
to the calculated wave functions, which incorporate electron correlations
effects at the CI level, is given in Fig. S5. The symbols $H$ and
$L$ represent the highest occupied molecular orbital (HOMO), and
the lowest unoccupied molecular orbital (LUMO), respectively. The
configurations included in the wave function are interpreted as single-,
double-,..., virtual excitations with respect to the restricted Hartree-Fock
wave function, and are denoted using arrows from the occupied to unoccupied
(virtual) orbitals.

\begin{figure}[H]
\includegraphics[width=9cm]{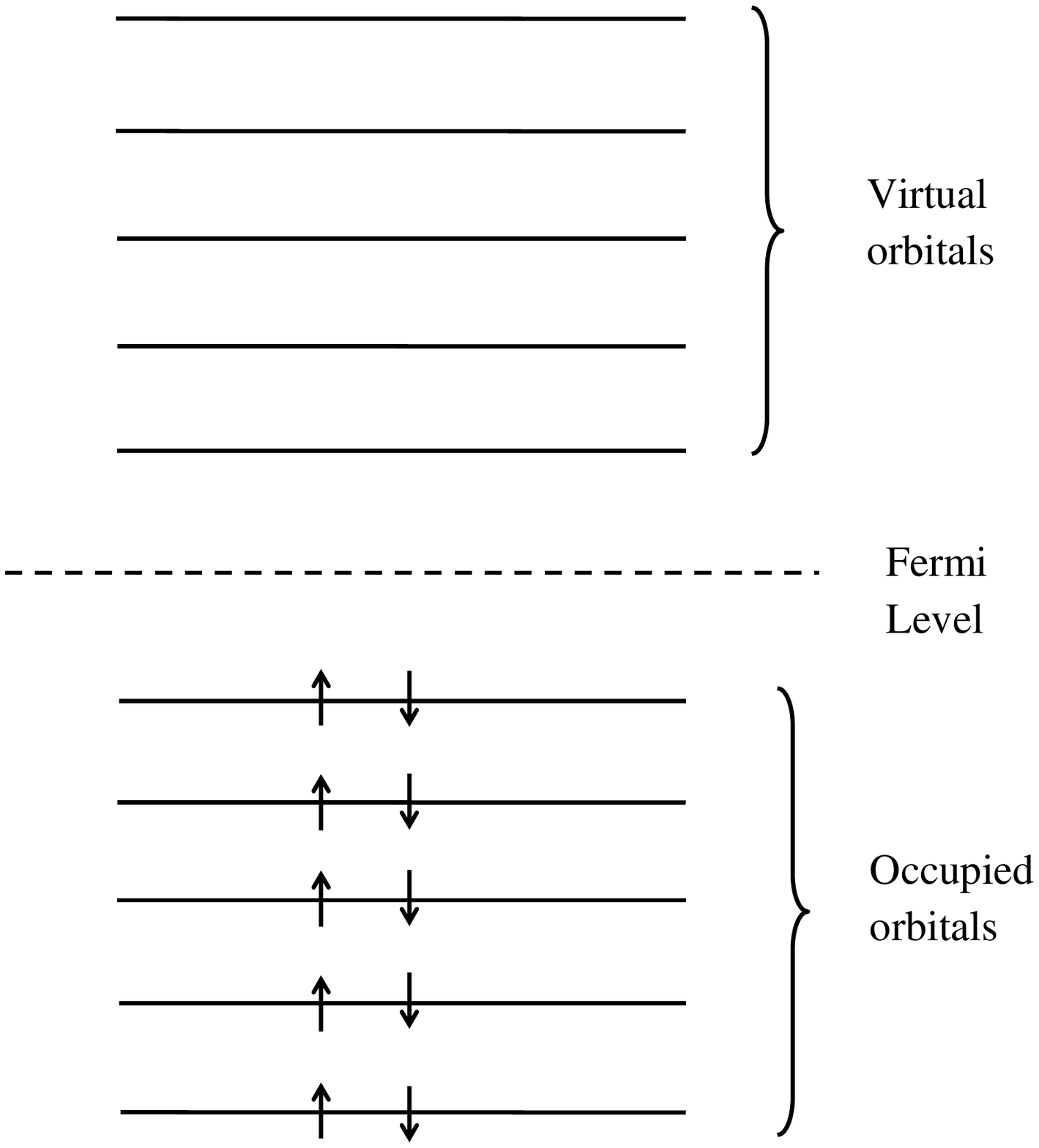}\caption{A pictorial representation of the restricted Hartree-Fock wave function
of a general molecule, consisting of a few doubly-occupied (filled),
and unoccupied (virtual) orbitals. Above, symbol H (L) indicates HOMO
(LUMO), while H-n (L+n) denotes n-th orbital below (above) HOMO (LUMO),
as shown in the figure.}
\end{figure}

\begin{figure}[H]
\subfloat{\includegraphics[width=7cm]{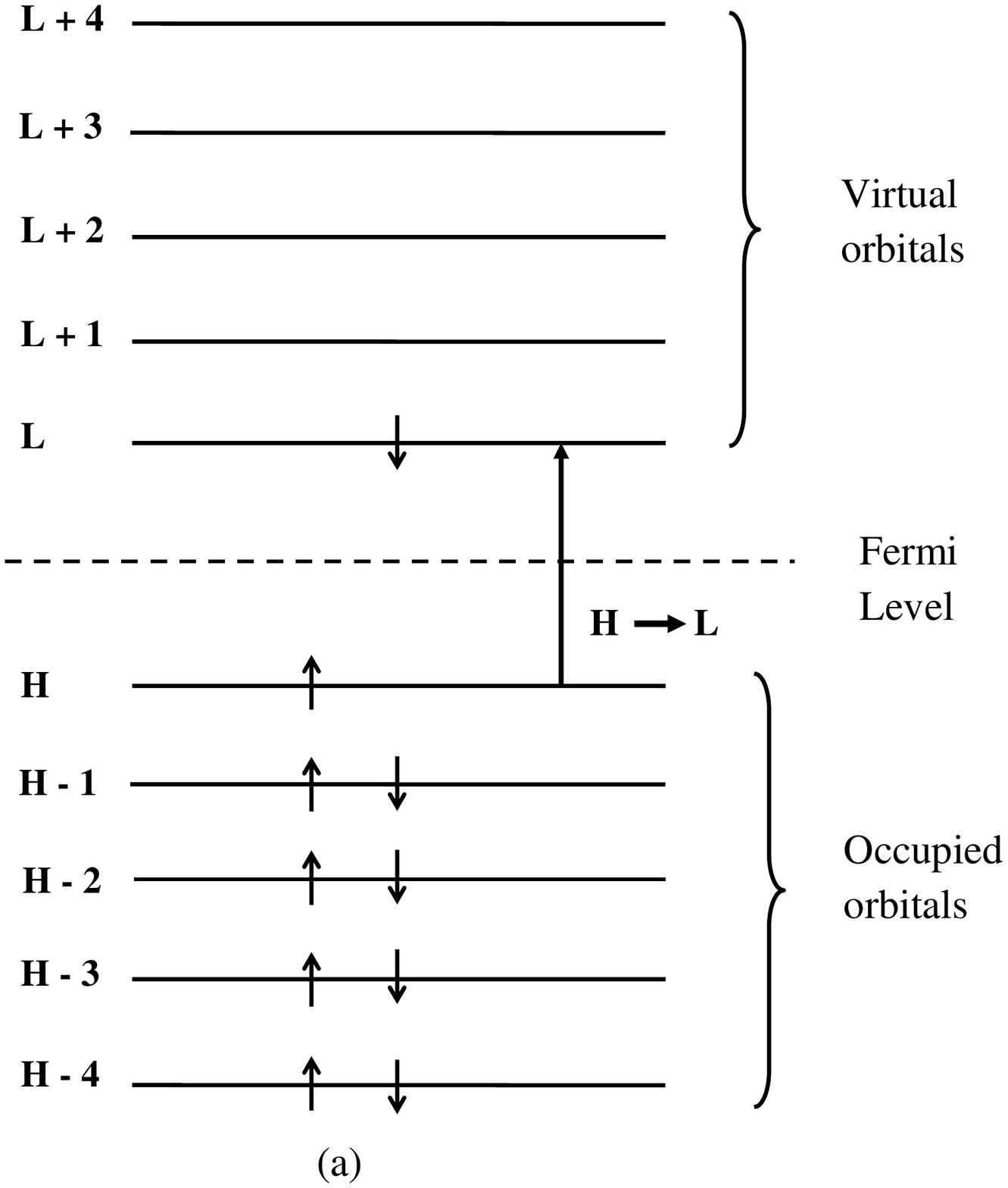}

}\subfloat{\includegraphics[width=7cm]{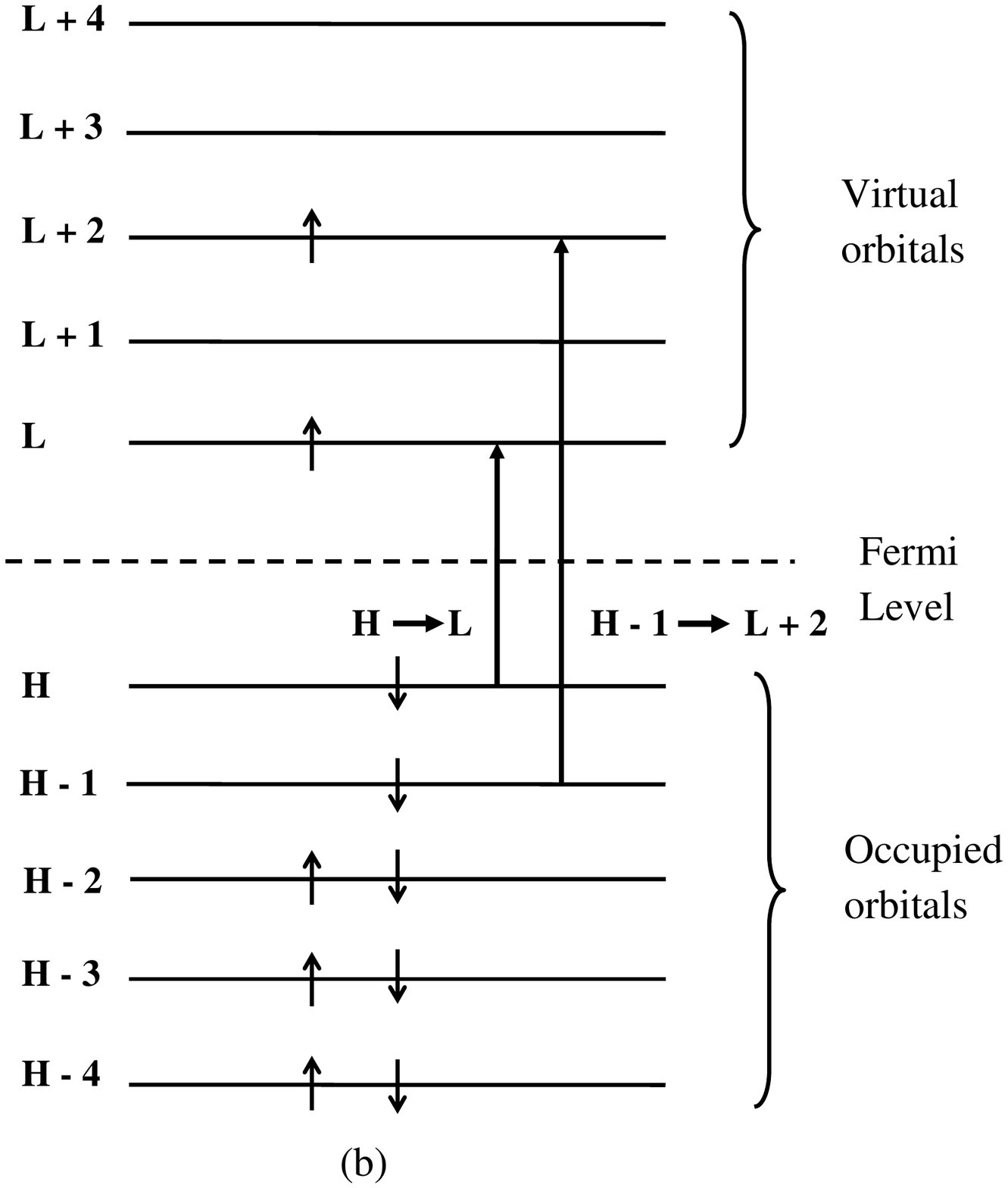}

}

\caption{A graphical representation of typical singly- and doubly-excited configurations
contributing to the CI wave function of a state (ground or excited)
of a molecule: (a) a singly excited configuration obtained by promoting
one electron from the HOMO to the LUMO, denoted as $|H\rightarrow L\rangle$
and (b) a doubly-excited configuration obtained by promoting one electron
from HOMO to LUMO, and another one from HOMO-1 to the LUMO+2 orbital,
and denoted as $|H\rightarrow L;H-1\rightarrow L+2\rangle$. The symbols
$H$ and $L$ represent the HOMO, and the LUMO, respectively. The
arrows represent virtual excitations from the occupied to virtual
orbitals. In reality, the configurations are composed of properly
anti-symmetrized Slater determinants.}

\end{figure}

\section{Dimensions of CI matrices}

In this section, we present the dimensions of the CI matrices employed
for the computation of excited state ordering and two-photon absorption
spectrum of DQD-16, DQD-30 and DQD-48. For excited state ordering
calculation, the first two lowest energy two-photon states ($1A_{g}$
and $2A_{g}$) and first one-photon ($1B_{2u}$) state was optimized
at the quadratic configuration interaction (QCI) level for DQD-16,
while multi-reference singles-doubles configuration interaction (
MRSDCI) methodology was adopted for DQD-30 and DQD-48. The dimensions
of the resultant CI matrices for $A_{g}$ and $B_{2u}$ symmetries
for DQD-16, DQD-30 and DQD-48 are given in Table S1.

\begin{table}[H]
Table S1: Dimensions of the CI matrices for $A_{g}$ and $B_{2u}$
symmetries for excited state ordering computation for DQD-16, DQD-30
and DQD-48.

\begin{tabular}{|c|c|c|}
\hline 
\textcolor{black}{System } & \multicolumn{2}{c|}{\textcolor{black}{Dimension of CI matrix}}\tabularnewline
\cline{2-3} 
 & \textcolor{black}{$A_{g}$ } & \textcolor{black}{$B_{2u}$ }\tabularnewline
\hline 
\textcolor{black}{DQD-16 } & \textcolor{black}{73857 } & \textcolor{black}{126279 }\tabularnewline
\hline 
\textcolor{black}{DQD-30 } & \textcolor{black}{2896621} & \textcolor{black}{2737260}\tabularnewline
\hline 
\textcolor{black}{DQD-48 } & \textcolor{black}{17903219} & \textcolor{black}{12424745}\tabularnewline
\hline 
\end{tabular}

\end{table}

For computing the two-photon absorption (TPA) and photoinduced absorption
(PA) spectra, the maximum energy limit of the excited states was restricted
to 10 eV for CI calculations. The QCI method was used for the $A_{g}$
and the $B_{2u}$ states of DQD-16. The MRSDCI approach was adopted
for the $B_{3u}$ and the $B_{1g}$ symmetries of DQD-16. Further,
the MRSDCI methodology was considered for optimizing all the symmetries
of DQD-30 and DQD-48. The dimensions of CI matrices of $A_{g}$, $B_{1g}$,
$B_{2u}$ and $B_{3u}$ symmetries considered for the computation
of TPA spectra of DQD-16 and DQD-30 are given in the table S2. In
case of the computation of PA spectra of DQD-16, DQD-30 and DQD-48,
the size of the CI matrix for computing the first optically excited
$1B\textrm{\ensuremath{_{2u}}}$ state is also given in the table.

\begin{table}[H]
Table S2: Dimensions of CI matrices of $A_{g}$, $B_{1g}$, $B_{2u}$
and $B_{3u}$ symmetries for the computation of TPA and PA spectra
of DQD-16, DQD-30 and DQD-48.

\textcolor{red}{}%
\begin{tabular}{|c|c|c|c|}
\hline 
\textcolor{black}{Symmetry } & \multicolumn{3}{c|}{\textcolor{black}{Dimension of CI matrix}}\tabularnewline
\cline{2-4} 
 & \textcolor{black}{DQD-16} & \textcolor{black}{DQD-30} & \textcolor{black}{DQD-48}\tabularnewline
\hline 
\textcolor{black}{$A_{g}$ } & \textcolor{black}{73857 } & \textcolor{black}{2274042} & \textcolor{black}{10252792}\tabularnewline
\hline 
\textcolor{black}{$B_{1g}$} & \textcolor{black}{153690} & \textcolor{black}{1803596} & \textcolor{black}{10337444}\tabularnewline
\hline 
\textcolor{black}{$B_{2u}$} & \textcolor{black}{126279} & \textcolor{black}{1564554} & \textemdash{}\tabularnewline
\hline 
\textcolor{black}{$B_{3u}$} & \textcolor{black}{127304} & \textcolor{black}{1359014} & \textemdash{}\tabularnewline
\hline 
$1$$B_{2u}$ & 126279 & 152637 & 702796\tabularnewline
\hline 
\end{tabular}

\end{table}

\section{Computed energies and wave functions of the excited states contributing
to the non- linear absorption spectra}

\textcolor{black}{In this section, the dominant wave-function and
energies of differen}t excited states contributing to $TPA_{xxxx}$,
$TPA_{yyyy}$ and $TPA_{xyxy}$ components of the two photon absorption
spectrum for DQD-16 and DQD-30 are listed in tables S3, S4, S5, S6,
S7 and S8. The calculations have been done at the QCI level for $A_{g}$
and $B_{2u}$ symmetry states, and MRSDCI level for $B_{1g}$ and
$B_{3u}$ states for DQD-16. However, for DQD-30 and DQD-48, the calculations
for all symmetries have been done at the MRSDCI level. The abbreviation
``c.c.'' represents coefficient of charge conjugate of a given singly
excited configuration while the sign (+/\textminus ) preceding \textquotedblleft c.c.\textquotedblright{}
implies that the two coefficients have (same/opposite) signs. In case
of possibility of more than one charge-conjugate counterparts, each
with its own sign, no +/\textminus{} sign precedes c.c. The symbol
$|H\rightarrow L\rangle$ implies singly excited configuration obtained
by promoting one electron from HOMO to LUMO, with respect to the closed-shell
Hartree-Fock reference state. Similarly, one can deduce the meaning
of other singly-excited configurations such as $|H-1\rightarrow L+2\rangle$
etc. The symbol $|H\rightarrow L;H-1\rightarrow L+1\rangle$ denotes
a doubly-excited configuration obtained by exciting two electrons
from the Hartree-Fock reference state, one from HOMO to LUMO, the
other one from HOMO-1 to LUMO+1. Nature of other doubly-excited configurations
can also be deduced, similarly.

\begin{table}[H]
Table S3: Excited states contributing to $TPA_{xxxx}$ component of
the two photon absorption spectrum of DQD-16, computed employing QCI
and MRSDCI approaches. \textcolor{black}{The $2^{1}A\textrm{\ensuremath{_{g}}}$
state does not give rise to a peak in the TPA spectrum. }

\begin{tabular}{|c|c|c|c|}
\hline 
Peak  & State  & E (eV)  & Dominant Configurations\tabularnewline
\hline 
- & $2^{1}A\textrm{\ensuremath{_{g}}}$  & 3.68 & $|H\rightarrow L;H\rightarrow L\rangle$$(0.5109)$\tabularnewline
 &  &  & $|H-3\rightarrow L\rangle-c.c.$$(0.4084)$\tabularnewline
\hline 
$I$ & $3^{1}A\textrm{\ensuremath{_{g}}}$  & $4.65$ & $|H-1\rightarrow L+2\rangle$$-c.c.(0.5294)$ \tabularnewline
 &  &  & $|H\rightarrow L;H-1\rightarrow L+1\rangle$$(0.2088)$ \tabularnewline
\hline 
$II$ & $5^{1}A\textrm{\ensuremath{_{g}}}$  & $5.49$  & $|H-1\rightarrow L+4\rangle$$+c.c.(0.4732)$\tabularnewline
 &  &  & $|H-1\rightarrow L+1;H-1\rightarrow L+1\rangle$$(0.2681)$\tabularnewline
\hline 
$III$ & $8^{1}A\textrm{\ensuremath{_{g}}}$  & $6.18$  & $|H\rightarrow L;H-1\rightarrow L+1\rangle$$(0.5066)$\tabularnewline
 &  &  & $|H\rightarrow L;H-5\rightarrow L\rangle$$+c.c.(0.2352)$\tabularnewline
\hline 
$IV$ & $9^{1}A\textrm{\ensuremath{_{g}}}$  & $6.44$  & $|H-2\rightarrow L+2;H\rightarrow L\rangle$$(0.3801)$\tabularnewline
 &  &  & $|H\rightarrow L;H\rightarrow L\rangle$$(0.2878)$\tabularnewline
 &  &  & $|H\rightarrow L;H-1\rightarrow L+1\rangle$$(0.2773)$\tabularnewline
\hline 
$V$ & $11^{1}A\textrm{\ensuremath{_{g}}}$  & $6.84$  & $|H\rightarrow L;H-2\rightarrow L+2\rangle$$(0.2734)$\tabularnewline
 &  &  & $|H\rightarrow L;H-1\rightarrow L+1\rangle$$(0.2448)$\tabularnewline
 &  &  & $|H-2\rightarrow L;H-2\rightarrow L\rangle$$+c.c.(0.2159)$\tabularnewline
\hline 
$VI$ & $14^{1}A\textrm{\ensuremath{_{g}}}$  & $7.27$  & $|H\rightarrow L;H-2\rightarrow L+2\rangle$$(0.3199)$\tabularnewline
 &  &  & $|H-1\rightarrow L+7\rangle$$+c.c.(0.2022)$\tabularnewline
 &  &  & $|H-2\rightarrow L+1;H-1\rightarrow L+2\rangle$$(0.1882)$\tabularnewline
 &  &  & $|H\rightarrow L;H-1\rightarrow L+1\rangle$$(0.1827)$\tabularnewline
\hline 
$VII$ & $15^{1}A\textrm{\ensuremath{_{g}}}$  & $7.43$  & $|H-1\rightarrow L+7\rangle$$+c.c.(0.3022)$\tabularnewline
 &  &  & $|H-2\rightarrow L+1;H-1\rightarrow L+2\rangle$$(0.2246)$\tabularnewline
 &  &  & $|H\rightarrow L;H-1\rightarrow L+1\rangle$$(0.2121)$\tabularnewline
\hline 
$VIII$ & $24^{1}A\textrm{\ensuremath{_{g}}}$  & $8.43$  & $|H\rightarrow L;H-1\rightarrow L+1\rangle$$(0.2572)$\tabularnewline
 &  &  & $|H\rightarrow L+3;H-3\rightarrow L\rangle$$(0.1979)$\tabularnewline
 &  &  & $|H-2\rightarrow L+1;H-1\rightarrow L+2\rangle$$(0.1932)$\tabularnewline
 &  &  & $|H\rightarrow L;H-4\rightarrow L+4\rangle$$(0.1767)$\tabularnewline
 &  &  & $|H-4\rightarrow L+2;H-2\rightarrow L+4\rangle$$(0.1736)$\tabularnewline
\hline 
$IX$ & $38^{1}A\textrm{\ensuremath{_{g}}}$  & $9.35$ & $|H-2\rightarrow L+2;H-2\rightarrow L+2\rangle$$(0.2020)$\tabularnewline
 &  &  & $|H-2\rightarrow L;H-2\rightarrow L\rangle$$+c.c.(0.1916)$\tabularnewline
 &  &  & $|H\rightarrow L;H-2\rightarrow L+2\rangle$$(0.1770)$\tabularnewline
 &  &  & $|H-2\rightarrow L+2;H-5\rightarrow L\rangle$$-c.c.(0.1683)$\tabularnewline
\hline 
$X$ & $40^{1}A\textrm{\ensuremath{_{g}}}$  & $9.46$ & $|H\rightarrow L;H-4\rightarrow L+4\rangle$$(0.2160)$\tabularnewline
 &  &  & $|H\rightarrow L+3;H-3\rightarrow L\rangle$$(0.2057)$\tabularnewline
 &  &  & $|H\rightarrow L+3;H-1\rightarrow L+4\rangle$$-c.c.(0.1580)$\tabularnewline
 &  &  & $|H-2\rightarrow L+3;H-1\rightarrow L\rangle$$+c.c.(0.1544)$\tabularnewline
\hline 
\end{tabular}
\end{table}

\begin{table}[H]
Table S4: Excited states contributing to $TPA_{yyyy}$ component of
the two photon absorption spectrum of DQD-16, computed employing QCI
and MRSDCI approaches. 

\begin{tabular}{|c|c|c|c|}
\hline 
Peak  & State  & E (eV)  & Dominant Configurations\tabularnewline
\hline 
$I$ & $4^{1}A\textrm{\ensuremath{_{g}}}$  & $5.06$ & $|H\rightarrow L;H\rightarrow L\rangle$$(0.4372)$ \tabularnewline
 &  &  & $|H-2\rightarrow L+2;H\rightarrow L\rangle$$(0.3465)$ \tabularnewline
 &  &  & $|H-3\rightarrow L\rangle$$-c.c.(0.3275)$ \tabularnewline
\hline 
$II$ & $9^{1}A\textrm{\ensuremath{_{g}}}$  & $6.44$  & $|H-2\rightarrow L+2;H\rightarrow L\rangle$$(0.3801)$\tabularnewline
 &  &  & $|H\rightarrow L;H\rightarrow L\rangle$$(0.2878)$\tabularnewline
 &  &  & $|H\rightarrow L;H-1\rightarrow L+1\rangle$$(0.2773)$\tabularnewline
\hline 
$III$ & $11^{1}A\textrm{\ensuremath{_{g}}}$  & $6.84$  & $|H\rightarrow L;H-2\rightarrow L+2\rangle$$(0.2734)$\tabularnewline
 &  &  & $|H\rightarrow L;H-1\rightarrow L+1\rangle$$(0.2448)$\tabularnewline
 &  &  & $|H-2\rightarrow L;H-2\rightarrow L\rangle$$+c.c.(0.2159)$\tabularnewline
\hline 
$IV$ & $13^{1}A\textrm{\ensuremath{_{g}}}$  & $6.98$  & $|H-1\rightarrow L;H-1\rightarrow L\rangle$$+c.c.(0.3367)$\tabularnewline
 &  &  & $|H\rightarrow L;H-1\rightarrow L+1\rangle$$(0.2960)$ \tabularnewline
 &  &  & $|H\rightarrow L;H-2\rightarrow L+2\rangle$$(0.2176)$\tabularnewline
\hline 
\end{tabular}
\end{table}

\begin{table}[H]
Table S5: Excited states contributing to $TPA_{xyxy}$ component of
the two photon absorption spectrum of DQD-16, computed employing QCI
and MRSDCI approaches. 

\begin{tabular}{|c|c|c|c|}
\hline 
Peak  & State  & E (eV)  & Dominant Configurations\tabularnewline
\hline 
$I$ & $10^{1}B\textrm{\ensuremath{_{1g}}}$  & $6.87$ & $|H-2\rightarrow L+5\rangle$$+c.c.(0.3527)$ \tabularnewline
 &  &  & $|H\rightarrow L+7\rangle$$+c.c.(0.2039)$ \tabularnewline
\hline 
$II$ & $12^{1}B\textrm{\ensuremath{_{1g}}}$  & $7.22$  & $|H-4\rightarrow L+5\rangle$$-c.c.(0.2980)$\tabularnewline
 &  &  & $|H\rightarrow L;H-6\rightarrow L\rangle$$+c.c.(0.2031)$\tabularnewline
\hline 
$III$ & $13^{1}B\textrm{\ensuremath{_{1g}}}$  & $7.38$  & $|H\rightarrow L+3;H\rightarrow L+2\rangle$$-c.c.(0.2672)$\tabularnewline
 &  &  & $|H\rightarrow L+2\rangle$$-c.c.(0.1797)$\tabularnewline
$IV$ & $27^{1}B\textrm{\ensuremath{_{1g}}}$  & $8.75$  & $|H-5\rightarrow L+4\rangle$$-c.c.(0.2345)$\tabularnewline
 &  &  & $|H\rightarrow L+2;H-3\rightarrow L\rangle$$c.c.(0.1815)$\tabularnewline
\hline 
\end{tabular}
\end{table}

\begin{table}[H]
Table S6: Excited states contributing to $TPA_{xxxx}$ component of
two photon absorption spectrum of DQD-30, computed employing MRSDCI
approach. 

\begin{tabular}{|c|c|c|c|}
\hline 
Peak  & State  & E (eV)  & Dominant Configurations\tabularnewline
\hline 
$I$ & $3^{1}A\textrm{\ensuremath{_{g}}}$  & $3.32$ & $|H-1\rightarrow L+1;H\rightarrow L\rangle$$(0.3786)$\tabularnewline
 &  &  & $|H\rightarrow L+3\rangle$$-c.c.(0.3669)$\tabularnewline
 &  &  & $|H\rightarrow L;H-7\rightarrow L\rangle$$+c.c.(0.2233)$\tabularnewline
\hline 
$II$ & $4^{1}A\textrm{\ensuremath{_{g}}}$  & $3.76$  & $|H-1\rightarrow L+1;H\rightarrow L\rangle$$(0.4052)$\tabularnewline
 &  &  & $|H-2\rightarrow L+1\rangle$$+c.c.(0.3708)$\tabularnewline
 &  &  & $|H-1\rightarrow L;H-1\rightarrow L\rangle$$+c.c.(0.2293)$\tabularnewline
\hline 
$III$ & $5^{1}A\textrm{\ensuremath{_{g}}}$  & $4.06$  & $|H-4\rightarrow L;H\rightarrow L\rangle$$-c.c.(0.3451)$\tabularnewline
 &  &  & $|H\rightarrow L;H\rightarrow L\rangle$$(0.2952)$\tabularnewline
 &  &  & $|H-1\rightarrow L+1;H\rightarrow L\rangle$$(0.2577)$\tabularnewline
\hline 
$IV$ & $8^{1}A\textrm{\ensuremath{_{g}}}$  & $4.51$  & $|H-10\rightarrow L\rangle$$-c.c.(0.3163)$\tabularnewline
 &  &  & $|H\rightarrow L;H\rightarrow L+7\rangle$$+c.c.(0.2521)$\tabularnewline
\hline 
$V$ & $10^{1}A\textrm{\ensuremath{_{g}}}$  & $4.81$  & $|H-1\rightarrow L+1;H\rightarrow L\rangle$$(0.3394)$\tabularnewline
 &  &  & $|H\rightarrow L;H\rightarrow L+4\rangle$$-c.c.(0.2895)$\tabularnewline
\hline 
$VI$ & $13^{1}A\textrm{\ensuremath{_{g}}}$  & $5.11$  & $|H-2\rightarrow L+5\rangle$$+c.c.(0.2628)$\tabularnewline
 &  &  & $|H-6\rightarrow L+1\rangle$$+c.c.(0.2231)$\tabularnewline
\hline 
$VII$ & $14^{1}A\textrm{\ensuremath{_{g}}}$  & $5.33$  & $|H-4\rightarrow L;H\rightarrow L+4\rangle$$(0.4188)$\tabularnewline
 &  &  & $|H\rightarrow L;H-1\rightarrow L+9\rangle$$+c.c.(0.2556)$\tabularnewline
\hline 
$VIII$ & $16^{1}A\textrm{\ensuremath{_{g}}}$  & $5.43$  & $|H\rightarrow L;H-2\rightarrow L+2\rangle$$(0.3961)$\tabularnewline
 &  &  & $|H-2\rightarrow L+5\rangle$$+c.c.(0.2383)$\tabularnewline
\hline 
$IX$ & $19^{1}A\textrm{\ensuremath{_{g}}}$  & $5.71$  & $|H\rightarrow L;H-2\rightarrow L+2\rangle$$(0.4582)$\tabularnewline
 &  &  & $|H-1\rightarrow L+1;H\rightarrow L\rangle$$(0.1968)$\tabularnewline
\hline 
$X$ & $22^{1}A\textrm{\ensuremath{_{g}}}$  & $5.99$  & $|H-5\rightarrow L+1;H\rightarrow L\rangle$$+c.c.(0.2582)$\tabularnewline
 &  &  & $|H-4\rightarrow L+3\rangle$$-c.c.(0.2561)$\tabularnewline
 &  &  & $|H\rightarrow L;H-2\rightarrow L+2\rangle$$(0.2557)$\tabularnewline
\hline 
$XI$ & $24^{1}A\textrm{\ensuremath{_{g}}}$  & $6.11$  & $|H-1\rightarrow L+1;H\rightarrow L+4\rangle$$+c.c.(0.3335)$\tabularnewline
 &  &  & $|H-1\rightarrow L+1;H-1\rightarrow L+1\rangle$$(0.3331)$\tabularnewline
\hline 
$XII$ & $28^{1}A\textrm{\ensuremath{_{g}}}$  & $6.30$  & $|H-6\rightarrow L+5\rangle$$+c.c.(0.2660)$\tabularnewline
 &  &  & $|H\rightarrow L+2;H\rightarrow L+6\rangle$$-c.c.(0.2158)$\tabularnewline
 & $30^{1}A\textrm{\ensuremath{_{g}}}$  & $6.31$  & $|H\rightarrow L;H-2\rightarrow L+6\rangle$$+c.c.(0.2592)$\tabularnewline
 &  &  & $|H\rightarrow L;H-5\rightarrow L+1\rangle$$+c.c.(0.1626)$\tabularnewline
\hline 
$XIII$ & $32^{1}A\textrm{\ensuremath{_{g}}}$ & $6.50$ & $|H\rightarrow L;H-4\rightarrow L+4\rangle$$(0.2656)$\tabularnewline
 &  &  & $|H-1\rightarrow L+1;H-1\rightarrow L+1\rangle$$(0.2007)$\tabularnewline
\hline 
$XIV$ & $36^{1}A\textrm{\ensuremath{_{g}}}$ & $6.64$ & $|H\rightarrow L;H-2\rightarrow L+6\rangle+c.c.$$(0.2281)$\tabularnewline
 &  &  & $|H-9\rightarrow L+2\rangle+c.c.$$(0.2196)$\tabularnewline
\hline 
\end{tabular}
\end{table}

\begin{table}[H]
Table S7: Excited states contributing to $TPA_{yyyy}$ component of
the two photon absorption spectrum of DQD-30, computed employing MRSDCI
approach. 

\begin{tabular}{|c|c|c|c|}
\hline 
Peak  & State  & E (eV)  & Dominant Configurations\tabularnewline
\hline 
$I$ & $2^{1}A\textrm{\ensuremath{_{g}}}$  & $2.25$ & $|H\rightarrow L;H\rightarrow L\rangle$$(0.6332)$\tabularnewline
 &  &  & $|H\rightarrow L+3\rangle$$-c.c.(0.2902)$\tabularnewline
\hline 
$II$ & $3^{1}A\textrm{\ensuremath{_{g}}}$  & $3.32$  & $|H-1\rightarrow L+1;H\rightarrow L\rangle$$(0.3786)$\tabularnewline
 &  &  & $|H\rightarrow L+3\rangle$$-c.c.(0.3669)$\tabularnewline
 &  &  & $|H\rightarrow L;H-7\rightarrow L\rangle$$+c.c.(0.2233)$\tabularnewline
\hline 
$III$ & $4^{1}A\textrm{\ensuremath{_{g}}}$  & $3.76$  & $|H-1\rightarrow L+1;H\rightarrow L\rangle$$(0.4052)$\tabularnewline
 &  &  & $|H-2\rightarrow L+1\rangle$$+c.c.(0.3708)$\tabularnewline
 &  &  & $|H-1\rightarrow L;H-1\rightarrow L\rangle$$+c.c.(0.2293)$\tabularnewline
\hline 
$IV$ & $5^{1}A\textrm{\ensuremath{_{g}}}$  & $4.06$  & $|H-4\rightarrow L;H\rightarrow L\rangle$$-c.c.(0.3451)$\tabularnewline
 &  &  & $|H\rightarrow L;H\rightarrow L\rangle$$(0.2952)$\tabularnewline
 &  &  & $|H-1\rightarrow L+1;H\rightarrow L\rangle$$(0.2577)$\tabularnewline
\hline 
$V$ & $6^{1}A\textrm{\ensuremath{_{g}}}$  & $4.23$  & $|H-1\rightarrow L+2\rangle$$+c.c.(0.3315)$\tabularnewline
 &  &  & $|H\rightarrow L;H-1\rightarrow L+1\rangle$$(0.2389)$\tabularnewline
 &  &  & $|H\rightarrow L+1;H\rightarrow L+1\rangle$$+c.c.(0.2243)$\tabularnewline
\hline 
$VI$ & $8^{1}A\textrm{\ensuremath{_{g}}}$  & $4.51$  & $|H-10\rightarrow L\rangle$$-c.c.(0.3163)$\tabularnewline
 &  &  & $|H\rightarrow L;H\rightarrow L+7\rangle$$+c.c.(0.2521)$\tabularnewline
\hline 
$VII$ & $17^{1}A\textrm{\ensuremath{_{g}}}$  & $5.62$  & $|H-1\rightarrow L+8\rangle$$+c.c.(0.2370)$\tabularnewline
 &  &  & $|H-4\rightarrow L+3\rangle$$-c.c.(0.1872)$\tabularnewline
 &  &  & $|H-5\rightarrow L;H-1\rightarrow L\rangle$$-c.c.(0.1803)$\tabularnewline
\hline 
$VIII$ & $34^{1}A\textrm{\ensuremath{_{g}}}$  & $6.63$  & $|H-1\rightarrow L+1;H\rightarrow L\rangle$$(0.2317)$\tabularnewline
 &  &  & $|H\rightarrow L+3;H-3\rightarrow L\rangle$$(0.2312)$\tabularnewline
 &  &  & $|H\rightarrow L;H-6\rightarrow L+6\rangle$$(0.2246)$\tabularnewline
 &  &  & $|H-9\rightarrow L+2\rangle$$+c.c.(0.2012)$\tabularnewline
 & $36^{1}A\textrm{\ensuremath{_{g}}}$  & $6.64$  & $|H\rightarrow L;H-2\rightarrow L+6\rangle$$+c.c.(0.2281)$\tabularnewline
 &  &  & $|H-9\rightarrow L+2\rangle$$+c.c.(0.2196)$\tabularnewline
 &  &  & $|H\rightarrow L+2;H\rightarrow L+2\rangle$$+c.c.(0.2058)$\tabularnewline
\hline 
\end{tabular}
\end{table}

\begin{table}[H]
Table S8: Excited states contributing to $TPA_{xyxy}$ component of
the two photon absorption spectrum of DQD-30, computed employing MRSDCI
approach.

\begin{tabular}{|c|c|c|c|}
\hline 
Peak  & State  & E (eV)  & Dominant Configurations\tabularnewline
\hline 
$I$ & $5^{1}B\textrm{\ensuremath{_{1g}}}$  & $4.43$ & $|H\rightarrow L;H-6\rightarrow L\rangle$$+c.c.(0.2754)$\tabularnewline
 &  &  & $|H-9\rightarrow L\rangle$$-c.c.(0.2634)$\tabularnewline
 &  &  & $|H\rightarrow L;H\rightarrow L+2\rangle$$+c.c.(0.2380)$\tabularnewline
\hline 
$II$ & $6^{1}B\textrm{\ensuremath{_{1g}}}$  & $4.58$  & $|H\rightarrow L;H\rightarrow L+6\rangle$$+c.c.(0.3054)$\tabularnewline
 &  &  & $|H\rightarrow L;H\rightarrow L+2\rangle$$+c.c.(0.2953)$\tabularnewline
 &  &  & $|H-3\rightarrow L+2\rangle$$+c.c.(0.1985)$\tabularnewline
\hline 
$III$ & $9^{1}B\textrm{\ensuremath{_{1g}}}$  & $4.86$  & $|H-3\rightarrow L+2\rangle$$+c.c.(0.3156)$\tabularnewline
 &  &  & $|H-1\rightarrow L+4\rangle$$-c.c.(0.2421)$\tabularnewline
 &  &  & $|H\rightarrow L+1\rangle$$-c.c.(0.2372)$\tabularnewline
\hline 
$IV$ & $12^{1}B\textrm{\ensuremath{_{1g}}}$  & $5.21$  & $|H-9\rightarrow L\rangle$$-c.c.(0.3308)$\tabularnewline
 &  &  & $|H-1\rightarrow L;H-2\rightarrow L+1\rangle$$-c.c.(0.1894)$\tabularnewline
\hline 
$V$ & $34^{1}B\textrm{\ensuremath{_{1g}}}$  & $6.65$  & $|H\rightarrow L+1;H-6\rightarrow L+1\rangle$$+c.c.(0.2532)$\tabularnewline
 &  &  & $|H-1\rightarrow L;H-6\rightarrow L+1\rangle$$c.c.(0.1818)$\tabularnewline
 &  &  & $|H-1\rightarrow L+7\rangle$$+c.c.(0.1718)$\tabularnewline
 &  &  & $|H-5\rightarrow L+7\rangle$$+c.c.(0.1564)$\tabularnewline
\hline 
\end{tabular}
\end{table}

\section{Computed energies and wave functions of the excited states contributing
to the photoinduced absorption spectra}

\textcolor{black}{In this section, the dominant wave-function and
energies of the differen}t excited states contributing to the photoinduced
absorption spectra for DQD-16, DQD-30 and DQD-48 are listed in tables
S9, S10 and S11. These tables represent the positions of the peaks
from the $1B\textrm{\ensuremath{_{2u}}}$ state and the dominant wave-functions
of the excited states contributing to these peaks. The abbreviation
``c.c.'' represents coefficient of charge conjugate of a given singly
excited configuration while the sign (+/\textminus ) preceding \textquotedblleft c.c.\textquotedblright{}
implies that the two coefficients have (same/opposite) signs. In case
of possibility of more than one charge-conjugate counterparts, each
with its own sign, no +/\textminus{} sign precedes c.c.

\begin{table}[H]
Table S9: Excited states contributing to the $1B\textrm{\ensuremath{_{2u}}}$
photoinduced absorption spectrum of DQD-16.

\begin{tabular}{|c|c|c|c|}
\hline 
Peak & State & E (eV) & Dominant Configurations\tabularnewline
\hline 
- & $2^{1}A\textrm{\ensuremath{_{g}}}$ &  & \tabularnewline
\hline 
$I_{x}$ & $1^{1}B\textrm{\ensuremath{_{1g}}}$ & $3.78$ & $|H\rightarrow L+2\rangle$$-c.c.(0.5495)$\tabularnewline
 &  &  & $|H\rightarrow L+2;H-3\rightarrow L\rangle$$c.c.(0.1573)$\tabularnewline
\hline 
$II_{x}$ & $2^{1}B\textrm{\ensuremath{_{1g}}}$ & $3.95$ & $|H\rightarrow L+4\rangle$$+c.c.(0.4862)$\tabularnewline
 &  &  & $|H\rightarrow L+1;H\rightarrow L\rangle$$-c.c.(0.2527)$\tabularnewline
\hline 
$III_{x\&y}$ & $4^{1}A\textrm{\ensuremath{_{g}}}$ & $5.06$ & $|H\rightarrow L;H\rightarrow L\rangle$$(0.4372)$\tabularnewline
 &  &  & $|H-2\rightarrow L+2;H\rightarrow L\rangle$$(0.3465)$\tabularnewline
 & $4^{1}B\textrm{\ensuremath{_{1g}}}$ & $5.07$ & $|H-1\rightarrow L+3\rangle$$-c.c.(0.5071)$\tabularnewline
 &  &  & $|H-1\rightarrow L;H\rightarrow L\rangle$$-c.c.(0.2065)$\tabularnewline
\hline 
$IV_{y}$ & $5^{1}A\textrm{\ensuremath{_{g}}}$ & $5.49$ & $|H-1\rightarrow L+4\rangle$$+c.c.(0.4732)$\tabularnewline
 &  &  & $|H-1\rightarrow L+1;H-1\rightarrow L+1\rangle$$(0.2681)$\tabularnewline
\hline 
$V_{x}$ & $6^{1}B\textrm{\ensuremath{_{1g}}}$ & $5.84$ & $|H-1\rightarrow L;H\rightarrow L\rangle$$-c.c.(0.3211)$\tabularnewline
 &  &  & $|H\rightarrow L+3;H\rightarrow L+2\rangle$$-c.c.(0.2153)$\tabularnewline
 &  &  & $|H-4\rightarrow L\rangle$$+c.c.(0.1931)$\tabularnewline
\hline 
$VI_{y}$ & $8^{1}A\textrm{\ensuremath{_{g}}}$ & $6.18$  & $|H\rightarrow L+1;H-1\rightarrow L\rangle$$(0.5066)$\tabularnewline
 &  &  & $|H-5\rightarrow L;H\rightarrow L\rangle$$+c.c.(0.2352)$\tabularnewline
\hline 
$VII_{y}$ & $9^{1}A\textrm{\ensuremath{_{g}}}$ & $6.44$ & $|H-2\rightarrow L+2;H\rightarrow L\rangle$$(0.3801)$\tabularnewline
 &  &  & $|H\rightarrow L;H\rightarrow L\rangle$$(0.2878)$\tabularnewline
\hline 
$VIII_{x}$ & $10^{1}B\textrm{\ensuremath{_{1g}}}$ & $6.87$ & $|H-2\rightarrow L+5\rangle$$+c.c.(0.3527)$\tabularnewline
 &  &  & $|H\rightarrow L+7\rangle$$+c.c.(0.2039)$\tabularnewline
 &  &  & $|H-5\rightarrow L+1;H\rightarrow L\rangle$$c.c.(0.1953)$\tabularnewline
\hline 
$IX_{y}$ & $13^{1}A\textrm{\ensuremath{_{g}}}$ & $6.98$ & $|H-1\rightarrow L;H-1\rightarrow L\rangle$$+c.c.(0.3367)$\tabularnewline
 &  &  & $|H\rightarrow L;H-1\rightarrow L+1\rangle$$(0.2960)$ \tabularnewline
 &  &  & $|H\rightarrow L;H-2\rightarrow L+2\rangle$$(0.2176)$\tabularnewline
\hline 
\end{tabular}
\end{table}

\begin{table}[H]
\begin{tabular}{|c|c|c|c|}
\hline 
Peak & State & E (eV) & Dominant Configurations\tabularnewline
\hline 
\hline 
$X_{x}$ & $13^{1}B\textrm{\ensuremath{_{1g}}}$ & $7.38$ & $|H\rightarrow L+3;H\rightarrow L+2\rangle$$-c.c.(0.2672)$\tabularnewline
 &  &  & $|H\rightarrow L+2\rangle$$-c.c.(0.1797)$\tabularnewline
 &  &  & $|H\rightarrow L+3;H-2\rightarrow L\rangle$$c.c.(0.1546)$\tabularnewline
\hline 
$XI_{x}$ & $16^{1}B\textrm{\ensuremath{_{1g}}}$ & $7.79$ & $|H\rightarrow L+1;H-1\rightarrow L+1\rangle$$-c.c.(0.2406)$\tabularnewline
 &  &  & $|H-4\rightarrow L+1;H\rightarrow L+2\rangle$$c.c.(0.2209)$\tabularnewline
 &  &  & $|H-2\rightarrow L+1;H-4\rightarrow L\rangle$$-c.c.(0.1581)$\tabularnewline
\hline 
$XII_{x}$ & $17^{1}B\textrm{\ensuremath{_{1g}}}$ & $8.02$ & $|H-1\rightarrow L;H-1\rightarrow L+1\rangle$$-c.c.(0.2721)$\tabularnewline
 &  &  & $|H-1\rightarrow L+2;H-2\rightarrow L\rangle$$c.c.(0.2037)$\tabularnewline
 &  &  & $|H\rightarrow L+2;H-3\rightarrow L\rangle$$c.c.(0.1693)$\tabularnewline
\hline 
$XIII_{y}$ & $24^{1}A\textrm{\ensuremath{_{g}}}$ & 8.43 & $|H-1\rightarrow L;H\rightarrow L+1\rangle$$(0.2572)$\tabularnewline
 &  &  & $|H-3\rightarrow L;H\rightarrow L+3\rangle$$(0.1979)$\tabularnewline
 &  &  & $|H-2\rightarrow L+1;H-1\rightarrow L+2\rangle$$(0.1932)$\tabularnewline
\hline 
$XIV_{x\&y}$ & $32^{1}A\textrm{\ensuremath{_{g}}}$ & $8.97$ & $|H-1\rightarrow L+1;H-3\rightarrow L+3\rangle$$(0.1826)$\tabularnewline
 &  &  & $|H\rightarrow L+1;H-3\rightarrow L+4\rangle$$-c.c.(0.1608)$\tabularnewline
 &  &  & $|H-1\rightarrow L+4;H-3\rightarrow L\rangle$$-c.c.(0.1540)$\tabularnewline
\hline 
 & $30^{1}B\textrm{\ensuremath{_{1g}}}$ & $8.97$ & $|H-2\rightarrow L+1;H-4\rightarrow L\rangle$$-c.c.(0.2226)$\tabularnewline
 &  &  & $|H-2\rightarrow L;H\rightarrow L+3\rangle$$c.c.(0.1738)$\tabularnewline
 &  &  & $|H-4\rightarrow L+5\rangle$$-c.c.(0.1180)$\tabularnewline
\hline 
\end{tabular}

\end{table}

\begin{table}[H]

\begin{tabular}{|c|c|c|c|}
\hline 
Peak & State & E (eV) & Dominant Configurations\tabularnewline
\hline 
\hline 
$XV_{y}$ & $33^{1}A\textrm{\ensuremath{_{g}}}$ & $9.04$ & $|H-3\rightarrow L+4;H\rightarrow L+1\rangle$$-c.c.(0.2192)$\tabularnewline
 &  &  & $|H-5\rightarrow L;H\rightarrow L\rangle$$+c.c.(0.1885)$\tabularnewline
\hline 
$XVI_{x\&y}$ & $38^{1}A\textrm{\ensuremath{_{g}}}$ & $9.35$ & $|H-2\rightarrow L+2;H-2\rightarrow L+2\rangle$$(0.2020)$\tabularnewline
 &  &  & $|H-2\rightarrow L;H-2\rightarrow L\rangle$$+c.c.(0.1916)$\tabularnewline
 &  &  & $|H-2\rightarrow L+2;H\rightarrow L\rangle$$(0.1770)$\tabularnewline
\hline 
 & $37^{1}B\textrm{\ensuremath{_{1g}}}$ & $9.38$ & $|H\rightarrow L+2;H\rightarrow L+2;H-2\rightarrow L\rangle$$-c.c.(0.2102)$\tabularnewline
 &  &  & $|H-2\rightarrow L+5\rangle$$+c.c.(0.1817)$\tabularnewline
 &  &  & $|H\rightarrow L;H\rightarrow L;H-5\rightarrow L+2\rangle$$+c.c.(0.1402)$\tabularnewline
\hline 
\end{tabular}

\end{table}

\begin{table}[H]
Table S10: Excited states contributing to the $1B\textrm{\ensuremath{_{2u}}}$
photoinduced absorption spectrum of DQD-30.

\begin{tabular}{|c|c|c|c|}
\hline 
Peak & State & E (eV) & Dominant Configurations\tabularnewline
\hline 
$I_{y}$ & $3^{1}A\textrm{\ensuremath{_{g}}}$ & $3.32$ & $|H-1\rightarrow L+1;H\rightarrow L\rangle$$(0.3786)$\tabularnewline
 &  &  & $|H\rightarrow L+3\rangle-$$c.c.(0.3669)$\tabularnewline
\hline 
$II_{y}$ & $4^{1}A\textrm{\ensuremath{_{g}}}$ & $3.76$ & $|H-1\rightarrow L+1;H\rightarrow L\rangle$$(0.4052)$\tabularnewline
 &  &  & $|H-2\rightarrow L+1\rangle+$$c.c.(0.3708)$\tabularnewline
\hline 
$III_{y}$ & $5^{1}A\textrm{\ensuremath{_{g}}}$ & $4.06$ & $|H-4\rightarrow L;H\rightarrow L\rangle-$$c.c.(0.3451)$\tabularnewline
 &  &  & $|H\rightarrow L;H\rightarrow L\rangle$$(0.2952)$\tabularnewline
\hline 
$IV_{y}$ & $8^{1}A\textrm{\ensuremath{_{g}}}$ & $4.51$ & $|H-10\rightarrow L\rangle-$$c.c.(0.3163)$\tabularnewline
 &  &  & $|H\rightarrow L+7;H\rightarrow L\rangle+$$c.c.(0.2521)$\tabularnewline
 &  &  & $|H\rightarrow L+3;H-3\rightarrow L\rangle$$(0.2076)$\tabularnewline
\hline 
$V_{y}$ & $10^{1}A\textrm{\ensuremath{_{g}}}$ & $4.81$ & $|H-1\rightarrow L+1;H\rightarrow L\rangle$$(0.3394)$\tabularnewline
 &  &  & $|H\rightarrow L+4;H\rightarrow L\rangle-$$c.c.(0.2895)$\tabularnewline
\hline 
$VI_{x}$ & $12^{1}B\textrm{\ensuremath{_{1g}}}$ & $5.21$ & $|H-9\rightarrow L\rangle-$$c.c.(0.3308)$\tabularnewline
 &  &  & $|H-1\rightarrow L;H-2\rightarrow L+1\rangle$$c.c.(0.1894)$\tabularnewline
 &  &  & $|H-11\rightarrow L\rangle+$$c.c.(0.1533)$\tabularnewline
\hline 
\end{tabular}
\end{table}

\begin{table}[H]
\begin{tabular}{|c|c|c|c|}
\hline 
Peak & State & E (eV) & Dominant Configurations\tabularnewline
\hline 
\hline 
$VII_{x}$ & $14^{1}B\textrm{\ensuremath{_{1g}}}$ & $5.46$ & $|H\rightarrow L+11\rangle+$$c.c.(0.2115)$\tabularnewline
 &  &  & $|H-1\rightarrow L;H-2\rightarrow L+1\rangle$$c.c.(0.2016)$\tabularnewline
 &  &  & $|H-1\rightarrow L+7\rangle+$$c.c.(0.1887)$\tabularnewline
\hline 
$VIII_{x\&y}$ & $17^{1}B\textrm{\ensuremath{_{1g}}}$ & $5.70$ & $|H-1\rightarrow L;H\rightarrow L+1;H\rightarrow L+1\rangle$$-c.c.(0.2382)$\tabularnewline
 &  &  & $|H-3\rightarrow L+2\rangle+$$c.c.(0.2271)$\tabularnewline
 &  &  & $|H-1\rightarrow L;H-3\rightarrow L\rangle-$$c.c.(0.2072)$\tabularnewline
 & $19^{1}A\textrm{\ensuremath{_{g}}}$  & $5.71$  & $|H\rightarrow L;H-2\rightarrow L+2\rangle$$(0.4582)$\tabularnewline
 &  &  & $|H-1\rightarrow L+1;H\rightarrow L\rangle$$(0.1968)$\tabularnewline
\hline 
$IX_{y}$ & $28^{1}A\textrm{\ensuremath{_{g}}}$ & $6.30$ & $|H-6\rightarrow L+5\rangle+$$c.c.(0.2660)$\tabularnewline
 &  &  & $|H\rightarrow L+2;H\rightarrow L+6\rangle-$$c.c.(0.2158)$\tabularnewline
 &  &  & $|H-4\rightarrow L;H\rightarrow L+4\rangle$$(0.1843)$\tabularnewline
\hline 
$X_{x\&y}$ & $34^{1}B\textrm{\ensuremath{_{1g}}}$ & $6.65$ & $|H\rightarrow L+1;H-6\rightarrow L+1\rangle+$$c.c.(0.2532)$\tabularnewline
 &  &  & $|H-1\rightarrow L;H-6\rightarrow L+1\rangle$$c.c.(0.1818)$\tabularnewline
 &  &  & $|H-1\rightarrow L+7\rangle+$$c.c.(0.1718)$\tabularnewline
\hline 
 & $34^{1}A\textrm{\ensuremath{_{g}}}$  & $6.63$  & $|H-1\rightarrow L+1;H\rightarrow L\rangle$$(0.2317)$\tabularnewline
 &  &  & $|H\rightarrow L+3;H-3\rightarrow L\rangle$$(0.2312)$\tabularnewline
 &  &  & $|H\rightarrow L;H-6\rightarrow L+6\rangle$$(0.2246)$\tabularnewline
 &  &  & $|H-9\rightarrow L+2\rangle$$+c.c.(0.2012)$\tabularnewline
\hline 
 & $36^{1}A\textrm{\ensuremath{_{g}}}$ & $6.64$ & $|H\rightarrow L;H-2\rightarrow L+6\rangle+$$c.c.(0.2281)$\tabularnewline
 &  &  & $|H-9\rightarrow L+2\rangle+$$c.c.(0.2196)$\tabularnewline
 &  &  & $|H\rightarrow L+2;H\rightarrow L+2\rangle+$$c.c.(0.2058)$\tabularnewline
\hline 
\end{tabular}

\end{table}

\begin{table}[H]
Table S11: Excited states contributing to the $1B\textrm{\ensuremath{_{2u}}}$
photoinduced absorption spectrum of DQD-48.

\begin{tabular}{|c|c|c|c|}
\hline 
Peak  & State  & E (eV)  & Dominant Configurations\tabularnewline
\hline 
$I_{y}$ & $4^{1}A\textrm{\ensuremath{_{g}}}$  & $2.60$  & $|H-1\rightarrow L+1\rangle$$(0.4402)$\tabularnewline
 &  &  & $|H\rightarrow L+1;L\rightarrow L+1\rangle$$+c.c.(0.3217)$\tabularnewline
\hline 
$II_{y}$ & $5^{1}A\textrm{\ensuremath{_{g}}}$  & $2.99$  & $|H\rightarrow L+2\rangle$$-c.c.(0.3116)$\tabularnewline
 &  &  & $|L\rightarrow L+4\rangle$$-c.c.(0.3050)$\tabularnewline
\hline 
$III_{y}$ & $7^{1}A\textrm{\ensuremath{_{g}}}$  & $3.29$  & $|H-1\rightarrow H;H-1\rightarrow L+1;L\rightarrow L+1\rangle$$(0.3717)$\tabularnewline
 &  &  & $|H\rightarrow L+2\rangle$$-c.c.(0.3177)$\tabularnewline
\hline 
$IV_{y}$ & $8^{1}A\textrm{\ensuremath{_{g}}}$  & $3.44$  & $|H-3\rightarrow H;L\rightarrow L+1\rangle$$-c.c.(0.3341)$\tabularnewline
 &  &  & $|H-2\rightarrow L+2\rangle$$(0.2758)$\tabularnewline
\hline 
$V_{y}$ & $11^{1}A\textrm{\ensuremath{_{g}}}$  & $3.64$  & $|H-2\rightarrow L+2\rangle$$(0.2575)$\tabularnewline
 &  &  & $|H-12\rightarrow H\rangle$$+c.c.(0.2432)$\tabularnewline
\hline 
$VI_{y}$ & $14^{1}A\textrm{\ensuremath{_{g}}}$  & $3.95$  & $|H-5\rightarrow H;L\rightarrow L+1\rangle$$-c.c.(0.2566)$\tabularnewline
 &  &  & $|H-1\rightarrow H;H-1\rightarrow L+1;L\rightarrow L+1\rangle$$(0.2385)$\tabularnewline
\hline 
$VII_{y}$ & $16^{1}A\textrm{\ensuremath{_{g}}}$  & $4.19$  & $|H\rightarrow L+8\rangle$$-c.c.(0.3148)$\tabularnewline
 &  &  & $|H-5\rightarrow H;L\rightarrow L+1\rangle$$-c.c.(0.2625)$\tabularnewline
\hline 
$VIII_{y}$ & $22^{1}A\textrm{\ensuremath{_{g}}}$  & $4.58$  & $|H-1\rightarrow H;H-1\rightarrow L+1;L\rightarrow L+1\rangle$$(0.2463)$\tabularnewline
 &  &  & $|H-1\rightarrow L+10\rangle$$-c.c.(0.2103)$\tabularnewline
\hline 
$IX_{y}$ & $27^{1}A\textrm{\ensuremath{_{g}}}$  & $4.85$  & $|H-1\rightarrow L+1\rangle$$(0.3106)$\tabularnewline
 &  &  & $|H-10\rightarrow L+1\rangle$$-c.c.(0.2118)$\tabularnewline
\hline 
$X_{x\&y}$ & $30^{1}B\textrm{\ensuremath{_{1g}}}$  & $4.98$  & $|H-1\rightarrow H;H-8\rightarrow L\rangle$$-c.c.(0.2416)$\tabularnewline
 &  &  & $|H\rightarrow L+1;H-2\rightarrow L+2\rangle$$c.c.(0.1882)$\tabularnewline
 & $32^{1}A\textrm{\ensuremath{_{g}}}$  & $4.98$  & $|H-7\rightarrow L+7\rangle$$(0.4544)$\tabularnewline
 &  &  & $|H-13\rightarrow L+2\rangle$$-c.c.(0.3240)$\tabularnewline
\hline 
$XI_{y}$ & $37^{1}A\textrm{\ensuremath{_{g}}}$  & $5.19$  & $|H-3\rightarrow L+3\rangle$$(0.2206)$\tabularnewline
 &  &  & $|H\rightarrow L+1;L\rightarrow L+7\rangle$$-c.c.(0.2085)$\tabularnewline
\hline 
\end{tabular}
\end{table}